\documentclass[mnsc,nonblindrev]{informs3modified} % current default for manuscript submission

\OneAndAHalfSpacedXI % current default line spacing
\usepackage{endnotes}
\let\footnote=\endnote
\let\enotesize=\normalsize
\def\notesname{Endnotes}%
\def\makeenmark{\hbox to1.275em{\theenmark.\enskip\hss}}
\def\enoteformat{\rightskip0pt\leftskip0pt\parindent=1.275em
  \leavevmode\llap{\makeenmark}}

\usepackage{amsfonts,mathrsfs,bm}
\usepackage{algpseudocode}
\usepackage{algorithm}
\usepackage{subcaption}
\usepackage{multirow}
\usepackage{epstopdf}
\usepackage{color}
\usepackage{accents}
\usepackage{tikz}

\newcommand{\prth}[1]{\left(#1\right)}
\newcommand{\set}[1]{\left\{ #1 \right\}}

\newcommand{\abs}[1]{\left \vert #1 \right\vert}
\newcommand{\brac}[1]{\left[ #1 \right]}
\newcommand{\R}{\mathbb{R}}

\newcommand{\dbhat}[1]{\hat{\vphantom{\rule[4pt]{1pt}{5.5pt}}\smash{\hat{#1}}}}
\newcommand{\dbwidehat}[1]{\widehat{\vphantom{\rule[4pt]{1pt}{5.5pt}}\smash{\widehat{#1}}}}

\graphicspath{{figures/}}
\newcommand{\tabincell}[2]{\begin{tabular}{@{}#1@{}}#2\end{tabular}}

\usepackage{natbib}
 \bibpunct[, ]{(}{)}{,}{a}{}{,}%
 \def\bibfont{\small}%
\TheoremsNumberedThrough     % Preferred (Theorem 1, Lemma 1, Theorem 2)
\ECRepeatTheorems

\EquationsNumberedThrough    % Default: (1), (2), ...
\begin{document}
%%%%%%%%%%%%%%%%

%\setcounter{page}{1}
%\DOI{10.1287/xxxx.0000.0000}%

\RUNAUTHOR{Henry Lam and Huajie Qian}

\RUNTITLE{Optimization-based Input Uncertainty via Empirical Likelihood}

% \TITLE{Bundling Information Goods of Decreasing Value}
\TITLE{Optimization-based Quantification of Simulation Input Uncertainty via Empirical Likelihood}

\ARTICLEAUTHORS{%
\AUTHOR{Henry Lam}
\AFF{Department of Industrial Engineering and Operations Research, Columbia University, New York, NY 10027, \EMAIL{khl2114@columbia.edu}} %, \URL{}}
\AUTHOR{Huajie Qian}
\AFF{Department of Industrial Engineering and Operations Research, Columbia University, New York, NY 10027, \EMAIL{hq2157@columbia.edu}}
} % end of the block

\ABSTRACT{%
We study an optimization-based approach to construct statistically accurate confidence intervals for simulation performance measures under nonparametric input uncertainty. This approach computes confidence bounds from simulation runs driven by probability weights defined on the data, which are obtained from solving optimization problems under suitably posited averaged divergence constraints. We illustrate how this approach offers benefits in computational efficiency and finite-sample performance compared to the bootstrap and the delta method. While resembling robust optimization, we explain the procedural design and develop tight statistical guarantees of this approach via a generalization of the empirical likelihood method.

%We compare our approach with established techniques like the bootstrap and the delta method in terms of variabilities and coverage performances. %also analyze the performances of our simulation optimization algorithm, which provides a computational substitute to the resampling effort used in the common bootstrap schemes. We demonstrate through numerical experiments the benefits of our optimization-based approach in reducing the variability of the constructed confidence intervals relative to the bootstrap.
% Enter your abstract
}%

%\KEYWORDS{deterministic inventory theory; infinite linear programming duality;
%  existence of optimal policies; semi-Markov decision process; cyclic schedule}

 %\HISTORY{This paper was
%first submitted on April 12, 1922 and has been with the authors for
%83 years for 65 revisions.}

\maketitle
%%%%%%%%%%%%%%%%%%%%%%%%%%%%%%%%%%%%%%%%%%%%%%%%%%%%%%%%%%%%%%%%%%%%%%

% Samples of sectioning (and labeling) in OPRE
% NOTE: (1) \section and \subsection do NOT end with a period
%       (2) \subsubsection and lower need end punctuation
%       (3) capitalization is as shown (title style).
%
%\section{Introduction.}\label{intro} %%1.
%\subsection{Duality and the Classical EOQ Problem.}\label{class-EOQ} %% 1.1.
%\subsection{Outline.}\label{outline1} %% 1.2.
%\subsubsection{Cyclic Schedules for the General Deterministic SMDP.}
%  \label{cyclic-schedules} %% 1.2.1
%\section{Problem Description.}\label{problemdescription} %% 2.

% Text of your paper here

\section{Introduction}
Stochastic simulation relies on the propagation of the input variates, through the simulation logic, to generate outputs for decision-making; see, e.g., \cite{banks2005discrete} for an array of applications. Given that in practice the models that govern the input variates are often not fully known but only observed from limited data, the generated simulation outputs can be subject to input errors or uncertainty that adversely affects the decision. Handling this important source of errors has long been advocated and has gathered a fast growth of studies in recent years (see, e.g., the surveys \citealt{barton2012tutorial}, \citealt{henderson2003input}, \citealt{chick2006bayesian}, \citealt{song2014advanced} and \citealt{lam2016tutorial}).

In this paper, we consider the fundamental task of constructing confidence intervals (CIs) for simulation outputs that account for the input uncertainty, in addition to the noises in generating the random variates in the simulation process (known commonly as the stochastic or simulation uncertainty). We focus particularly on the nonparametric regime that makes no assumption on the specific parametric form of the input models. A common approach is the bootstrap (e.g., \citealt{barton1993uniform,barton2001resampling}), which repeatedly generates resampled distributions to drive simulation runs and uses the quantiles of the simulated outputs to construct the CIs. Another approach is the delta method (e.g., \citealt{asmussen2007stochastic}, Chapter III) that estimates the asymptotic variance in the central limit theorem (CLT) directly. The latter has been considered mostly in the parametric setting (e.g., \citealt{cheng1997sensitivity,cheng1998two,cheng2004calculation}) but bears a straightforward analog in our considered nonparametric scenario (as we will illustrate later). Estimating this variance can also be conducted by bootstrapping (e.g., \citealt{cheng1997sensitivity,song2015quickly}).

Our focus in this paper is a new approach to construct input-induced CIs by using optimization as an underpinning tool. Our approach looks for a set of ``maximal" and a set of ``minimal" probability weights on the input data, obtained by solving a pair of convex optimization problems with constraints involving a suitably averaged statistical divergence. These weights can be viewed as ``worst-case" representations of the input distributions which are then used to generate the input variates to drive the simulation, giving rise to upper and lower bounds that together form a CI on the performance measure of interest.

% of the (percentile) bootstrap , for which there is no known rigorous guidance as far as we know at least two aspects. First, the bootstrapis more rigid and  such as higher moments of the considered or other performance measures. They can also be used to
We will illustrate how this optimization-based approach offers benefits relative to the bootstrap and the delta method. The bootstrap typically involves nested simulation due to the resampling step before simulation runs,  which leads to a multiplicative computational requirement that can be substantial. At the same time, its performance can also be sensitive to the simulation budget allocation in the nested procedure. A key element of our approach is to use convex optimization to replace the resampling step. With the tractabiltiy of our optimization problem via standard solvers, our approach offers a lighter computational requirement, and also does not succumb to the multiplicative budget allocation problem. On the other hand, the bootstrap possesses more flexibility as the resampled simulation replications can  be used to approximate many statistics and to construct CIs at different confidence levels, without re-running the bootstrap procedure again. On the contrary, our approach needs re-optimization and a re-evaluation step for each new confidence level or statistic of interest. Nonetheless, we will see that the re-optimization needs only be run once for each problem, while the re-evaluation step only requires a sample size for standard output analysis that is free of input uncertainty.

Our method is closer to the delta method than the bootstrap in that, like the former, we need to estimate gradient information. While our approach and the delta method have similar asymptotic behaviors, we will demonstrate situations where our approach tends to outperform in finite sample. Roughly speaking, this outperformance arises since the delta method relies solely on a linear approximation in constructing CIs, whereas using the weighted distributions to drive simulation runs in our approach can introduce nonlinearity that naturally follows the boundaries of a given problem, which in turn alleviates the under-coverage issue experienced in the delta method. 
% We will see how this benefit is elicited with a small computational overhead.

% prominently when the performance and curb the under-coverage issues often encountered in the delta method, with only a small computational overhead.

% ,  does not require the estimation of gradient information, whereas our approach requires so and in this sense is more closely related to the delta method.
% However, akin to the delta method, our approach needs approximating gradient information. 

As our main technical contributions, we design and analyze procedures to achieve tight statistical coverage guarantees for the resulting optimization-based CIs. Our approach aligns with the recent surge of robust optimization (\citealt{ben2002robust,bertsimas2011theory}) in handling decision-making under uncertainty, where decisions are chosen to perform well under the worst-case scenario among a so-called uncertainty or ambiguity set of possibilities. Our approach particularly resembles distributionally robust optimization (DRO) (e.g., \citealt{ben2013robust,delage2010distributionally,goh2010distributionally,wiesemann2014distributionally}) where the uncertainty of the considered problem lies in the probability distributions, as our involved optimization formulation contains decision variables that are probability weights of the input distributions. However, contrary to the DRO rationale that postulates the uncertainty sets to contain the truth (including those studied recently in the simulation literature; \citealt{hu2012robust,glasserman2014robust,lam2013robust,lam2016serialdependency,ghosh2015computing}), we will explain our procedures by viewing the constraints as log-likelihoods on the input data, and develop the resulting statistical guarantees from a multi-sample generalization of the empirical likelihood (EL) method (\citealt{owen2001empirical}), a nonparametric analog of the celebrated maximum likelihood method in parametric statistics.  Consequently, the form of our proposed constraint (i.e., the averaged statistical divergence constraint) differs drastically from previous DRO suggestions, and the guarantee is provably tight asymptotically. We mention that, though EL has appeared in statistics for a long time, its use in operations research has appeared only recently and is limited to optimization problems (e.g., \citealt{lamzhou2016,dgn2016,lam2016recovering,bk2016,bkm2016}). We therefore contribute by showing that a judicious use of this idea can offer new benefits in the equally important area of simulation analysis.

The rest of this paper is as follows. Section \ref{sec:review} reviews some related literature. Section \ref{sec:EL} presents our procedure and main results on statistical guarantees. Section \ref{sec:theory} explains the underlying theory giving rise to our approach and statistical results.
% Section \ref{sec:bootstrap} discusses some strengths and weaknesses of our approach relative to the bootstrap. 
Section \ref{sec:numerics} shows some numerical results and compares with previous approaches. The Appendix contains all technical proofs.

\section{Related Literature} \label{sec:review}
We briefly survey three areas of related work, one on the problem domain and two on methodologies. The input uncertainty problem in simulation aims to compute CIs or closely related output  variance decompositions. In the parametric case, \cite{cheng1997sensitivity} studies both the delta method and the basic bootstrap for computing the variance due to the input noise. \cite{cheng1998two} and \cite{cheng2004calculation} study the so-called two-point method that reduces the total number of simulation runs in estimating the gradient, or the sensitivity coefficients, in applying the delta method. Under the Bayesian framework, \cite{zouaoui2003accounting} studies the variance decomposition and sampling of posterior output distribution. \cite{barton2013quantifying,xie2014bayesian,xie2016multivariate} further study the construction of CIs built on Gaussian process metamodels. Beyond parametric uncertainty, \cite{chick2001input} and \cite{zouaoui2004accounting} study Bayesian model averaging (BMA) under the choice of several candidate input parametric models. In the nonparametric regime (our focus in this paper), \cite{barton1993uniform,barton2001resampling} propose direct resampling (similar to sectioning; \citealt{asmussen2007stochastic}, Chapter III), bootstrap resampling and the Bayesian bootstrap to construct quantile-based CIs, where they use a single simulation run per bootstrap resample motivated from the overwhelming input noise in their problem setting. \cite{yi2017efficient} studies an approach based on ranking and selection to efficiently allocate budget in bootstrapping quantile estimates. \cite{song2015quickly} studies a mean-variance model to capture the effect of input uncertainty and uses the bootstrap to approximate the input variance component. Finally, some recent work utilizes a risk perspective with respect to model or distributional uncertainty (e.g., \citealt{glasserman2014robust,zhu2015risk,lam2013robust,lam2016serialdependency}).

Our methodologies are related to several tools in statistics. First is the EL method. Initially proposed by \cite{owen1988empirical} as a nonparametric counterpart of the maximum likelihood theory, the EL method has been widely studied in statistical problems like regression and hypothesis testing etc.  (e.g., \citealt{qin1994empirical,owen2001empirical,hjort2009extending}). Its use in operations research is relatively recent and is limited to optimization. \cite{lamzhou2016} investigates the use of EL in quantifying uncertainty in sample average approximation. \cite{lam2016recovering} uses EL to derive uncertainty sets for DRO that guarantees feasibility for stochastic constraints. \cite{dgn2016} generalizes the EL method to Hadamard differentiable functions and obtains tight optimality bounds for stochastic optimization problems. \cite{bk2016,bkm2016} generalize the EL method to inference using the Wasserstein distance. In addition, our work also utilizes the influence function, which captures nonparametric sensitivity information of a statistic, and is first proposed by \cite{hampel1974influence} in the context of robust statistics (\citealt{huber2011robust,hampel2011robust}) as a heuristic tool to measure the effect of data contamination. Influence function is also used in deriving asymptotic results for von Mises differentiable functionals which have profound applications in  $U$-statistics (\citealt{serfling2009approximation}).

Lastly, our approach resembles DRO, which utilizes worst-case perspectives in stochastic decision-making problems under ambiguous probability distributions. In particular, our optimization posited over the space of input probability distributions has a similar spirit as the search for the worst-case distribution in the inner optimization in DRO. The DRO framework has been applied in various disciplines such as economics (\citealt{hansen2008robustness}), finance (\citealt{glasserman2013robust,glasserman2014robust}), stochastic control (\citealt{petersen2000minimax,iyengar2005robust,nilim2005robust,doi:10.1287/moor.1120.0540}), queueing (\citealt{jain2010optimality}) and dynamic pricing (\citealt{lim2007relative}). Among them, constraints in terms of $\phi$-divergences, which include the Burg-entropy divergence appearing in our approach, have been considered in, e.g. \cite{ben2013robust,bayraksan2015data,jiang2012data}, so are other types of statistical distances such as Renyi divergence (e.g., \citealt{atar2015robust,dey2012incorporating,blanchet2016distributionally}) and the Wasserstein distance (e.g., \citealt{esfahani2015data,blanchet2016quantifying,gao2016distributionally}), and other constraint types including moments and support (e.g., \citealt{delage2010distributionally,goh2010distributionally,hu2012robust,wiesemann2014distributionally}). In simulation, the DRO idea has appeared in \cite{glasserman2014robust,lam2013robust,lam2016serialdependency,ghosh2015computing} in quantifying model risks. Nonetheless, although our involved optimization looks similar to DRO, the underpinning statistical guarantees of our approach stem from the EL method. As we will explain, our constraints possess properties that are dramatically different from those studied in DRO, and their precise forms also deviate from any known DRO suggestions.

\section{Optimization-based Confidence Intervals}\label{sec:EL}
This section presents our main procedure and statistical guarantees. We start with our problem setting and some notations.

\subsection{Problem Setting}
We consider a performance measure in the form
\begin{equation}
Z^*=Z(P_1,\ldots,P_m)=\mathbb E_{P_1,\ldots,P_m}\brac{h(\mathbf X_1,\ldots,\mathbf X_m)},\label{perfm}
\end{equation}
where $P_1,\ldots,P_m$ are the distributions governing $m$ independent input models, $\mathbf X_i=(X_{i}(1),\ldots,X_{i}(T_i))$ is a sequence of $T_i$ i.i.d. random variables/vectors each distributed under $P_i$, and $T_i$ is a deterministic run length. The distribution $P_i$ has (possibly multivariate) domain $\mathfrak X_i$. The function $h$ mapping from $\mathfrak X_1^{T_1}\times\cdots\times\mathfrak X_m^{T_m}$ to $\R$ is assumed computable given the inputs $\mathbf X_i$'s. In other words, given the sequence $\mathbf X_1,\ldots,\mathbf X_m$, the value of $h(\mathbf X_1,\ldots,\mathbf X_m)$ can be evaluated by the computer. The notation $\mathbb E_{P_1,\ldots,P_m}[\cdot]$ is a shorthand for $\mathbb E_{P_1^{T_1}\times\cdots\times P_m^{T_m}}[\cdot]$, the expectation taken over all the independent i.i.d. sequences $\mathbf X_1,\ldots,\mathbf X_m$, i.e., under the product measure $P_1^{T_1}\times\cdots\times P_m^{T_m}$. We use $X_i$ to denote a generic random variable/vector distributed under $P_i$.

As a simple example, $\mathbf X_1$ and $\mathbf X_2$ can represent respectively the sequences of inter-arrival times and service times in a queueing system. $P_1$ and $P_2$ represent the corresponding input distributions. $h$ denotes the indicator function of the exceedance of some waiting time above a threshold. Then $Z(P_1,P_2)$ becomes the waiting time tail probability.

Our premise is that there exists a true $P_i$ that is unknown for each $i$, but a sample of $n_i$ i.i.d.~observations $\{X_{i,1},\ldots,X_{i,n_i}\}$ is available from each $P_i$. The true value of \eqref{perfm} is therefore unknown even under abundant simulation runs. Our goal is to find an asymptotically accurate  $(1-\alpha)$-level CI for the true performance measure $Z^*$. To be more precise, we call a CI asymptotically exact if it consists of two numbers $\mathscr L, \mathscr U$, derived from the data and the simulation, such that
\begin{equation*}\label{exact CI}
\lim_{\text{each }n_i\,\text{and}\,R\to\infty} P(\mathscr L\leq Z^*\leq \mathscr U)= 1-\alpha
\end{equation*}
% \,\text{in some way}
where $R$ is the total number of simulation replications involved in generating the CI, and the probability $P$ is taken with respect to the joint randomness in the data and the simulation. The asymptotic above is qualified by certain growth rates of $n_i$ and $R$ that we will detail. 

Along our development will also arise cases in which a coverage guarantee is provided as a lower bound, i.e.,
% In the case where only a correct asymptotic lower bound is guaranteed (e.g., Theorems \ref{erfreeCI2} and \ref{erfreeCI3} in Section \ref{sec:stat guarantees}) for the coverage probability, or formally\,\text{in some way}
\begin{equation*}\label{valid CI}
\liminf_{\text{each }n_i\,\text{and}\,R\to\infty} P(\mathscr L\leq Z^*\leq \mathscr U)\geq 1-\alpha
\end{equation*}
We call $[\mathscr L,\mathscr U]$ an asymptotically valid $(1-\alpha)$-level CI. The CIs constructed from our procedures will be either asymptotically exact or, asymptotically valid and accompanied with an associated upper bound that quantities the tightness of the coverage. Lastly, our developments fix the number of independent input models $m$ and the run lengths $T_i$'s, i.e., we focus primarily on transient performance measures with a moderate number of input models relative to the data and simulation sizes.

% , the true input models $P_i$'s and the performance function $h$, and focuses on the regime where only the number of observations and the simulation effort grow.
% \begin{equation}\label{exact CI}system specifications to be fixed, including 
% \lim_{n\to\infty,R\to\infty} P(\mathscr L\leq Z(P_1,\ldots,P_m)\leq \mathscr U)= 1-\alpha
% \end{equation}
% where $R$ measures the simulation effort involved, and the probability $P$ is taken with respect to the randomness in both the data and the simulation. 

\subsection{Main Procedure}\label{sec:main procedure}
Algorithm \ref{algo1} gives a step-by-step description of our basic procedure for computing $\mathscr L$ and $\mathscr U$. The quantity $\dbhat{G}_i(X_{i,j})$ for each $i=1,\ldots,m,j=1,\ldots,n_i$ introduced in Step 1 is the sample estimate of the so-called influence function of $Z$, which can be viewed as the gradient of $Z$ taken with respect to the input distributions (see Assumption \ref{non-degeneracy} and the subsequent discussion).  This sample estimate of the influence function is obtained from $R_1$ simulation runs.

Step 2 in Algorithm \ref{algo1} outputs a minimizer and a maximizer of the optimization \eqref{ciopt4min} in which ``$\min/\max$" denotes a pair of minimization and maximization, and the calibrating constant $\mathcal X^2_{1,1-\alpha}$ is the $1-\alpha$ quantile of the chi-square distribution with degree of freedom one. Optimization  \eqref{ciopt4min} can be viewed as a sample average approximation (SAA) (\citealt{shapiro2014lectures}) on the influence function (expressible as an expectation), with decision variables being the probability weights $w_{i,j},i=1,\ldots,m,j=1,\ldots,n_i$ on the influence function evaluated at each observation $X_{i,j}$ of input model $i$. For convenience, we denote $\mathbf w_i=(w_{i,j})_{j=1,\ldots,n_i}$ as the weight vector associated with input model $i$, and $\mathbf w=(\mathbf w_i)_{i=1,\ldots,m}$ be the aggregate weight vector.

\begin{algorithm}
   \caption{Basic Empirical-Likelihood-Based Procedure (BEL)}
  \textbf{Input: }Data $\{X_{i,1},\ldots,X_{i,n_i}\}$ for each input model $i=1,\ldots,m$. A target confidence level $1-\alpha$, and numbers of simulation replications, $R_1,2R_2$, to be used in Step 1 and Step 3 respectively.

  \textbf{Procedure: }
  \begin{algorithmic}

\State \textbf{1. Influence Function Estimation:} For each $i=1,\ldots,m,j=1,\ldots,n_i$ compute estimate of the influence function evaluated at $X_{i,j}$ 
\begin{equation}\label{gradientest}
    \hat{\vphantom{\rule[4pt]{1pt}{5.5pt}}\smash{\hat{G}}}_{i}(X_{i,j})=\frac{1}{R_1}\sum_{r=1}^{R_1}\big[(h(\mathbf X_1^r,\ldots,\mathbf X_m^r)-\hat{Z}(\hat P_1,\ldots,\hat P_m))(n_i\sum_{t=1}^{T_i}\mathbf{1}\{X_i^r(t)=X_{i,j}\}-T_i)\big]
\end{equation}
where for each $r=1,\ldots,R_1$, $\mathbf X_i^r=(X_{i}^r(1),\ldots,X_{i}^r(T_i))$ are i.i.d. variates drawn independently from the uniform distribution on $\set{X_{i,1},\ldots,X_{i,n_i}}$ for each $i$, $\mathbf{1}\{\cdot\}$ is the indicator function, and $\hat{Z}(\hat P_1,\ldots,\hat P_m)=\sum_{r=1}^{R_1}h(\mathbf X_1^r,\ldots,\mathbf X_m^r)/R_1$ is the sample mean of the outputs.

\State \textbf{2. Optimization:} Compute respective optimal solutions $(\mathbf w_1^{\text{min}},\ldots,\mathbf w_m^{\text{min}})$ and $(\mathbf w_1^{\text{max}},\ldots,\mathbf w_m^{\text{max}})$ of the following pair of programs
\begin{equation}
\begin{aligned}
&\min/\max &&\sum_{i=1}^m\sum_{j=1}^{n_i}\dbhat{G}_i(X_{i,j})w_{i,j}\\
& \text{subject to} && -2\sum_{i=1}^{m}\sum_{j=1}^{n_i}\log(n_iw_{i,j}) \leq \mathcal X_{1,1-\alpha}^2\\
& && \sum_{j=1}^{n_i}w_{i,j}=1,\text{ for all }i=1,\ldots,m\\
&&&w_{i,j}\geq 0,\text{ for all }i=1,\ldots,m,j=1,\ldots,n_i
\end{aligned}\label{ciopt4min}
\end{equation}

\State \textbf{3. Evaluation:} Compute
$$
\mathscr L^{BEL} =\frac{1}{R_2}\sum_{r=1}^{R_2}h(\mathbf X_1^{r,\min},\ldots,\mathbf X_m^{r,\min}),\ \mathscr U^{BEL} =\frac{1}{R_2}\sum_{r=1}^{R_2}h(\mathbf X_1^{r,\max},\ldots,\mathbf X_m^{r,\max})
$$
where for each $r=1,\ldots,R_2$, $\mathbf X_i^{r,\min}=(X_{i}^{r,\min}(1),\ldots,X_{i}^{r,\min}(T_i))$ and $\mathbf X_i^{r,\max}=(X_{i}^{r,\max}(1),\ldots,X_{i}^{r,\max}(T_i))$ are i.i.d. variates drawn independently from a weighted distribution on  $\set{X_{i,1},\ldots,X_{i,n_i}}$, according to weights $\mathbf w_{i}^{\min}$ and $\mathbf w_{i}^{\max}$, respectively for each $i$.
\end{algorithmic}\label{algo1}
\textbf{Output: }The CI $[\mathscr L^{BEL},\mathscr U^{BEL}]$.
\end{algorithm}

Optimization \eqref{ciopt4min} can be interpreted as two worst-case optimization problems over $m$ independent input distributions, each on support $\{X_{i,1},\ldots,X_{i,n_i}\}$, subject to a weighted average of individual statistical divergences (\citealt{pardo2005statistical}). To explain, the quantity $D_{n_i}(\mathbf w_i)=-(1/n_i)\sum_{j=1}^{n_i}\log(n_iw_{i,j})$ is the Burg-entropy divergence (\citealt{ben2013robust}) (or the Kullback-Leibler (KL) divergence) between the probability weights $\mathbf w_i$ and the uniform weights. Thus, letting $N=\sum_{i=1}^mn_i$ be the total number of observations from all input models, we have
$$-\frac{1}{N}\sum_{i=1}^{m}\sum_{j=1}^{n_i}\log(n_iw_{i,j})=\sum_{i=1}^{m}\frac{n_i}{N}\left(-\frac{1}{n_i}\sum_{j=1}^{n_i}\log(n_iw_{i,j})\right)=\sum_{i=1}^m\frac{n_i}{N}D_{n_i}(\mathbf w_i)$$
which is an average of the Burg-entropy divergences imposed on different input models, each weighted by the proportion of the respective observations, $n_i/N$. The first constraint in \eqref{ciopt4min} can thus be written as
$$\sum_{i=1}^m\frac{n_i}{N}D_{n_i}(\mathbf w_i)\leq\frac{\mathcal X^2_{1,1-\alpha}}{2N}$$
which constitutes a neighborhood ball of size $\mathcal X^2_{1,1-\alpha}/(2N)$ measured by the averaged Burg-entropy divergence.

Finally, Step 3 in Algorithm \ref{algo1} uses the obtained optimal probability weights $\mathbf w_{i}^{\min}$ and $\mathbf w_{i}^{\max}$ to form two weighted empirical distributions on $\{X_{ij}\}_{j=1,\ldots,n_i}$ for input model $i$, which are used to drive two independent sets of simulation runs, each of size $R_2$, in order to output the lower and upper confidence bounds respectively.

An efficient method to solve optimization \eqref{ciopt4min} is discussed in the following proposition:
\begin{proposition}\label{optroutine}
For each $i$ and every $\beta>0$ define $\lambda_i(\beta)$ to be the unique solution of the equation
\begin{equation}\label{eqn:lambda}
    \sum_{j=1}^{n_i}\frac{2\beta}{\hat{\vphantom{\rule[4pt]{1pt}{5.5pt}}\smash{\hat{G}}}_{i}(X_{i,j})+\lambda_i}=1
\end{equation}
on the interval $(-\min_j\dbhat{G}_i(X_{i,j}),\infty)$. Let $\beta^*>0$ solve the equation
\begin{equation}\label{eqn:beta}
    2\sum_{i=1}^{m}\sum_{j=1}^{n_i}\log \frac{2n_i\beta}{\hat{\vphantom{\rule[4pt]{1pt}{5.5pt}}\smash{\hat{G}}}_{i}(X_{i,j})+\lambda_i(\beta)}+\mathcal X_{1,1-\alpha}^2=0.
\end{equation}

If there exist some $i_0\in\{1,\ldots,m\}$ and $j_1,j_2\in\{1,\ldots,n_{i_0}\}$ such that $\dbhat{G}_{i_0}(X_{i_0,j_1})\neq \dbhat{G}_{i_0}(X_{i_0,j_2})$, then $\beta^*\in\big(0,D/\big(2\big(1-e^{-\frac{\mathcal X^2_{1,1-\alpha}}{2N}}\big)\min_in_i\big)\big)$ and is unique, where $D=\max\{\max_j\dbhat{G}_i(X_{i,j})-\min_j\dbhat{G}_i(X_{i,j})\vert i=1,\ldots,m\},N=\sum_{i=1}^mn_i$, and the minimizer $(\mathbf w_1^{\min},\ldots,\mathbf w_m^{\min})$ of \eqref{ciopt4min} can be obtained by
\begin{equation*}
w_{i,j}^{\min}=\frac{2\beta^*}{\hat{\vphantom{\rule[4pt]{1pt}{5.5pt}}\smash{\hat{G}}}_{i}(X_{i,j})+\lambda_i(\beta^*)}.
\end{equation*}
% where $(\beta^*,\lambda_1^*,\ldots,\lambda_m^*)$ is the unique solution, in the domain $\{(\beta,\lambda_1,\ldots,\lambda_m):\beta>0,\lambda_i>-\min_{j} \hat{\vphantom{\rule[4pt]{1pt}{5.5pt}}\smash{\hat{G}}}_{i}(X_{i,j})\text{ for all }i\}$, of the system of equations 
% \begin{equation}\label{kkteqn}
% 2\sum_{i=1}^{m}\sum_{j=1}^{n_i}\log \frac{2n_i\beta}{\hat{\vphantom{\rule[4pt]{1pt}{5.5pt}}\smash{\hat{G}}}_{i}(X_{i,j})+\lambda_i}+\mathcal X_{1,1-\alpha}^2=0,\;\sum_{j=1}^{n_i}\frac{2\beta}{\hat{\vphantom{\rule[4pt]{1pt}{5.5pt}}\smash{\hat{G}}}_{i}(X_{i,j})+\lambda_i}=1\text{ for all }i.
% \end{equation}
% where $\beta^*,\lambda_1^*,\ldots,\lambda_m^*)$ is the unique solution, in the domain $\{(\beta,\lambda_1,\ldots,\lambda_m):\beta>0,\lambda_i>-\min_{j} \hat{\vphantom{\rule[4pt]{1pt}{5.5pt}}\smash{\hat{G}}}_{i}(X_{i,j})\text{ for all }i\}$, of the system of equations 
% \begin{equation}\label{kkteqn}
% 2\sum_{i=1}^{m}\sum_{j=1}^{n_i}\log \frac{2n_i\beta}{\hat{\vphantom{\rule[4pt]{1pt}{5.5pt}}\smash{\hat{G}}}_{i}(X_{i,j})+\lambda_i}+\mathcal X_{1,1-\alpha}^2=0,\;\text{ for all }i.
% \end{equation}
The maximizer $(\mathbf w_1^{\max},\ldots,\mathbf w_m^{\max})$ can be computed in the same way except that each $\dbhat{G}_i(X_{i,j})$ is replaced by $-\dbhat{G}_i(X_{i,j})$.

Otherwise, if for each $i=1,\ldots,m$ the coefficient $\dbhat{G}_i(X_{i,j})$ takes the same value across all $j=1,\ldots,n_i$, then \eqref{ciopt4min} has a constant objective hence becomes trivial.
\end{proposition}
The proof of Proposition \ref{optroutine} uses the Karush-Kuhn-Tucker (KKT) conditions of \eqref{ciopt4min}, and can be found in Section \ref{proof:section 3} of the Appendix. To implement what Proposition \ref{optroutine} suggests, given a value of $\beta$ we can efficiently evaluate each $\lambda_i(\beta)$ by solving \eqref{eqn:lambda} with Newton's method. Then, $\beta^*$ is obtained by running a bisection on \eqref{eqn:beta} over the interval $(0,D/\big(2\big(1-e^{-\frac{\mathcal X^2_{1,1-\alpha}}{2N}}\big)\min_in_i\big)\big)$, and finally each $w_{i,j}^{\min}$ or $w_{i,j}^{\max}$ is computed from $\beta^*$, $\lambda_i(\beta^*)$'s and $\dbhat{G}_i(X_{i,j})$'s. Note that for any $\beta>0$ the left hand side of \eqref{eqn:lambda} is monotonically decreasing and convex in $\lambda_i$, hence Newton's method is guaranteed to converge to $\lambda_i(\beta)$ as long as it starts within $(-\min_j\dbhat{G}_i(X_{i,j}),\lambda_i(\beta))$, say at $2\beta-\min_j\dbhat{G}_i(X_{i,j})$. The advantage of this approach over directly solving the convex optimization \eqref{ciopt4min} is that we reduce the dimension of the decision space, from linear in the sample sizes to only solving univariate equations in \eqref{eqn:lambda} and \eqref{eqn:beta}, which is much more favorable when the sample sizes are large.

% in place of directly dealing with optimization \eqref{ciopt4min}whereas the approach proposed here only requires 

Next we provide two variants of Algorithm \ref{algo1}, depicted as Algorithms \ref{algo2} and \ref{algo3}, which differ only by the last step. 
% are variants of only in the last step so we simply highlight the differences to avoid repetition.
The motivation (with more details in Section \ref{sec:evaluation}) is that Algorithm \ref{algo1} tends to under-cover the true performance value because its last step only outputs the sample mean of the simulation replications and does not take full account of the stochastic uncertainty. Algorithm \ref{algo2} takes care of this uncertainty by outputting the standard normal lower and upper confidence bounds in the last step. However, this simple adjustment does not account for the joint variances from the input data and the  stochasticity in a tight manner, and tends to generate conservative CIs that over-cover the truth. This motivates the refined adjustment in Algorithm \ref{algo3} that is designed to match the CI inflation from combined input and stochastic uncertainties, by taking into account the asymptotic form of the joint variance, and subsequently leads to accurate coverage performances. The $\hat{\sigma}_I^2$ in Algorithm \ref{algo3} estimates the input-induced variance. In the expression of $\hat\sigma_I^2$, the sample variance $\sum_{j=1}^{n_i}\big(\dbhat{G}_i(X_{i,j})\big)^2/n_i$ for input model $i$ is upward biased due to the simulation noise in each $\dbhat{G}_i(X_{i,j})$, which is removed by introducing the term $n_iT_i\hat{\sigma}^2/R_1$. The positive-part operation is to handle small $R_1$ situations where such a variance estimate could yield negative values due to the bias correction, in which case we reset it to zero.
\begin{algorithm}
   \caption{Evaluation-Adjusted Empirical Likelihood (EEL)}
  \begin{algorithmic}
\State Follow Algorithm \ref{algo1} until Step 3. Replace Step 3 by
$$
\mathscr L^{EEL} =\hat{Z}^{\min}-z_{1-\alpha/2}\frac{\hat{\sigma}_{\min}}{\sqrt{R_2}},\;\mathscr U^{EEL} =\hat{Z}^{\max}+z_{1-\alpha/2}\frac{\hat{\sigma}_{\max}}{\sqrt{R_2}}
$$
where
$$
\hat{Z}^{\min}=\frac{1}{R_2}\sum_{r=1}^{R_2}h(\mathbf X_1^{r,\min},\ldots,\mathbf X_m^{r,\min}),\ \hat{\sigma}_{\min}^2=\frac{1}{R_2-1}\sum_{r=1}^{R_2}(h(\mathbf X_1^{r,\min},\ldots,\mathbf X_m^{r,\min})-\hat{Z}^{\min})^2
$$
are the sample mean and variance of the $R_2$ simulation runs driven by distributions on  $\set{X_{i,1},\ldots,X_{i,n_i}}$ with weights $\mathbf w_1^{\min},\ldots,\mathbf w_m^{\min}$, and $\hat{Z}^{\max},\hat{\sigma}_{\max}^2$ are defined accordingly. $z_{1-\alpha/2}$ is the $1-\alpha/2$ quantile of the standard normal.
\end{algorithmic}\label{algo2}
\textbf{Output: }The CI $[\mathscr L^{EEL},\mathscr U^{EEL}]$.
\end{algorithm}

\begin{algorithm}
   \caption{Fully Adjusted Empirical Likelihood (FEL)}
  \begin{algorithmic}
  
\State  Follow Algorithm \ref{algo1} until Step 3. Replace Step 3 by
$$
\mathscr L^{FEL} =\hat{Z}^{\min}-z_{1-\alpha/2}\Big(\sqrt{\hat{\sigma}_I^2+\frac{\hat{\sigma}_{\min}^2}{R_2}}-\hat{\sigma}_I\Big),\;\mathscr U^{FEL} =\hat{Z}^{\max}+z_{1-\alpha/2}\Big(\sqrt{\hat{\sigma}_I^2+\frac{\hat{\sigma}_{\max}^2}{R_2}}-\hat{\sigma}_I\Big)
$$
where $z_{1-\alpha/2},\hat{Z}^{\min},\hat{\sigma}_{\min}^2,\hat{Z}^{\max},\hat{\sigma}_{\max}^2$ are the same as in Algorithm \ref{algo2}, and
\begin{equation}\label{input_var_est}
\hat{\sigma}_I^2=\max\Big\{\sum_{i=1}^m\frac{1}{n_i}\Big[\sum_{j=1}^{n_i}\frac{\big(\dbhat{G}_i(X_{i,j})\big)^2}{n_i}-\frac{n_iT_i\hat{\sigma}^2}{R_1}\Big],0\Big\},\text{ with }\hat{\sigma}^2=\frac{1}{R_1-1}\sum_{r=1}^{R_1}(h(\mathbf X_1^r,\ldots,\mathbf X_m^r)-\hat{Z})^2
\end{equation}
is computed from the $R_1$ replications generated in Step 1.
\end{algorithmic}\label{algo3}
\textbf{Output: }The CI $[\mathscr L^{FEL},\mathscr U^{FEL}]$.
\end{algorithm}

\subsection{Statistical Guarantees}\label{sec:stat guarantees}
We present statistical guarantees of Algorithms \ref{algo1}, \ref{algo2} and \ref{algo3}. We assume the following:
\begin{assumption}\label{balanced data}
There exist constants $0<\underline{c},\overline{c}<\infty$ such that $\underline{c}\leq \frac{n_i}{n}\leq \overline{c}$ for all $i=1,\ldots,m$ as all $n_i\to\infty$, where $n=\frac{1}{m}\sum_{i=1}^mn_i$ is the averaged data size.
\end{assumption}
Assumption \ref{balanced data} postulates that data sizes across different input models grow at the same rate. For convenience, we shall use the averaged size $n$ to represent the overall scale of the data size throughout the paper.
\begin{assumption}\label{non-degeneracy}
At least one of $\mathrm{Var}(G_i(X_i)),i=1,\ldots,m$ is non-zero, where
$$G_i(x)=\sum_{t=1}^{T_i}\mathbb E_{P_1,\ldots,P_m}[h(\mathbf X_1,\ldots,\mathbf X_m)\vert X_{i}(t)=x]-T_iZ(P_1,\ldots,P_m).$$
\end{assumption}
\begin{assumption}\label{8th moment}
For each $i$ let $I_{i}=(I_{i}(1),\ldots,I_{i}(T_i))$ be a sequence of indices such that $1\leq I_i(t)\leq T_i$, and $\mathbf X_{i,I_i}=\prth{X_{i}(I_i(1)),\ldots,X_{i}(I_i(T_i))}$. Assume $\mathbb E_{P_1,\ldots,P_m}[\abs{h(\mathbf X_{1,I_1},\ldots,\mathbf X_{m,I_m})}^8]$ is finite for all such $I_{i}$'s.
\end{assumption}

The function $G_i(x)$ in Assumption \ref{non-degeneracy} is the influence function (\citealt{hampel1974influence,hampel2011robust}) of the performance measure $Z(P_1,\ldots,P_m)$ with respect to the input distribution $P_i$, which measures the infinitesimal effect caused by perturbing $P_i$ and represents the Gateaux derivative of $Z$ in the sense
\begin{equation}
\frac{d}{d\epsilon}Z(P_1,\ldots,P_{i-1},(1-\epsilon)P_i+\epsilon Q_i,P_{i+1},\ldots,P_m)\Big|_{\epsilon=0^+}=\int G_i(x)dQ_i(x)\label{Gateaux derivative}
\end{equation}
for any distribution $Q_i$ on $\mathfrak X_i$. Assumption \ref{non-degeneracy} entails that at least one of the influence functions is non-degenerate at the true input distributions $P_i$'s, or in other words, at least one of these distributions would exert a first-order effect on the performance measure. This assumption is essential in ensuring a normality asymptotic for the output performance measure. In lack of this assumption, the output performance measure will satisfy a $\chi^2$ or even higher-order asymptotic behavior as the input data size grows, which has never been observed in the simulation literature to our best knowledge (the parametric analog of this would be to say that the first-order sensitivities to all input parameters are zero). 
% We note that, while this assumption may look obscure, it has been implicitly made in essentially all frequentist approaches in input uncertainty quantification (e.g., \cite{for each input model $i$ and observation $X_{i,j}$

Note that the $\dbhat{G}_i(X_{i,j})$ in Step 1 of Algorithm \ref{algo1} is a sample version of $G_i(X_{i,j})$. Assumption \ref{8th moment} is a moment condition that, as we will see, controls the magnitude of the linearization error in Step 2 and the simulation error in Steps 1 and 3 of our algorithms. It holds if, for instance, $h$ is bounded.

We have the following statistical guarantees in using the three proposed algorithms to construct input-induced CIs:
% , whose proofs are deferred to Section \ref{proof:section 3} of the Appendix:
\begin{theorem}\label{erfreeCI1}
Suppose Assumptions \ref{balanced data}, \ref{non-degeneracy} and \ref{8th moment} hold. If the simulation sizes $R_1,R_2$ are chosen such that $\frac{R_1}{n}\to\infty,\frac{R_2}{n}\to\infty$, then the outputs $\mathscr L^{BEL},\mathscr U^{BEL}$ of Algorithm \ref{algo1} constitute an asymptotically exact $(1-\alpha)$-level CI, i.e.,
%\begin{description}
%\item[1.]\,\text{and}\
\begin{equation}\label{exactness_algo1}
\lim_{n,R_1,R_2\to \infty:\ \frac{R_1}{n}\to\infty, \frac{R_2}{n}\to\infty}P\prth{\mathscr L^{BEL}\leq Z^*\leq \mathscr U^{BEL}}= 1-\alpha.
\end{equation}
\end{theorem}
\begin{theorem}\label{erfreeCI2}
Suppose Assumptions \ref{balanced data}, \ref{non-degeneracy} and \ref{8th moment} hold. If the simulation sizes $R_1,R_2$ are chosen such that $\frac{R_1}{n}\to\infty,\frac{R_2}{n}\leq M$ for some constant $M>0$, then the outputs $\mathscr L^{EEL},\mathscr U^{EEL}$ of Algorithm \ref{algo2} constitute an asymptotically valid $(1-\alpha)$-level CI, i.e.,
\begin{align*}
\liminf_{n,R_1,R_2\to \infty:\ \frac{R_1}{n}\to\infty, \frac{R_2}{n}\,\text{bounded}}P\prth{\mathscr L^{EEL}\leq Z^*\leq \mathscr U^{EEL}}&\geq 1-\alpha\\
\limsup_{n,R_1,R_2\to \infty:\ \frac{R_1}{n}\to\infty, \frac{R_2}{n}\,\text{bounded}}P\prth{\mathscr L^{EEL}\leq Z^*\leq \mathscr U^{EEL}}&\leq 1-\tilde\alpha+\frac{\tilde\alpha^2}{4}
\end{align*}
where $1-\frac{\tilde\alpha}{2} = \Phi(\sqrt{2}z_{1-\alpha/2})$ with $\Phi$ being the distribution function of the standard normal. Moreover, if $\frac{R_2}{n}\to\infty$ like in Theorem \ref{erfreeCI1}, then the CI is asymptotically exact, i.e., \eqref{exactness_algo1} holds for $\mathscr L^{EEL},\mathscr U^{EEL}$.
\end{theorem}
\begin{theorem}\label{erfreeCI3}
% Under the same assumptions and the same conditions on $R_1,R_2$ as in Theorem \ref{erfreeCI2}, 
Suppose Assumptions \ref{balanced data}, \ref{non-degeneracy} and \ref{8th moment} hold. If the simulation sizes $R_1,R_2$ are chosen such that $\frac{R_1}{n}\to\infty,\frac{R_2}{n}\leq M$ for some constant $M>0$, then the outputs $\mathscr L^{FEL},\mathscr U^{FEL}$ of Algorithm \ref{algo3} constitute an asymptotically valid $(1-\alpha)$-level CI, i.e.,
\begin{align*}
\liminf_{n,R_1,R_2\to \infty:\ \frac{R_1}{n}\to\infty, \frac{R_2}{n}\,\text{bounded}}P\prth{\mathscr L^{FEL}\leq Z^*\leq \mathscr U^{FEL}}&\geq 1-\alpha\\
\limsup_{n,R_1,R_2\to \infty:\ \frac{R_1}{n}\to\infty, \frac{R_2}{n}\,\text{bounded}}P\prth{\mathscr L^{FEL}\leq Z^*\leq \mathscr U^{FEL}}&\leq 1-\alpha+\frac{\alpha^2}{4}.
\end{align*}
Moreover, if $\frac{R_2}{n}\to\infty$ like in Theorem \ref{erfreeCI1}, then the CI is asymptotically exact, i.e., \eqref{exactness_algo1} holds for $\mathscr L^{FEL},\mathscr U^{FEL}$.
\end{theorem}

Theorem \ref{erfreeCI1} states that Algorithm \ref{algo1} generates an asymptotically exact CI for the true performance measure, when the simulation budgets available to both Step 1 and Step 3 dominate the data size.
% These errors consist of three terms.
% The first term corresponds to the combination of input errors with the approximation of the performance measure using its influence function. The second term corresponds to the sample average approximation undertaken in Steps 1 and 2. The third term corresponds to the simulation errors from the final evaluation in Step 3.
Theorems \ref{erfreeCI2} and \ref{erfreeCI3} show that in Algorithms \ref{algo2} and \ref{algo3} the simulation effort for Step 3 can be reduced to grow independent of the data size. This is thanks to the adjustment in the evaluation of the confidence bounds that accounts for the stochastic uncertainty in Step 3. The CI from Algorithm \ref{algo2} tends to be conservative and can over-cover the truth with a level of $1-\tilde\alpha+\tilde\alpha^2/2$. To get a sense of this conservativeness, when the desired coverage level $1-\alpha=90\%$, the guaranteed level can be as high as $1-\tilde{\alpha}+\tilde\alpha^2/2\approx 98\%$. On the other hand, the further refinement in Algorithm \ref{algo3} is able to recover the exact coverage up to an error of $\alpha^2/4$, which is negligible for most purposes (e.g., when $\alpha=5\%$, $\alpha^2/4=0.0625\%$).

\section{Theory on Statistical Guarantees}\label{sec:theory}
This section further elaborates on Algorithms \ref{algo1}, \ref{algo2} and \ref{algo3}, and explains the underlying theories leading to Theorems \ref{erfreeCI1}, \ref{erfreeCI2} and \ref{erfreeCI3}. Section \ref{sec:DRO to input} starts with an initial interpretation of our approach from a distributionally robust optimization (DRO) perspective. The subsequent subsections then discuss the guarantees in several steps.  Section \ref{sec:linearization} first presents a linear approximation on the performance measures to bypass some statistical and computational bottlenecks. Sections \ref{sec:EL theorem} and \ref{sec:EL-based CI} develop the EL method for the linearized problem and CI construction. Section \ref{sec:nonconvexity} incorporates the simulation errors. Lastly, Section \ref{sec:evaluation} discusses the last evaluation steps in our procedures and links them to the conclusions of Theorems  \ref{erfreeCI1}, \ref{erfreeCI2} and \ref{erfreeCI3}.

%  to Input Uncertainty Quantification
\subsection{An Initial Interpretation from DRO}\label{sec:DRO to input}
On a high level, our algorithms in Section \ref{sec:main procedure} can be interpreted as attempting to solve the following problem. Given the observations $\{X_{i,1},\ldots,X_{i,n_i}\}$ for input model $i$, we consider the weighted empirical distribution $(1/n_i)\sum_{j=1}^{n_i}w_{i,j}\delta_{X_{i,j}}(x)$, where $\delta_{X_{i,j}}$ denotes the delta measure on $X_{i,j}$. Slightly abusing notations to denote $Z(\mathbf w_1,\ldots,\mathbf w_m)$ as the performance measure evaluated at these weighed distributions, we consider
\begin{equation}\label{original_pair}
\begin{aligned}
\mathscr L/\mathscr U:=&\min/\max &&Z(\mathbf w_1,\ldots,\mathbf w_m)\\
& \text{subject to} && \mathbf w\in\mathcal U_\alpha
\end{aligned}
\end{equation}
where \begin{equation}\label{uncertainty set}
    \mathcal U_{\alpha}=\set{(\mathbf w_1,\ldots,\mathbf w_m)\in \R^N\Bigg\vert\begin{array}{l} % or pmatrix or bmatrix or Bmatrix or ...
       -2\sum_{i=1}^{m}\sum_{j=1}^{n_i}\log(n_iw_{i,j}) \leq \mathcal X_{1,1-\alpha}^2 \\
      \sum_{j=1}^{n_i}w_{i,j}=1,\text{ for all }i=1,\ldots,m \\
      w_{i,j}\geq 0,\text{ for all }i,j
   \end{array}}
\end{equation}
This problem resembles DRO, which is a special class of robust optimization whose uncertainty is on the probability distribution. More specifically, robust optimization considers decision-making under uncertainty or ambiguity of the underlying parameters, and advocate optimizing the objective under the worst-case scenario, where the worst-case is over all parameters within the so-called uncertainty set or ambiguity set. In DRO, the uncertain quantities are the probability distributions that govern a stochastic optimization, so that the uncertainty set lies in the space of distributions. From this view, optimization \eqref{original_pair} calculates the worst-case performance measure subject to the uncertainty set $\mathcal U_\alpha$. In particular, as discussed in Section \ref{sec:main procedure}, the constraint in \eqref{uncertainty set} resembles an averaged Burg-entropy divergence, comprising of $m$ terms each being the divergence between the distribution weighted by $\mathbf w_i$ and the uniform distribution, on the support generated by the empirical data $\{X_{i,j}\}_{j=1,\ldots,n_i}$. 

Despite this Burg-entropy divergence interpretation that ties the optimal weights in \eqref{original_pair} to ``worst-case" distributions, the conceptual reasoning of $\mathcal U_\alpha$ that we present below is fundamentally different from DRO, the latter advocates the use of uncertainty sets that contain the true distribution with a certain confidence. To this end, a divergence ball used as an uncertainty set must use a ``baseline" distribution that is absolutely continuous to the true distribution, in order to have an overwhelming (or at least non-zero) probability of containing the truth (\citealt{jiang2012data,esfahani2015data}). This condition is violated in formulation \eqref{original_pair} when the true input distribution is continuous. As the baseline distribution in our divergence (namely the empirical distribution) is supported only on the data, the resulting ball does not contain any continuous distributions. Moreover, the use of weighted average and its particular weights put on each of these empirically defined divergences is also an unnatural choice from a DRO perspective. 

Thus, instead of arguing the statistical behaviors of \eqref{original_pair} through the conventional reasoning of DRO, we will explain them using a generalization of the empirical likelihood (EL) method, which is a nonparametric analog of maximum likelihood and endows a tight statistical confidence guarantee in using \eqref{original_pair} that can be translated to our procedures. Moreover, we also note that, from a computational viewpoint, \eqref{original_pair} is non-convex and intractable in general. Our procedures as well as statistical developments thus rely on a linearization of the objective function in \eqref{original_pair}. Furthermore, estimating the objective (i.e., the performance measure) and its linearization involves running simulation and incurs the associated errors. The next several subsections detail the linearization, the EL method development, and the sampling error control.

\subsection{Linearization of Performance Measure}\label{sec:linearization}
% Using the formulation \eqref{original_pair} to construct CIs involves both theoretical and algorithmic challenges. On the theory side, standard DRO tools can no longer be exploited to guarantee statistical validity of the CI given by \eqref{original_pair}, as discussed at the end of Section \ref{sec:DRO to input}. We will instead explain our procedure using the EL method and demonstrate that the CI is in fact asymptotically exact. However, due to the potential highly complicated structure of the performance measure $Z$ in relation to the input models, a direct analysis of \eqref{original_pair} seems difficult.

% On the algorithm side, the optimization pair \eqref{original_pair} is computationally intractable because the performance measure $Z$, as a function of the input models $\mathbf w_1,\ldots,\mathbf w_m$, is usually highly nonlinear and nonconvex. Even if $Z$ happens to be convex or concave, at least one optimization problem of the pair remains intractable unless $Z$ is linear. This leaves us no choice but to find a linear surrogate for the nonlinear performance measure of interest, in order to make both the minimization and maximization efficiently solvable.

% Fortunately, both challenges can be overcome by linearizations of the performance measure $Z$. Specifically we use the influence function of $Z$ as the coefficients of the linear approximation. 
We first state a property related to a more general notion of the influence function in \eqref{Gateaux derivative} that shows up in Assumption \ref{non-degeneracy}:
% The finite-horizon structure of the performance measure allows us to derive a precise expression of its influence function, depicted as:
\begin{proposition}\label{derivative}
Let $(Q_1^1,\ldots,Q_m^1),(Q_1^2,\ldots,Q_m^2)$ be two sets of distributions such that for any $s_{i,t}\in \set{1,2}$ with $i=1,\ldots,m$ and $t=1,\ldots,T_i$
\begin{equation*}
\int \abs{h(\mathbf x_1,\ldots,\mathbf x_m)}\prod_{i=1}^m\prod_{t=1}^{T_i}dQ_{i}^{s_{i,t}}(x_{i,t})<+\infty,
\end{equation*}
where $\mathbf x_i=(x_{i,t})_{t=1,\ldots,T_i}$, then
\begin{equation}
\lim_{\epsilon\rightarrow 0+}\frac{1}{\epsilon}\prth{Z((1-\epsilon)Q_1^1+\epsilon Q_1^2,\ldots,(1-\epsilon)Q_m^1+\epsilon Q_m^2)-Z(Q_1^1,\ldots,Q_m^1)}=\sum_{i=1}^m\mathbb E_{Q_i^2} [G_i^{Q_1^1,\ldots,Q_m^1}(X_i)],\label{derivative new}
\end{equation}
where $\mathbb E_{Q_i^2}[\cdot]$ denotes the expectation with respect to $Q_i^2$ that governs $X_i$, and $G_i^{Q_1^1,\ldots,Q_m^1}$ is the influence function of $Z(Q_1^1,\ldots,Q_m^1)$ with respect to the distribution $Q_i^1$, given by
\begin{equation*}
G_i^{Q_1^1,\ldots,Q_m^1}(x)=\sum_{t=1}^{T_i}\mathbb E_{Q_1^1,\ldots,Q_m^1}[h(\mathbf X_1,\ldots,\mathbf X_m)\vert X_{i}(t)=x]-T_iZ(Q_1^1,\ldots,Q_m^1).
\end{equation*}
Moreover, $\mathbb E_{Q_i^1} [G_i^{Q_1^1,\ldots,Q_m^1}(X_i)]=0$ for all $i=1,\ldots,m$.
\end{proposition}
% The detailed proof is left to Section \ref{proof:section 4.2} of the Appendix. 
Proposition \ref{derivative} can be shown by using techniques in the asymptotic analysis of von Mises statistical functionals (e.g., \citealt{serfling2009approximation}). It suggests the following linear approximation of $Z(Q_1^2,\ldots,Q_m^2)$ around $(Q_1^1,\ldots,Q_m^1)$
\begin{equation}\label{linear approximation}
Z(Q_1^1,\ldots,Q_m^1)+\sum_{i=1}^m\mathbb E_{Q_i^2} [G_i^{Q_1^1,\ldots,Q_m^1}(X)]
\end{equation}
where the sum consists of expectations of influence functions under $Q_i^2$ and hence is linear in $Q_i^2$. In particular, when $Q_i^1=P_i$, i.e., the true input distribution, and $Q_i^2=\mathbf w_i$ (abusing notations slightly to denote $\mathbf w_i$  as the weighted distribution supported on the observations $\{X_{i,j}\}_{j=1,\ldots,n_i}$), \eqref{linear approximation} suggests a linear approximation of $Z(\mathbf w_1,\ldots,\mathbf w_m)$ given by
\begin{equation}\label{linear approximation truth}
Z_L(\mathbf w_1,\ldots,\mathbf w_m):=Z^*+\sum_{i=1}^m\sum_{j=1}^{n_i}G_{i}(X_{i,j})w_{i,j}
\end{equation}
where the $G_i$'s are defined in Assumption \ref{non-degeneracy} and correspond to the influence functions of $Z$ at the true input distributions.

% To overcome the theoretical challenge mentioned before, we take Z(Q_1^2,\ldots,Q_m^2)\approx 
% However, the linearization $Z_L$ does not help with the tractability issue because its evaluation requires knowledge of the influence function $G_i(X_{i,j})$ which is under the unknown true input distributions. Instead, to resolve the algorithmic issue 
Furthermore, taking $Q_i^1=\hat{P}_i$, i.e., the empirical input distribution, and $Q_i^2=\mathbf w_i$ in \eqref{linear approximation}, we arrive at the linearization of  $Z(\mathbf w_1,\ldots,\mathbf w_m)$ around the uniform weights $w_{i,j}=1/n_i$
\begin{equation}\label{linear approximation empirical}
\widehat{Z_L}(\mathbf w_1,\ldots,\mathbf w_m):=Z(\hat{P}_1,\ldots,\hat{P}_m)+\sum_{i=1}^m\sum_{j=1}^{n_i}\hat{G}_{i}(X_{i,j})w_{i,j}
\end{equation}
where the $\hat{G}_i$'s are the influence functions of $Z$ at the empirical input distributions, defined by
\begin{equation}
\hat{G}_i(x)=\sum_{t=1}^{T_i}\mathbb E_{\hat{P}_1,\ldots,\hat{P}_m}[h(\mathbf X_1,\ldots,\mathbf X_m)\vert X_{i}(t)=x]-T_iZ(\hat{P}_1,\ldots,\hat{P}_m).\label{empirical influence function}
\end{equation}

The following result characterizes the quality of the above two linear approximations:
\begin{proposition}\label{linearization error}
Under Assumptions \ref{balanced data} and \ref{8th moment}, as the input data size $n\to\infty$ we have
\begin{align}
\mathbb E\big[\sup_{(\mathbf w_1,\ldots,\mathbf w_m)\in \mathcal U_{\alpha}}\big\lvert Z(\mathbf w_1,\ldots,\mathbf w_m)-Z_L(\mathbf w_1,\ldots,\mathbf w_m)\big\rvert^2\big]&=O\big(\frac{1}{n^2}\big)\label{error:truth approximation}\\
\mathbb E\big[\sup_{(\mathbf w_1,\ldots,\mathbf w_m)\in \mathcal U_{\alpha}}\big\lvert Z(\mathbf w_1,\ldots,\mathbf w_m)-\widehat{Z_L}(\mathbf w_1,\ldots,\mathbf w_m)\big\rvert^2\big]&=O\big(\frac{1}{n^2}\big)\label{error:empirical approximation}
\end{align}
where $\mathcal U_{\alpha}$ is defined in \eqref{uncertainty set}.
\end{proposition}
Proposition \ref{linearization error} suggests that, restricting to $\mathcal U_\alpha$, the maximal deviations of the linear approximations from the true performance measure vanish as fast as $1/n$. Next we will build the theories and explain our procedures for a linearized performance measure, and relate them back to the original nonlinear performance measure $Z$ through Proposition \ref{linearization error}.

% algorithms for \eqref{original_pair} with $Z$ replaced by the linearized performance measures $Z_L,\widehat{Z_L}$, and then show that the same results carry over to the original formulation with nonlinear $Z$ thanks to the rapidly vanishing errors in \eqref{error:truth approximation}\eqref{error:empirical approximation}.

% The proof is in Section \ref{proof:section 4.2} of the Appendix. 
% , This observation is in line with the notion of ``empirical" uncertainty set recently studied in \cite{lam2016recovering} in the context of feasibility guarantees for stochastic constraints.

% This step is introduced to hammer the effect of linearization on the final evaluation of the bounds, and allows us to obtain numerical performance superior to the delta method.

\subsection{Empirical Likelihood Theory for Sums of Means}\label{sec:EL theorem}
% In this subsection and the next, we justify the formulation \eqref{original_pair} using the EL method. 
First proposed by \cite{owen1988empirical}, the EL method can be viewed as a nonparametric counterpart of the maximum likelihood theory. Here we will develop this method for the linear approximation $Z_L$. Note that the second term in \eqref{linear approximation truth} can be expressed as a sum of means, i.e., $\sum_{i=1}^m\mathbb E_{\mathbf w_i}[G_{i}(X_i)]$. Therefore, to ease notation and emphasize its generality, we will present our EL method as a generic inference tool for estimating sums of means. %This will be the key to obtaining our formulation \eqref{zcilimit}.  %is a nonparametric counterpart of the Wilks' theorem in parameter inference, which states that the logarithmic nonparametric likelihood ratio converges to a chi-squared distribution.

Suppose we are given $m$ independent samples of i.i.d.~observations $\{Y_{i,1},\ldots,Y_{i,n_i}\}, i=1,\ldots,m$, with each $Y_{i,j}$ distributed according to a common distribution $F_i$. For the $i$-th sample, we define its nonparametric likelihood, in terms of the probability weights $\mathbf w_i$ over the support points of the data, to be $\prod_{j=1}^{n_i}w_{i,j}$. The multi-sample likelihood is $\prod_{i=1}^{m}\prod_{j=1}^{n_i}w_{i,j}$. By a simple convexity argument, it can be shown that assigning uniform weights $w_{i,j}=1/n_i$ for each sample yields the maximal value $\prod_{i=1}^m(1/n_i)^{n_i}$. Moreover, uniform weights still maximize even if one allows putting weights outside the support of data, in which case $\sum_{j=1}^{n_i}w_{i,j}<1$ for some $i$, making $\prod_{j=1}^{n_i}w_{i,j}$ even smaller. Therefore, the uniform weights $w_{i,j}=1/n_i$ for all $j=1,\ldots,n_i$ can be viewed as the nonparametric maximum likelihood estimate for the $i$-th distribution $F_i$, and $w_{i,j}=1/n_i$ for all $i,j$ is the multi-sample counterpart.

% To proceed, we need to define a parameter of interest that is determined by the input models. In our case, the natural parameter of interest is the performance measure $Z(P_1,\ldots,P_m)$. This is in general a complex nonlinear function of $P_i$'s. As a building block, we focus here on the special case that $h(\mathbf X_1,\ldots,X_m)$ is linearly separable among different input models, and each separated component is a simple function of $X_i$. In other words, $h(\mathbf X_1,\ldots,\mathbf X_m)=\sum_{i=1}^mh_i(X_{i}(1))$ for some $h_i:\mathfrak X_i\rightarrow \R$, with $T_i=1$ for all $i$. The parameter of interest is therefore simply the sum of means of random variables $h_i(X_{i}(1))$. For convenience, we treat $h_i(X_i(1))$ as an elementary random variable and upon notational replacement we consider estimating the true parameter $\mu_0=\sum_{i=1}^m\mathbb EX_i$ for a generic variable $X_i$ under $P_i$.

To proceed, we need to define a parameter of interest that is determined by the distributions $F_i$'s. In our case, the parameter of interest is the sum of means $\mu_0:=\sum_{i=1}^m\mathbb EY_{i}$ where each $Y_i$ is distributed under $F_i$.

The key of the EL method is to establish limit theorems analogous to the celebrated Wilks' Theorem (\cite{wilks1938large}) in the maximum likelihood theory, which stipulates that a suitably defined logarithmic likelihood ratio converges to a $\mathcal X^2$ random variable. In the EL setting, we use the so-called profile nonparametric likelihood ratio to carry out inference on parameters. To explain this, first, the nonparametric likelihood ratio is defined as the ratio between the nonparametric likelihood of a given set of weights and the uniform weights (i.e.,~the nonparametric maximum likelihood estimate).
%empirically produce a given value of the parameter, and the uniform weights . , which is however possibly nonlinear in $P_i$'s and hence not easy to work with. Fortunately, it turns out that there isn't much loss to deal with the special case
%We will focus on this case for now. To simplify further, let us consider estimating $\sum_{i=1}^m\mathbb EX_i$, where for convenience we write $X_i$ in place of $X_i(1)$.
The profile nonparametric likelihood ratio is defined as the maximal ratio among all probability weights giving rise to a particular value $\mu$ for the sum of means, i.e.,
\begin{equation}
R(\mu)=\max\set{\prod_{i=1}^m\prod_{j=1}^{n_i}n_iw_{i,j}\bigg\vert \sum_{i=1}^m\sum_{j=1}^{n_i}Y_{i,j}w_{i,j}=\mu,\ \sum_{j=1}^{n_i}w_{i,j}=1\text{\ for all\ }i,\ w_{i,j}\geq 0\text{ for all }i,j},\label{profileratio}
\end{equation}
and is defined to be $0$ if the optimization problem in \eqref{profileratio} is infeasible. Profiling here refers to the categorization of weights that lead to the same value $\mu$.

The quantity $R(\mu)$ satisfies the following asymptotic property:%Now we state our theorem.
\begin{theorem}
Let $Y_i$ be a random variable distributed under $F_i$. Assume $\mathrm{Var}(Y_{i})<\infty$ for all $i=1,\ldots,m$ and at least one of them is non-zero, and that the sample sizes $n_i$'s satisfy Assumption \ref{balanced data}. Then $-2\log R(\mu_0)$, where $\mu_0$ is the sum of the true means, converges in distribution to $\mathcal X^2_1$, the chi-square distribution with degree of freedom one, as $n\to \infty$.\label{elt}
\end{theorem}
% \begin{theorem}
% Let $X_i$ be a random variable distributed under $P_i$. Assume $0<\sum_{i=1}^m\mathrm{Var}(X_i)<\infty$, and $\max_{i\in I}n_i\leq \gamma\min_{i\in I}n_i$ always holds for some constant $\gamma>0$, where $I=\{i\vert\mathrm{Var}(X_i)>0\}$. Then $-2\log R(\mu_0)$, where $\mu_0=\sum_{i=1}^m\mathbb EX_i$ is the true parameter, converges in distribution to $\mathcal X^2_1$, the chi-square distribution with degree of freedom one, as $n_i\to \infty$ for $i\in I$.\label{elt}
% \end{theorem}
In other words, the logarithmic profile nonparametric likelihood ratio at the true value asymptotically follows a chi-square distribution with degree of freedom one. This degree of freedom is the effective number of parameters to be estimated which, in this case, is one since there is only a single target parameter $\mu_0$. Note that this is independent of the number of input distributions $m$.

% This degree of freedom is independent of the number of distributions $m$, but is equal to the number of non-probability-simplex constraints, e.g., $\sum_{i=1}^m\sum_{j=1}^{n_i}Y_{i,j}w_{i,j}=\mu$ in \eqref{profileratio}.
% As another interpretation,
% From another perspective, treating the weights $\mathbf w_i$ as parameters, the degree of freedom is the difference between the dimensions of the full and the constrained parameter space, and one constraint results in a loss of one dimension of the parameter space.

% In Theorem \ref{elt} only the sample sizes of those distributions having positive variances are required to grow to infinity. To see the reason for this, note that the data of inputs with zero variance are always equal to the true mean so we can always assign them the uniform weights in \eqref{profileratio}. Other than these zero-variance inputs, the condition $\max_{i\in I}n_i\leq \gamma\min_{i\in I}n_i$ forces the sample sizes to grow at the same rate.
Theorem \ref{elt} is a sum-of-mean generalization of the well-known empirical likelihood theorem (ELT) for single-sample mean:%, which can be found in Chapter 2 of \cite{owen2001empirical}. %We state it as a special case where $m=1$.
\begin{theorem}[\cite{owen2001empirical} Theorem 2.2]
Consider only the first sample $\{Y_{1,1},\ldots,Y_{1,n_1}\}$. Assume $0<\mathrm{Var}(Y_{1})<\infty$. Then $-2\log R(\mathbb EY_1)$ converges in distribution to $\mathcal X^2_{1}$, as $n_1\rightarrow \infty$. The function $R(\cdot)$ here is the same as that in \eqref{profileratio} but with $m=1$.
\end{theorem}
% \begin{theorem}[\cite{owen2001empirical} Theorem 2.2]
% Let $Y_{1,1},Y_{1,2},\ldots,Y_{1,n}$ be i.i.d.~random variables distributed under $F_1$, with $0<\mathrm{Var}(Y_{1,1})<\infty$. Then $-2\log R(\mathbb EY_1)$ converges in distribution to $\mathcal X^2_{1}$, as $n\rightarrow \infty$. The function $R(\cdot)$ here is the same as that in \eqref{profileratio} with $m=1,n_1=n,X_{1,j}=Y_j$.
% \end{theorem}
Extensions of this theorem have been studied in the literature (e.g., \citealt{owen1990empirical,owen1991empirical,qin1994empirical,hjort2009extending}). The most relevant one is in the context of analysis-of-variance (ANOVA), in which the logarithmic profile nonparametric likelihood ratio at the true means of multiple independent samples are shown to converge to $\mathcal X^2_m$, where $m$ is the number of samples (or groups). However, the argument for this result relies on viewing the multiple samples as a collection of heteroscedastic data and applying the triangular array ELT (\citealt{owen1991empirical}), which does not apply obviously to our case. Another related extension is the plug-in EL (\citealt{hjort2009extending}) which entails that, under $p$ estimating functions that possibly involve unknown nuisance parameters, the associated logarithmic profile likelihood ratio converges to a weighted sum of $p$ independent $\mathcal X^2_1$'s, if ``good enough" estimators of the unknown nuisance parameters are used in evaluating the profile likelihood ratio. However, \cite{hjort2009extending} focuses on the single-sample case, thus is not directly applicable. There have also been studies on applying EL to hypothesis testing of two-sample mean differences (\citealt{liu2008empirical,wu2012empirical}), but it appears that a fully rigorous proof is not available for our general multi-sample sum-of-means setting. In view of these, we provide a detailed proof of Theorem \ref{elt} in Section \ref{sec:proof elt} of the Appendix.

A sketch of the key idea is as follows. We first introduce the auxiliary variables $\mu_i$ that represent the means of individual samples, so that the constraint $\sum_{i=1}^m\sum_{j=1}^{n_i}Y_{i,j}w_{i,j}=\mu$ in \eqref{profileratio} is replaced by $\sum_{j=1}^{n_i}Y_{i,j}w_{i,j}=\mu_i,i=1,\ldots,m$ and $\sum_{i=1}^m\mu_i=\mu$. %reformulate \eqref{profileratio} as
%\begin{equation}
%\begin{aligned}
%& \underset{\mathbf w_1,\ldots,\mathbf w_m,\bm{\mu}}{\text{min}}& & -\sum_{i=1}^m\sum_{j=1}^{n_i}\log(n_iw_{i,j}) \\
%& \text{subject to}
%& & \sum_{j=1}^{n_i}w_{i,j}X_{i,j}=\mu_i,i=1,\ldots,m\\
%&&&\sum_{j=1}^{n_i}w_{i,j}=1,i=1,\ldots,m\\
%&&&\sum_{i=1}^m\mu_i=0\\
%\end{aligned}\label{keyopt1}
%\end{equation}
%where $\mathbf w_i=(w_{i,1},\ldots,w_{i,n_i})$ and $\bm{\mu}=(\mu_1,\ldots,\mu_m)$. The non-negativity constraints $w_{i,j}\geq 0$ are dropped since they are implicitly imposed in the objective function.
The KKT conditions then enforce the optimal weights to be
$$w_{i,j}^*=\frac{1}{n_i+\lambda^*(Y_{i,j}-\mu_i^*)}$$
where $\lambda^*$ is the Lagrange multiplier for the constraint $\sum_{i=1}^m\mu_i=\mu$ and $\mu_i^*$ is the optimal solution for $\mu_i$. When $\mu$ is the true value $\mu_0$, an asymptotic analysis on the KKT conditions approximates $\lambda^*$ as
$$\lambda^*\approx\frac{\sum_{i=1}^m(\bar Y_i-\mathbb EY_i)}{\sum_{i=1}^m\frac{\sigma_i^2}{n_i}}$$
where $\bar Y_i=(1/n_i)\sum_{j=1}^{n_i}Y_{i,j}$ is the sample mean and $\sigma_i^2$ is the variance of $Y_i$. Moreover, we have the approximation $\mu_i^*\approx\mathbb EY_i$. By Taylor's expansion, the logarithmic profile nonparametric likelihood ratio can be approximated as
\begin{align*}
-2\log R(\mu_0)&=2\sum_{i=1}^m\sum_{j=1}^{n_i}\log\left(1+\frac{\lambda^*}{n_i}(Y_{i,j}-\mu_i^*)\right)\\
&\approx2\sum_{i=1}^m\sum_{j=1}^{n_i}\prth{\frac{\lambda^*}{n_i}(Y_{i,j}-\mu_i^*)-\frac{{\lambda^*}^2}{2n_i^2}(Y_{i,j}-\mu_i^*)^2}\\
&\approx2\sum_{i=1}^m\lambda^*(\bar Y_{i}-\mu_i^*)-\sum_{i=1}^m\frac{{\lambda^*}^2\sigma_i^2}{n_i}\\
&\approx\left(\frac{\sum_{i=1}^m(\bar Y_i-\mathbb EY_i)}{\sqrt{\sum_{i=1}^m\frac{\sigma_i^2}{n_i}}}\right)^2\\
&\Rightarrow\mathcal X^2_1
\end{align*}
where ``$\Rightarrow$" denotes convergence in distribution. This gives our result in Theorem \ref{elt}.

\subsection{Duality and Optimization-based Confidence Interval}\label{sec:EL-based CI}
From Theorem \ref{elt}, a duality-type argument will give rise to a pair of optimization problems whose optimal values will serve as confidence bounds for the sum of the true means. We have the following:
%We discuss how to construct confidence interval for the quantity $Z(P_1,\ldots,P_m)$ based on the EL theory presented in the last section, and thus justify the validity of the formulation \eqref{zcilimit}. We will first study the linear output case, and then discuss the general performance measure in \eqref{perfm}.
% Suppose that, as in the last section, the output function takes the simple separable form $h(\mathbf X_1,\ldots,\mathbf X_m)=\sum_{i=1}^mh_i(X_{i})$, where $X_i$ is distributed under $P_i$. 
% The following asymptotically exact confidence guarantee is a consequence of Theorem \ref{elt}:% implies that the optimization pair \eqref{zcilimit} gives an asymptotically correct confidence interval with coverage probability $1-\alpha$, namely
\begin{theorem}\label{opt2gua}
Under the same conditions of Theorem \ref{elt}, we have
\begin{equation*}
\lim_{n\to\infty}P\prth{\underline{\mu}\leq \mu_0\leq \overline{\mu}}= 1-\alpha
\end{equation*}
where
\begin{equation}
\underline{\mu}/\overline{\mu}:=\min/\max\Big\{\sum_{i=1}^m\sum_{j=1}^{n_i}Y_{i,j}w_{i,j}\Big\vert(\mathbf w_1,\ldots,\mathbf w_m)\in \mathcal U_{\alpha}\Big\}.\label{linear opt}
\end{equation}
% \begin{equation}\label{ciopt2}
% \mathscr L_{L}/\mathscr U_{L}:=\underset{(\mathbf w_1,\ldots,\mathbf w_m)\in \mathcal U_{\alpha}}{\min/\max}Z_L(\mathbf w_1,\ldots,\mathbf w_m).
% \end{equation}
\end{theorem}
% \begin{theorem}\label{cisum}
% Assume $0<\sum_{i=1}^m\mathrm{Var}(h_i(X_i))<\infty$, and $\max_{i\in I}n_i\leq \gamma\min_{i\in I}n_i$ for some constant $\mu>0$, where $I=\{i\vert \mathrm{Var}(h_i(X_i))>0\}$. Then
% \begin{equation*}
% P\prth{L_{\alpha}\leq \sum_{i=1}^m\mathbb Eh_i(X_i)\leq U_{\alpha}}\rightarrow 1-\alpha,\text{ as }n_i\rightarrow \infty\text{ for }i\in I,
% \end{equation*}
% where
% \begin{equation*}
% L_{\alpha}/U_{\alpha}:=\underset{(\mathbf w_1,\ldots,\mathbf w_m)\in\mathcal U_{\alpha}}{\min/\max}\sum_{i=1}^{m}\sum_{j=1}^{n_i}w_{i,j}h_i(X_{i,j}).
% \end{equation*}
% \end{theorem}
Theorem \ref{opt2gua} thus translates the asymptotic convergence in Theorem \ref{elt} into an asymptotically exact confidence bound. This is argued by a duality argument that turns the first constraint in \eqref{profileratio} into objective and vice versa. The concept is similar to Wilks' Theorem for maximum likelihood, but with the profiling that leads to the resulting optimization problems in \eqref{linear opt}.

% The concept is similar to the construction of CI for maximum likelihood estimator in the parametric case, but here the CI is expressed in terms of optimization programs because of the profiling in the definition of $R(\cdot)$. The proof of Theorem \ref{opt2gua} is deferred to Section \ref{proof:section 4.4} of the Appendix.

Moreover, in terms of the positions, the optimization-based confidence bounds $\underline\mu$ and $\overline\mu$ are equivalent to the standard normality-based confidence bounds up to negligible errors, as described below:
\begin{proposition}\label{connection to normal CI}
Under the same conditions of Theorem \ref{elt}, the confidence bounds $\underline{\mu},\overline{\mu}$ from Theorem \ref{opt2gua} satisfy
\begin{align*}
\underline{\mu}&=\sum_{i=1}^m\bar Y_i-z_{1-\alpha/2}\sqrt{\sum_{i=1}^m\frac{\sigma_i^2}{n_i}}+o_p\big(\frac{1}{\sqrt n}\big)\\
\overline{\mu}&=\sum_{i=1}^m\bar Y_i+z_{1-\alpha/2}\sqrt{\sum_{i=1}^m\frac{\sigma_i^2}{n_i}}+o_p\big(\frac{1}{\sqrt n}\big)
\end{align*}
where $\bar{Y}_i=\sum_{j=1}^{n_i}Y_{i,j}/n_i$ is the sample mean of $\{Y_{i,1},\ldots,Y_{i,n_i}\}$ and $\sigma_i^2$ is the true variance of $Y_i$, and $z_{1-\alpha/2}$ is the $1-\alpha/2$ quantile of the standard normal.
\end{proposition}
The errors between $\underline\mu$, $\overline\mu$ and the normality-based bounds $\sum_{i=1}^m\bar Y_i\pm z_{1-\alpha/2}\sqrt{\sum_{i=1}^m\frac{\sigma_i^2}{n_i}}$ are negligible in the sense that they are of smaller order than the width of the resulting CI, which is of order $1/\sqrt n$. 
% Again, see Section \ref{proof:section 4.4} of the Appendix for the proof.

% \begin{proposition}
% \begin{equation*}
% \mathcal L_{\alpha}/\mathcal U_{\alpha}=Z(P_1,\ldots,P_m)+\sum_{i=1}^m\bar{G}_i\mp z_{1-\alpha/2}\sqrt{\sum_{i=1}^m\frac{\mathrm{\widehat{Var}}(G_i)}{n_i}}+o_p\Big(\sqrt{\sum_{i\in I}\frac{1}{n_i}}\Big),\text{ as }n_i\to \infty\text{ for }i\in I,
% \end{equation*}
% where $\bar{G}_i=\sum_{j=1}^{n_i}G_i(X_{i,j})/n_i$ and $\mathrm{\widehat{Var}}(G_i)=\sum_{j=1}^{n_i}(G_i(X_{i,j})-\bar{G}_i)^2/(n_i-1)$ are the sample mean and variance, $I=\{i\vert \mathrm{Var}(G_i(X_i))>0\}$, and $z_{1-\alpha/2}$ is the $1-\alpha/2$ critical value of the standard normal.
% \end{proposition}

Applying the above two results to the linear approximation $Z_L$, we have the following:
\begin{corollary}\label{CI: linear approximation truth}
Under Assumptions \ref{balanced data}, \ref{non-degeneracy} and \ref{8th moment}, we have
\begin{equation}
\lim_{n\to\infty}P\prth{\mathscr L_{L}\leq Z^*\leq \mathscr U_{L}}= 1-\alpha\label{confidence guarantee linear}
\end{equation}
where
\begin{equation}\label{oracle CI}
\mathscr L_{L}/\mathscr U_{L}:=\min/\max\Big\{Z_L(\mathbf w_1,\ldots,\mathbf w_m)\Big\vert(\mathbf w_1,\ldots,\mathbf w_m)\in \mathcal U_{\alpha}\Big\}.
\end{equation}
Moreover
\begin{equation}\label{equivalence to normal CI}
\begin{aligned}
    \mathscr L_{L}&=Z^*+\sum_{i=1}^m\bar{G}_i- z_{1-\alpha/2}\sigma_I+o_p\big(\frac{1}{\sqrt{n}}\big)\\
    \mathscr U_{L}&=Z^*+\sum_{i=1}^m\bar{G}_i+ z_{1-\alpha/2}\sigma_I+o_p\big(\frac{1}{\sqrt{n}}\big)
\end{aligned}
\end{equation}
where each $\bar G_i = \sum_{j=1}^{n_i}G_i(X_{i,j})/n_i$ is the sample mean of $\{G_i(X_{i,1}),\ldots,G_i(X_{i,n_i})\}$,  $\sigma_I^2=\sum_{i=1}^m\mathrm{Var}(G_i(X_i))/n_i$, and $z_{1-\alpha/2}$ is the $1-\alpha/2$ quantile of the standard normal.
\end{corollary}
% the input-induced variance

% our optimization formulation \eqref{original_pair}. The first result deals with the linearized version of \eqref{original_pair}, i.e., when the objective $Z$ is replaced by its linear approximation $Z_L$. Notice
Note that the influence functions in \eqref{linear approximation truth} satisfy $\sum_{i=1}^m\mathbb E_{P_i}[G_i(X_{i})]=0$ due to the last claim in Proposition \ref{derivative}. Thus, letting $Y_{i,j}=G_i(X_{i,j})$ in Theorem \ref{opt2gua} and Proposition \ref{connection to normal CI}, and noting that the $Z^*$ in \eqref{confidence guarantee linear} and \eqref{equivalence to normal CI} can be cancelled out, we arrive at the conclusion in Corollary \ref{CI: linear approximation truth}. 

Next, combining Corollary \ref{CI: linear approximation truth} and the linearization error \eqref{error:truth approximation}, we can establish similar results for $\mathscr L,\mathscr U$ that arise in \eqref{original_pair}: 
% for the original formulation \eqref{original_pair}, namely
\begin{theorem}\label{ideal CI}
Under Assumptions \ref{balanced data}, \ref{non-degeneracy} and \ref{8th moment}, the minimal and maximal values $\mathscr L,\mathscr U$ of \eqref{original_pair} satisfy
\begin{equation*}
\lim_{n\to\infty}P\prth{\mathscr L\leq Z^*\leq \mathscr U}= 1-\alpha,
\end{equation*}
and the asymptotic equivalence \eqref{equivalence to normal CI} holds true with $\mathscr L_{L}$, $\mathscr U_{L}$ replaced by $\mathscr L$, $\mathscr U$.
\end{theorem}
The proof of Theorem \ref{ideal CI} consists of first approximating the discrepancies between the optimal values, i.e., $\mathscr L=\mathscr L_L+O_p(1/n)$ and $\mathscr U=\mathscr U_L+O_p(1/n)$, using \eqref{error:truth approximation}, and then showing that any quantities that equal \eqref{equivalence to normal CI}, up to a small order of discrepancies, deliver an interval with asymptotically exact coverage probability by a standard application of Slutsky's Theorem. 

\subsection{Estimating Influence Function}\label{sec:nonconvexity}
% The linear approximation $Z_L$ is useful for establishing the theoretical guarantees, its evaluation however requires the unknown input models as pointed out in . In order to make both the minimization and the maximization computationally tractable, 
Our proposed CIs in Algorithms \ref{algo1}, \ref{algo2} and \ref{algo3} use a combination of the intervals suggested in Corollary \ref{CI: linear approximation truth} and Theorem \ref{ideal CI}. Before we explain this concretely, note that directly using the definition of $\mathscr L,\mathscr U$ in \eqref{original_pair} will encounter computational difficulties due to the general intractability of the optimization. Thus, we consider using optimization \eqref{oracle CI} or expression \eqref{equivalence to normal CI} (either $\mathscr L_L,\mathscr U_L$ in Corollary \ref{CI: linear approximation truth} or  $\mathscr L,\mathscr U$ in Theorem \ref{ideal CI}) as our confidence bounds. In either case, we need to estimate the influence function represented by $G_i(X_{i,j})$'s.

% , which has asymptotic exact coverage guaranteed 
% When solving the optimization \eqref{original_pair} in practice, we optimize the linear approximation $\widehat{Z_L}$ defined by \eqref{linear approximation empirical} in place of the nonlinear performance measure $Z$, as discussed in Section \ref{sec:linearization}.

% Note that the first term in $\widehat{Z_L}$, i.e., $Z(\hat{P}_1,\ldots,\hat{P}_m)$, does not depend on the weights, hence can be ignored if only the extreme weights are sought for but not the extreme objective values. This is indeed the case, because once the extreme weights are obtained in Step 2 of Algorithm \ref{algo1} the confidence bounds are computed in Step 3 from running two independent sets of simulation each driven by the minimal or maximal extreme weights.We discuss how to compute the coefficients $\hat{G}_i(X_{i,j})$'s 

There are two sources of errors in estimating $G_i(X_{i,j})$. First, since we do not know the true distribution $P_i$, we approximate it by the influence function evaluated at the empirical distribution, namely $\hat G_i(X_{i,j})$ defined in \eqref{empirical influence function} (which in turn forms the coefficient in $\widehat{Z_L}$). Second, $\hat G_i(X_{i,j})$, like $G_i(X_{i,j})$, is a sum of conditional expectations, which needs to be estimated by simulation. \cite{ghosh2015computing,ghosh2015mirror} propose an unbiased estimator for such quantities where the input distributions have arbitrary weights $w_{i,j}$ on their support points. Here we use their scheme for the special case of uniform weights. Similar approaches also arise in the so-called infinitesimal jackknife for bagging estimators (e.g., \cite{efron2014estimation,wager2014confidence}). Proposition \ref{gradcomputing} shows the scheme (see \citealt{ghosh2015computing} for the proof). 
\begin{proposition}\label{gradcomputing}
Given input data $\{X_{i,j}\}$, the empirical influence function $\hat{G}_i$ evaluated at data point $X_{i,j}$ satisfies
\begin{equation*}
\hat{G}_{i}(X_{i,j})=\mathrm{Cov}_{\hat{P}_1,\ldots,\hat{P}_m}(h(\mathbf X_1,\ldots,\mathbf X_m),S_{i,j}(\mathbf X_i)),
\end{equation*}
where $\mathrm{Cov}_{\hat{P}_1,\ldots,\hat{P}_m}$ denotes the covariance under the empirical input distributions, and
\begin{equation*}
S_{i,j}(\mathbf X_i)=\sum_{t=1}^{T_i}n_i\mathbf{1}\{X_i(t)=X_{i,j}\}-T_i.
\end{equation*}
\end{proposition}
Such a covariance interpretation of the influence function leads us to the Monte Carlo estimate \eqref{gradientest} of $\hat{G}_i(X_{i,j})$ in Step 1, denoted $\hat{\vphantom{\rule[4pt]{1pt}{5.5pt}}\smash{\hat{G}}}_{i}(X_{i,j})$, that takes the form of a sample covariance from $R_1$ simulation runs.
% of the $R_1$ simulation runs driven by the empirical input distributions.
% \begin{equation}\label{gradientest}
% \dbhat{G}_{i}(X_{i,j})=\frac{1}{R_1}\sum_{r=1}^{R_1}\big[(h(\mathbf X_1^r,\ldots,\mathbf X_m^r)-\hat{Z}(\hat{P}_1,\ldots,\hat{P}_m))(\sum_{t=1}^{T_i}n_i\mathbf{1}\{X_i^r(t)=X_{i,j}\}-T_i)\big],
% \end{equation}
% where $\mathbf X_1^r,\ldots,\mathbf X_m^r,r=1,\ldots,R_1$ are $R_1$ independent replications of the $m$ input processes generated under the uniform weights, and $\hat{Z}(\hat{P}_1,\ldots,\hat{P}_m)=\sum_{r=1}^{R_1}h(\mathbf X_1^r,\ldots,\mathbf X_m^r)/R_1$ is the sample mean.
Next, we introduce a sampled linear approximation for $Z(\mathbf w_1,\ldots,\mathbf w_m)$ given by
\begin{equation}\label{linear approximation empirical simulated}
    \dbwidehat{Z_L}(\mathbf w_1,\ldots,\mathbf w_m):=\hat{Z}(\hat{P}_1,\ldots,\hat{P}_m)+\sum_{i=1}^m\sum_{j=1}^{n_i}\dbhat{G}_i(X_{i,j})w_{i,j}.
\end{equation}
% Now using \eqref{gradientest} as an estimate of the coefficient $\hat G_i(X_{i,j})$ in $\widehat{Z_L}$ and the sample mean $\hat{Z}(\hat{P}_1,\ldots,\hat{P}_m)$ of the $R_1$ output replications as an estimate of $Z(\hat{P}_1,\ldots,\hat{P}_m)$, we arrive at the following sample average linear approximation of $Z(\mathbf w_1,\ldots,\mathbf w_m)$ as an estimate of $Z(\hat{P}_1,\ldots,\hat{P}_m)$
where $\hat{Z}(\hat{P}_1,\ldots,\hat{P}_m)$ is the sample mean of the $R_1$ replications. Optimization \eqref{ciopt4min} in Step 2 of the procedures uses $\dbwidehat{Z_L}(\mathbf w_1,\ldots,\mathbf w_m)$ as the objective function. But since $\hat{Z}(\hat{P}_1,\ldots,\hat{P}_m)$ does not depend on the weights $w_{i,j}$'s, it is dropped from the expression. 
The quality of the sample linear approximation \eqref{linear approximation empirical simulated} is quantified as:
\begin{proposition}\label{empirical_error}
Under Assumptions \ref{balanced data} and \ref{8th moment}, as the input data size $n\to\infty$ and simulation effort $R_1\to\infty$ we have $\mathbb E\big[\sup_{(\mathbf w_1,\ldots,\mathbf w_m)\in \mathcal U_{\alpha}}\big\lvert \widehat{Z_L}(\mathbf w_1,\ldots,\mathbf w_m)-\dbwidehat{Z_L}(\mathbf w_1,\ldots,\mathbf w_m)\big\rvert^2\big]=O\big( \frac{1}{R_1}\big)$,
where the expectation is taken with respect to the joint randomness of the data and the simulation. Hence together with \eqref{error:empirical approximation} we have
\begin{equation}\label{error:linear approximation empirical simulated}
\mathbb E\big[\sup_{(\mathbf w_1,\ldots,\mathbf w_m)\in \mathcal U_{\alpha}}\big\lvert Z(\mathbf w_1,\ldots,\mathbf w_m)-\dbwidehat{Z_L}(\mathbf w_1,\ldots,\mathbf w_m)\big\rvert^2\big]=O\big(\frac{1}{n^2}+ \frac{1}{R_1}\big).
\end{equation}
\end{proposition}
The uniform error \eqref{error:linear approximation empirical simulated} of $\dbwidehat{Z_L}$ as an approximation to $Z$ then implies the following guarantee on the difference between the weights $\{\mathbf w_i^{\min}\}_{i=1}^m,\{\mathbf w_i^{\max}\}_{i=1}^m$ obtained in Step 2 of Algorithm \ref{algo1}, and the optimal weights for the optimization pair \eqref{original_pair}, measured in terms of their evaluations of the performance measure $Z$:
\begin{theorem}\label{mainresult}
Let $Z^{\min} := Z(\mathbf w_1^{\min},\ldots,\mathbf w_m^{\min})$ and $Z^{\max}:=Z(\mathbf w_1^{\max},\ldots,\mathbf w_m^{\max})$. Under Assumptions \ref{balanced data} and \ref{8th moment}, as the input data size $n\to\infty$ and simulation effort $R_1\to\infty$ we have
\begin{equation*}
\mathbb E[(Z^{\min}-\mathscr L)^2]=O\big(\frac{1}{n^2}+ \frac{1}{R_1}\big),\ 
\mathbb E[(Z^{\max}-\mathscr U)^2]=O\big(\frac{1}{n^2}+ \frac{1}{R_1}\big)
\end{equation*}
where $\mathscr L,\mathscr U$ are defined in \eqref{original_pair}, and the expectation is taken with respect to the joint randomness of the data and the simulation.
\end{theorem}
% Proofs of Proposition \ref{empirical_error} and Theorem \ref{mainresult} can be found in Section \ref{proof:section 4.5} of the Appendix. 
Theorem \ref{mainresult} justifies using $\{\mathbf w_i^{\min}\}_{i=1}^m,\{\mathbf w_i^{\max}\}_{i=1}^m$ to evaluate the performance measure, which give rise to the asymptotically exact confidence bounds $\mathscr L,\mathscr U$ up to a small-order error. Step 3 of the algorithms utilizes this implication. However, we need to properly control the simulation error in evaluating the performance measure, which is detailed in the next subsection.

% Algorithm \ref{algo1} to evaluate the extreme performance measures $Z^{\min}$ and $Z^{\max}$, as a respective approximation to the ideal confidence bounds $\mathscr L$ and $\mathscr U$, via direct simulation.

%  The minimal and maximal weights are then used in because of Proposition \ref{empirical_error} and Theorem \ref{ideal CI}, alternatively we can compute a CI by evaluating the simulated linear approximation

As a side note, we can also use the linear approximation $\dbwidehat{Z_L}$ evaluated at the weights $\{\mathbf w_i^{\min}\}_{i=1}^m,\{\mathbf w_i^{\max}\}_{i=1}^m$ directly as our confidence interval. This forms another asymptotically exact CI  (see Theorem \ref{algoerror1} in Appendix \ref{proof:section 4.5}). Moreover, this approach would require less simulation effort than our procedures ($R_1$ versus $R_1+2R_2$). However, like the delta method, this approach relies heavily on the linear approximation to construct the CI. In contrast, the CIs in our procedures are constructed from simulating the (nonlinear) performance measure, under the carefully chosen empirical weights $\{\mathbf w_i^{\min}\}_{i=1}^m,\{\mathbf w_i^{\max}\}_{i=1}^m$. As a result, they conform more closely to the boundaries of a given problem and in turn can lead to better coverages. For example, when the performance measure is within a range (e.g., a probability that is between 0 and 1), using only the linear approximation frequently incurs under-coverage as the CIs can lie significantly outside the meaningful range (note that truncating at the boundaries would not solve the issue, which is intrinsic in the linear approximation), whereas our procedures would generate confidence bounds that much more often lie within the range and consequently offer better coverages.
% evaluates the original performance measure $Z$ rather than the linear approximation $\dbwidehat{Z_L}$ is less susceptible to this concern.

% Algorithm \ref{algo1}
% is more beneficial in terms of finite-sample coverage accuracy because it alleviates the effect of linearization. approach in Theorem \ref{algoerror1}, which behaves like the delta method,
% obtained from the optimization pair \eqref{ciopt4min}. Even though the approach suggested in Theorem \ref{algoerror1} 

% whereas Algorithm \ref{algo1} that outputs the final simulation evaluation is less susceptible to this concern.
% The reason is that the linearized outputs in the modified approach resemble those of the delta method, which consist of a number of terms that grows linearly with the number of input models. Thus, for a number of input models that is moderate or large relative to the sample size, the performance of the resulting CI can deteriorate. For this reason we choose Algorithm \ref{algo1} as our suggested procedure.

\subsection{Evaluation of CI Bounds}\label{sec:evaluation}
This section explains and compares Step 3 in Algorithms \ref{algo1}, \ref{algo2} and \ref{algo3} to evaluate the final confidence bounds, and relates these to the justify Theorems \ref{erfreeCI1}, \ref{erfreeCI2} and \ref{erfreeCI3}.
Algorithm \ref{algo1} constructs CIs by taking averages of $R_2$ independent simulation runs driven by the weighted empirical input distributions, with weights being  $\{\mathbf w_i^{\min}\}_{i=1}^m,\{\mathbf w_i^{\max}\}_{i=1}^m$, to evaluate the lower and upper bounds respectively. Note that by Theorem  \ref{mainresult}, the performance measures evaluated at the weighted empirical distributions, $Z^{\min}$ and $Z^{\max}$, are close to $\mathscr L$ and $\mathscr U$, which in turn by Theorem \ref{ideal CI} satisfy exact coverage guarantees. Step 3 of Algorithm \ref{algo1} adds simulation noises from the $R_2$ simulation runs in estimating $Z^{\min}$ and $Z^{\max}$. This results in the following discrepancies between the outputs of Algorithm \ref{algo1} and $\mathscr L$, $\mathscr U$:
\begin{proposition}\label{algoerror}
Under Assumptions \ref{balanced data} and \ref{8th moment}, as the input data size $n\to\infty$ and simulation effort $R_1\to\infty,R_2\to\infty$, the outputs $\mathscr L^{BEL},\mathscr U^{BEL}$ of Algorithm \ref{algo1} satisfy
\begin{equation*}
\mathbb E[(\mathscr L^{BEL}-\mathscr L)^2]=O\big(\frac{1}{n^2}+ \frac{1}{R_1}+\frac{1}{R_2}\big),\ \mathbb E[(\mathscr U^{BEL}-\mathscr U)^2]=O\big(\frac{1}{n^2}+ \frac{1}{R_1}+\frac{1}{R_2}\big)
\end{equation*}
where the expectation is taken with respect to the joint randomness of the data and the simulation.
\end{proposition}
Proposition \ref{algoerror} implies that, when the simulation sizes $R_1$ and $R_2$ both dominate the input data size $n$, the root-mean-square discrepancies between the outputs from Algorithm \ref{algo1}, $\mathscr L^{BEL}$, $\mathscr U^{BEL}$, and the asymptotically exact CIs formed by $\mathscr L$, $\mathscr U$, become $o(1/\sqrt{n})$, which is of smaller order than the width of the CI that is of order $1/\sqrt n$. This then leads to the asymptotic exactness of $[\mathscr L^{BEL},\mathscr U^{BEL}]$ in Theorem \ref{erfreeCI1}.
% See Section \ref{proof:section 4.6} of the Appendix for the proof. asymptotic equivalence \eqref{equivalence to normal CI} between $[\mathscr L^{BEL},\mathscr U^{BEL}]$ and the normal-based CI, and subsequently its which in turn allows for a smaller $R_2$. at the cost of slightly weaker coverage guarantees The following asymptotic representations of the confidence bounds by Algorithms \ref{algo2} and \ref{algo3} can be derived from the equivalence \eqref{equivalence to normal CI} and Theorem \ref{mainresult} (see  Section \ref{proof:section 4.6} of the Appendix for the formal proof)When $R_2$ is small, the estimates of $Z^\min$ and $Z^\max$ would incur significant simulation errors that are not accounted for. 

Algorithm \ref{algo1} requires both $R_1$ and $R_2$ to be large relative to $n$. Algorithms \ref{algo2} and \ref{algo3}, on the other hand, are designed to work well for smaller $R_2$. To explain, note that the reason of needing $R_2$ to be large in Algorithm \ref{algo1} is to wash away the simulation noises to a smaller magnitude than the CI width in Step 3. Instead of simply washing them away, Algorithms \ref{algo2} and \ref{algo3} suitably enlarge the CI to incorporate these errors in Step 3, so that $R_2$ can now be chosen as in standard CI construction (instead of depending on $n$). The key to this argument uses the following decomposition:
\begin{proposition}\label{algo2/3 representation}
Let $Z^{\min} := Z(\mathbf w_1^{\min},\ldots,\mathbf w_m^{\min})$ and $Z^{\max}:=Z(\mathbf w_1^{\max},\ldots,\mathbf w_m^{\max})$, and recall $\hat Z^{\min}$ and $\hat Z^{\max}$ in Step 3 of Algorithms \ref{algo2} and \ref{algo3}. Under Assumptions \ref{balanced data}, \ref{non-degeneracy} and \ref{8th moment}, as the input data size $n\to\infty$ and simulation effort $\frac{R_1}{n}\to\infty,R_2\to\infty$, the outputs $\mathscr L^{EEL},\mathscr U^{EEL}$ of Algorithm \ref{algo2} satisfy
\begin{equation*}
    \begin{aligned}
    \mathscr L^{EEL}&=Z^*+\sum_{i=1}^m\bar{G}_i+(\hat{Z}^{\min}-Z^{\min})- z_{1-\alpha/2}\Big(\sigma_I+\frac{\sigma}{\sqrt{R_2}}\Big)+o_p\big(\frac{1}{\sqrt{n}}+\frac{1}{\sqrt{R_2}}\big)\\
    \mathscr U^{EEL}&=Z^*+\sum_{i=1}^m\bar{G}_i+(\hat{Z}^{\max}-Z^{\max}) +z_{1-\alpha/2}\Big(\sigma_I+\frac{\sigma}{\sqrt{R_2}}\Big)+o_p\big(\frac{1}{\sqrt{n}}+\frac{1}{\sqrt{R_2}}\big)
    \end{aligned}
\end{equation*}
whereas the outputs $\mathscr L^{FEL},\mathscr U^{FEL}$ of Algorithm \ref{algo3} satisfy
\begin{equation*}
    \begin{aligned}
    \mathscr L^{FEL}&=Z^*+\sum_{i=1}^m\bar{G}_i+(\hat{Z}^{\min}-Z^{\min})- z_{1-\alpha/2}\sqrt{\sigma_I^2+\frac{\sigma^2}{R_2}}+o_p\big(\frac{1}{\sqrt{n}}+\frac{1}{\sqrt{R_2}}\big)\\
    \mathscr U^{FEL}&=Z^*+\sum_{i=1}^m\bar{G}_i+(\hat{Z}^{\max}-Z^{\max}) +z_{1-\alpha/2}\sqrt{\sigma_I^2+\frac{\sigma^2}{R_2}}+o_p\big(\frac{1}{\sqrt{n}}+\frac{1}{\sqrt{R_2}}\big)
    \end{aligned}
\end{equation*}
where $\sigma_I^2=\sum_{i=1}^m\mathrm{Var}(G_i(X_i))/n_i$ is as defined in Corollary \ref{CI: linear approximation truth}, $\sigma^2=\mathrm{Var}_{P_1,\ldots,P_m}(h(\mathbf X_1,\ldots,\mathbf X_m))$ is the output variance, and the $o_p$ is with respect to the joint randomness of the data and the simulation.
\end{proposition}
To see how these decompositions arise, we can write the outputs of Algorithm \ref{algo2} as (for the lower bound, say) $\hat Z^{\min}-z_{1-\alpha/2}\hat\sigma_{\min}/\sqrt{R_2}=Z^{\min}+(\hat Z^{\min}-Z^{\min})-z_{1-\alpha/2}\hat\sigma_{\min}/\sqrt{R_2}$ , where $Z^{\min}$, by Theorem \ref{mainresult}, is close to $\mathscr L$ that is in turn representable as $Z^*+\sum_{i=1}^m\bar G_i-z_{1-\alpha/2}\sigma_I$ up to a small error, by Theorem \ref{ideal CI}. Noting that $\hat\sigma_{\min}$ approximates $\sigma$, these together show the representation for $\mathscr L^{EEL}$ in Proposition \ref{algo2/3 representation}. The other expressions for $\mathscr U^{EEL}$, and $\mathscr L^{FEL}$, $\mathscr U^{FEL}$, follow analogously using the adjustments shown in Algorithms \ref{algo2} and \ref{algo3}.

We briefly discuss how Proposition \ref{algo2/3 representation} leads to Theorems \ref{erfreeCI2} and \ref{erfreeCI3}. Note that for FEL, the term $z_{1-\alpha/2}\sqrt{\sigma_I^2+\sigma^2/R_2}$ in $\mathscr L^{FEL}$ or $\mathscr U^{FEL}$ is the standard error term in a normality-based CI that comprises the uncertainties from two independent sources with variances $\sigma_I^2$ and $\sigma^2/R_2$. The two terms $Z^*+\sum_{i=1}^m\bar G_i$ and $\hat{Z}^{\min}-Z^{\min}$ in the expressions of $\mathscr L^{FEL}$,  $\mathscr U^{FEL}$, which contain the input error and the simulation error in Step 3 respectively, possess variances that are approximately $\sigma_I^2$ and $\sigma^2/R_2$. Thus the representations of $\mathscr L^{FEL}$ and $\mathscr U^{FEL}$  each matches the lower and upper bound of a normality-based CI. This almost gives an asymptotically exact CI, except that the quantities $\mathscr L^{FEL}$ and $\mathscr U^{FEL}$ contain some common, and some independent, sources of randomness in their construction that slightly corrupts the coverage. This leads to Theorem \ref{erfreeCI3}. The argument for EEL in Theorem \ref{erfreeCI2} follows similarly, but with the standard error term in $\mathscr L^{EEL}$ or $\mathscr U^{EEL}$ overestimating the uncertainty by a factor as large as $\sqrt 2$ (because  $1\leq \frac{\sigma_I+\sigma/\sqrt{R_2}}{\sqrt{\sigma_I^2+\sigma^2/R_2}}\leq \sqrt{2}$, where $\sqrt{2}$ is attained when $\sigma_I^2=\sigma^2/R_2$). In fact, under a coupling between all the simulation runs in Algorithms \ref{algo2} and \ref{algo3}, $\hat{\sigma}_{\min}/\sqrt{R_2}$ always upper bounds $\sqrt{\hat{\sigma}_I^2+\hat{\sigma}_{\min}^2/R_2}-\hat{\sigma}_I$ and hence Algorithm \ref{algo2} always generates wider CIs than Algorithm \ref{algo3}.

\section{Numerical Experiments}\label{sec:numerics}
%In this section we compare our approach with several existing methods through simulation experiments. 
We present some numerical results for Algorithm \ref{algo1} (BEL), Algorithm \ref{algo2} (EEL) and Algorithm \ref{algo3} (FEL). These include coverage probabilities and the statistical indicators, such as mean and standard deviation, of the positions or widths of the resulting CIs. We conduct experiments on two settings, a queueing model  in Section \ref{sec:queue} and stochastic activity networks in Section  \ref{sec:stochastic}. We consider various levels of simulation budgets, data sizes, and problem dimensions (i.e., number of estimated input models). Throughout this section we set the target confidence level  to $95\%$.

% The first one to compare with is the widely used bootstrap resampling in constructing input-induced CIs. The bootstrap is a general technique to approximate sampling distribution of a statistic from which CIs or other statistical quantities can be constructed (\cite{efron1992bootstrap}). One common approach, in the context of simulation input uncertainty, is The percentile bootstrap used in \cite{barton1993uniform,barton2001resampling}. with each variate from input model $i$ uniformly generated over $\{X_{i,1}^l,\ldots,X_{i,n_i}^l\}$, and take their average $Z^l$. In other words, drive the $R_b$ simulation runs using the bootstrapped empirical distributions 

We also compare our procedures with three methods: 
\begin{enumerate}
\item Percentile bootstrap resampling (``standard BT"): This scheme is suggested in \cite{barton1993uniform,barton2001resampling}. Given $m$ input data sets $\{X_{1,1},\ldots,X_{1,n_1}\},\ldots,\{X_{m,1},\ldots,X_{m,n_m}\}$, it proceeds as follows. First choose $B$, the number of bootstrap resamples of the input empirical distributions, and $R_b$, the number of simulation replications for each bootstrap resample. For each $l=1,2,\ldots,B$, draw a simple random sample of size $n_i$ with replacement, denoted by $\{X^l_{i,1},\ldots,X^l_{i,n_i}\}$, for each input model $i$, then generate $R_b$ simulation replications driven by the empirical distributions formed by $\{X^l_{i,1},\ldots,X^l_{i,n_i}\},i=1,\ldots,m$, and take their average to obtain $Z^l$. Finally output the $0.025(B+1)$-th and $0.975(B+1)$-th order statistics of $\{Z^l\}_{l=1}^B$.
% \item  $Z_{(\lfloor 0.025(B+1)\rfloor)}$ and $Z_{(\lfloor0.975(B+1)\rfloor)}$, as the lower and upper limits of the CI. % Section \ref{sec:accuracy} investigates the validity and statistical accuracy of the proposed approach. Section \ref{sec:comparison} compares our performance with the bootstrap and the delta method. Throughout this section, we consider the following two queueing systems. and less to the rest  to generate more accurate estimates of the quantiles.  splits a given simulation budget evenly to different resamples, some simulation runs are used to  (which are closer to the target quantiles)The detailed procedure is described in Algorithm \ref{adap_BT} in Section \ref{sec:adap_BT_procedure} of the Appendix. in order to obtain accurate CIs each randomly pickedin the tables below 
\item Adaptive percentile bootstrap (``adaptive BT''): Proposed by \cite{yi2017efficient}, this approach adaptively allocates simulation budget in order to obtain percentile bootstrap CIs more efficiently than the standard percentile bootstrap. It aims to allocate more simulation runs to the resamples whose corresponding performance measures are closer to the $0.025$ or $0.975$ quantiles. The procedure consists of two phases. The first phase uses simulation to sequentially screen out bootstrap resamples that will less likely give the target quantiles. The second phase allocates the remaining simulation budget to the surviving resamples to more accurately estimate their performance measures. For a given simulation budget, the tuning parameters $B,n_0,r,M$ (see \cite{yi2017efficient}) are needed. In our subsequent comparisons we offer it some advantages by randomly drawing $10$ different combinations of these parameters from a broad enough range of values, and reporting results on the top combinations ranked by the closeness of the coverage level to the nominal level.
\item The nonparametric delta method: This method has not been explicitly suggested in the simulation literature (in the nonparametric regime), and here we provide a heuristic version inspired from our analyses. The CI takes the form $\hat{Z}\pm z_{1-\frac{\alpha}{2}}\sqrt{\text{input-induced variance}+\text{stochastic variance}}$ where $\hat Z$ is an estimate of the performance measure under the empirical input distributions. We estimate the stochastic variance using the sample variance of the generated simulation replications, and estimate the input-induced variance using the $\hat\sigma_I$ in Algorithm \ref{algo3}. To be specific, we carry out Step 1 of Algorithm \ref{algo1} with $R_1=R_d$, and then construct the CI
%empirical input distributions to generate $R_d$ i.i.d.~replications of the output $h^r=h(\mathbf X_1^r,\ldots,\mathbf X_m^r)$, use them to estimate the gradient $\hat{G}_i(X_{i,j})$, and then compute the CI that accounts for both input and simulation uncertainty
\begin{equation*}
\hat{Z}\pm z_{1-\frac{\alpha}{2}}\sqrt{\frac{\hat{\sigma}^2}{R_d}+\sum_{i=1}^m\frac{1}{n_i}\Big(\frac{1}{n_i}\sum_{j=1}^{n_i}\big(\dbhat{G}_{i}(X_{i,j})\big)^2-\frac{n_iT_i\hat{\sigma}^2}{R_d}\Big)}
\end{equation*}
where $\hat{Z}$ and $\hat{\sigma}^2$ are respectively the sample mean and variance of the $R_d$ simulation replications.
\end{enumerate}

We will detail our comparisons under various problem and algorithmic configurations in the two experimental setups that follow. After that, in Section \ref{sec:bootstrap}, we summarize some highlights and provide further comparisons with the bootstrap.
% \textcolor{red}{\textbf{Move the following discussion to the introduction?}:\cite{barton1993uniform,barton2001resampling} also proposed what they called the method of uniformly randomized empirical distribution function (EDF), which entails building resamples of input empirical distributions by allocating probability weights on all data points that are randomly assigned from an ordered sequence of uniform distributions. This technique, equivalent to the so-called Bayesian bootstrap (\cite{rubin1981bayesian}), is structurally more similar to our method since they both involve assigning probability weights on the support of the data. \cite{barton1993uniform,barton2001resampling} used $R_b=1$, as the scale of the input noise dominated that of the stochastic noises in the example systems they considered. They concluded from their experiments that the percentile bootstrap and the uniformly randomized EDF gave rise to similar numerical performances and computational loads.}

%reporting top few whose estimated coverage probabilities are the closest to the nominal level in the tables below.
%$B\in\{100,200,300,400,500\},n_0\in\{5,10,15,20,25\},r\in\{1.2,1.4,1.6,1.8,2\},M\in\{1,3,5,7,9\}$

\subsection{Mean Waiting Time of an M/M/1 Queue}\label{sec:queue}
We first consider a canonical M/M/1 queue with arrival rate $0.95$ and service rate $1$. The system is empty when the first customer comes in. We set our target performance measure as the expected waiting time of the $10$-th customer. To put it in the form of \eqref{perfm}, let $A_t$ be the inter-arrival time between the $t$-th and $(t+1)$-th customers, $S_t$ be the service time of the $t$-th customer, and
\begin{equation*}
h(A_1,A_2,\ldots,A_{9},S_1,S_2,\ldots,S_{9})=W_{10},\label{waitprob}
\end{equation*}
where the waiting time $W_{10}$ is calculated via the Lindley recursion
\begin{equation*}
W_1=0,W_{t+1}=\max \{W_t+S_t-A_t,0\},\text{ for }t=1,\ldots,9.
\end{equation*}
Both the inter-arrival time distribution and the service time distribution are assumed unknown. Table \ref{mm1_small} shows the results of all the methods under a simulation budget $2000$ and input data sizes $n_1=30,n_2=25$. Table \ref{mm1_large} summarizes results under a budget $8000$ and data sizes $n_1=120,n_2=100$. For each row of the tables, $1000$ i.i.d.~input data sets are drawn from the true input distributions, and then a CI is constructed from each of them, from which the coverage probability, mean CI length and standard deviation of CI length are estimated. The word ``overshoot" means that the CI limits exceed the natural bounds of the performance measure, i.e., the lower bound being negative given that waiting time must be non-negative.

We test the coverage probabilities of the optimization-based CIs. For each of Tables \ref{mm1_small} and \ref{mm1_large}, we compute a ``benchmark" coverage of each method by generating $5000$ CIs each of which consumes $5\times 10^4$ simulation runs, to approximate the simulation-error-free coverage for comparison (the bracketed number underneath the name of each method in the tables). We observe first that the benchmark coverage of our optimization-based CIs are close to the nominal value $95\%$ in both tables (roughly $92\%$ in Table \ref{mm1_small} and $94\%$ in Table \ref{mm1_large}), which provides a sanity check for the validity of the EL method in our setting. Moreover, consistent with the asymptotic results, the benchmark coverage is closer to $95\%$ when the data size is bigger (Table  \ref{mm1_large}). Second, under the simulation budget of the experiments, Tables \ref{mm1_small} and \ref{mm1_large} show that in general BEL under-covers compared to the benchmark, EEL over-covers, whereas FEL is accurate (note that a performance close to the benchmark, instead of the nominal level, indicates the power of the procedure to jointly handle input and simulation errors, as the benchmark provides in a sense the best performance that is free of the simulation errors). For instance, in Table \ref{mm1_large} where the benchmark coverage of the EL method is $93.7\%$, BEL varies from $90\%$ to $92\%$, EEL ranges from $96\%$ to $99\%$, whereas FEL stays around $94\%$. This phenomenon is in line with Theorems \ref{erfreeCI1}, \ref{erfreeCI2} and \ref{erfreeCI3}  since, as we have discussed in Sections \ref{sec:stat guarantees} and \ref{sec:evaluation}, BEL does not take into account the stochastic uncertainty in the final evaluation, EEL captures the stochastic uncertainty but in a conservative manner, while FEL is designed to tightly match the magnitude of the uncertainty. The under-coverage issue of BEL and the over-coverage issue of EEL, especially for the larger-data case (Table \ref{mm1_large}), become more severe when $R_2$ is chosen small, while FEL delivers accurate coverage for all considered parameter values. Thus FEL seems to be more reliable over the other two procedures when the user has a limited simulation budget.

\begin{table}[h]
\caption{$M/M/1$ queue. $n_1=30,n_2=25$. Total simulation budget $2000$. Run times (second/CI): three EL methods $1.1\times 10^{-2}$, the bootstrap $1.2\times 10^{-2}$, delta method $1.0\times 10^{-2}$.}
% \caption{System $\sharp 1$. $n_1=30,n_2=25$. Total simulation budget $2000$. Run times (second/CI): delta method $1.0\times 10^{-2}$, bootstrap $(0.8\sim1.6)\times 10^{-2}$, three EL methods $(1.0\sim1.2)\times 10^{-2}$.}
\begin{center}
\begin{tabular}{|l|l|llll|}
\hline
\multicolumn{2}{|c|}{\tabincell{l}{methods \&\\parameters}}&\tabincell{l}{coverage\\ estimate}&\tabincell{l}{mean CI\\length}&\tabincell{l}{std. CI\\length}&\tabincell{l}{\% of\\overshoot}\\\hline
\multirow{4}{*}{\tabincell{l}{BEL\\ $(91.8\%^*)$}}&$R_1=1000,R_2=500$&$89.6\%$&$4.76$&$2.17$&$0\%$\\\cline{2-2}
&$R_1=1500,R_2=250$&$90.7\%$&$4.72$&$1.99$&$0\%$\\\cline{2-2}
&$R_1=1800,R_2=100$&$88.7\%$&$4.76$&$2.15$&$0\%$\\\cline{2-2}
&$R_1=1900,R_2=50$&$89.2\%$&$4.79$&$2.24$&$0\%$\\\hline
\multirow{4}{*}{\tabincell{l}{EEL\\ $(91.8\%^*)$}}&$R_1=1000,R_2=500$&$93.1\%$&$5.21$&$2.19$&$0\%$\\\cline{2-2}
&$R_1=1500,R_2=250$&$94.1\%$&$5.38$&$2.21$&$0\%$\\\cline{2-2}
&$R_1=1800,R_2=100$&$95.1\%$&$5.67$&$2.42$&$0\%$\\\cline{2-2}
&$R_1=1900,R_2=50$&$96.0\%$&$6.16$&$2.64$&$0.1\%$\\\hline
\multirow{4}{*}{\tabincell{l}{FEL\\ $(91.8\%^*)$}}&$R_1=1000,R_2=500$&$90.5\%$&$4.72$&$2.06$&$0\%$\\\cline{2-2}
&$R_1=1500,R_2=250$&$91.9\%$&$4.83$&$2.07$&$0\%$\\\cline{2-2}
&$R_1=1800,R_2=100$&$91.9\%$&$4.93$&$2.08$&$0\%$\\\cline{2-2}
&$R_1=1900,R_2=50$&$91.5\%$&$5.06$&$2.20$&$0\%$\\\hline
\multirow{4}{*}{\tabincell{l}{standard BT\\ $(91.0\%^*)$}}&$B=50,R_b=40$&$91.2\%$&$4.90$&$2.23$&$0\%$\\\cline{2-2}
&$B=100,R_b=20$&$93.5\%$&$4.98$&$2.02$&$0\%$\\\cline{2-2}
&$B=400,R_b=5$&$96.9\%$&$6.09$&$2.28$&$0\%$\\\cline{2-2}
&$B=1000,R_b=2$&$99.2\%$&$7.74$&$2.82$&$0\%$\\\hline
\multirow{4}{*}{\tabincell{l}{adaptive BT \\(4 best combinations)\\ $(91.0\%^*)$}}&$B=100,n_0=10,r=1.2,M=3$&$92.7\%$&$5.01$&$2.18$&$0\%$\\\cline{2-2}
&$B=100,n_0=10,r=1.2,M=1$&$92.0\%$&$5.02$&$2.22$&$0\%$\\\cline{2-2}
&$B=100,n_0=10,r=1.4,M=1$&$92.3\%$&$4.93$&$2.08$&$0\%$\\\cline{2-2}
&$B=100,n_0=10,r=1.8,M=1$&$92.5\%$&$5.00$&$2.24$&$0\%$\\\hline
\tabincell{l}{nonparametric delta\\method\\ $(86.6\%^*)$}&$R_d=2000$&$84.9\%$&$4.66$&$2.08$&$54\%$\\\hline
\multicolumn{6}{c}{$*$ denotes the benchmark coverage with negligible simulation noise.}
\end{tabular}
\end{center}
\label{mm1_small}
\end{table}

\begin{table}[h]
\caption{$M/M/1$ queue. $n_1=120,n_2=100$. Total simulation budget $8000$. Run times (second/CI): three EL methods $4.0\times 10^{-2}$, the bootstrap $3.4\times 10^{-2}$, delta method $5.3\times 10^{-2}$.}
% \caption{System $\sharp 1$. $n_1=120,n_2=100$. Total simulation budget $8000$. Run times (second/CI): delta method $5.3\times 10^{-2}$, bootstrap $(2.9\sim3.8)\times 10^{-2}$, three EL methods $(3.0\sim5.0)\times 10^{-2}$}
\begin{center}
\begin{tabular}{|l|l|llll|}
\hline
\multicolumn{2}{|c|}{\tabincell{l}{methods \&\\parameters}}&\tabincell{l}{coverage\\ estimate}&\tabincell{l}{mean CI\\length}&\tabincell{l}{std. CI\\length}&\tabincell{l}{\% of\\overshoot}\\\hline
\multirow{4}{*}{\tabincell{l}{BEL\\ $(93.7\%^*)$}}&$R_1=4000,R_2=2000$&$92.6\%$&$2.47$&$0.597$&$0\%$\\\cline{2-2}
&$R_1=7000,R_2=500$&$92.4\%$&$2.46$&$0.606$&$0\%$\\\cline{2-2}
&$R_1=7800,R_2=100$&$91.9\%$&$2.48$&$0.713$&$0\%$\\\cline{2-2}
&$R_1=7900,R_2=50$&$89.6\%$&$2.45$&$0.787$&$0\%$\\\hline
\multirow{4}{*}{\tabincell{l}{EEL\\ $(93.7\%^*)$}}&$R_1=4000,R_2=2000$&$95.7\%$&$2.66$&$0.626$&$0\%$\\\cline{2-2}
&$R_1=7000,R_2=500$&$97.7\%$&$2.90$&$0.678$&$0\%$\\\cline{2-2}
&$R_1=7800,R_2=100$&$98.0\%$&$3.50$&$0.870$&$0\%$\\\cline{2-2}
&$R_1=7900,R_2=50$&$98.8\%$&$3.94$&$1.04$&$0\%$\\\hline
\multirow{4}{*}{\tabincell{l}{FEL\\ $(93.7\%^*)$}}&$R_1=4000,R_2=2000$&$93.6\%$&$2.45$&$0.591$&$0\%$\\\cline{2-2}
&$R_1=7000,R_2=500$&$94.3\%$&$2.45$&$0.594$&$0\%$\\\cline{2-2}
&$R_1=7800,R_2=100$&$94.1\%$&$2.74$&$0.705$&$0\%$\\\cline{2-2}
&$R_1=7900,R_2=50$&$94.3\%$&$2.90$&$0.865$&$0\%$\\\hline
\multirow{4}{*}{\tabincell{l}{standard BT\\ $(94.2\%^*)$}}&$B=50,R_b=160$&$92.7\%$&$2.56$&$0.675$&$0\%$\\\cline{2-2}
&$B=100,R_b=80$&$96.4\%$&$2.64$&$0.613$&$0\%$\\\cline{2-2}
&$B=400,R_b=20$&$98.8\%$&$3.19$&$0.658$&$0\%$\\\cline{2-2}
&$B=1000,R_b=8$&$100\%$&$4.19$&$0.800$&$0\%$\\\hline
\multirow{4}{*}{\tabincell{l}{adaptive BT\\(4 best combinations)\\ $(94.2\%^*)$}}&$B=200,n_0=20,r=1.6,M=1$&$93.6\%$&$2.64$&$0.657$&$0\%$\\\cline{2-2}
&$B=200,n_0=15,r=2,M=1$&$95.0\%$&$2.68$&$0.687$&$0\%$\\\cline{2-2}
&$B=200,n_0=5,r=1.6,M=3$&$94.5\%$&$2.71$&$0.688$&$0\%$\\\cline{2-2}
&$B=400,n_0=10,r=1.8,M=1$&$94.5\%$&$2.72$&$0.654$&$0\%$\\\hline
\tabincell{l}{nonparametric delta\\method\\ $(91.5\%^*)$}&$R_d=8000$&$92.0\%$&$2.45$&$0.560$&$0\%$\\\hline
\multicolumn{6}{c}{$*$ denotes the benchmark coverage with negligible simulation noise.}
\end{tabular}
\end{center}
\label{mm1_large}
\end{table}

We compare our methods with the percentile bootstrap procedures in terms of coverage accuracy and algorithmic configuration. The benchmark coverages of our methods and the bootstrap appear to be quite similar in all considered cases (within $1\%$ in both Tables \ref{mm1_small} and \ref{mm1_large}). Moreover, the bootstrap methods perform competitively in terms of the actual coverages, when the budget allocation or tuning parameters are optimally chosen. Nonetheless, FEL appears to show more robust performance with respect to these tuning needs. In the standard bootstrap, when $R_b$ is chosen large relative to the data size and $B$ is set around $50$, the coverages of the CIs are close to the benchmark coverages in all cases. However, as $R_b$ decreases, the coverage probabilities of bootstrap CIs quickly rise towards $100\%$. This over-coverage issue can be attributed to the higher variability caused by small $R_b$ that is not properly accounted for, as discussed in \cite{barton2002panel} and \cite{barton2007presenting}. The adaptive bootstrap appears to mitigate this issue by more efficient allocation of the budget. It requires, however, a careful selection of the best parameter configurations (while the tables show the top four configurations, the worst case among our randomly selected $10$ choices has a coverage of $80\%$). In practice these parameters needs to be obtained via discrete simulation optimization \citep{yi2017efficient}. In contrast, the coverage probabilities of FEL stay almost unchanged under various budget allocations (including the case that $R_2$ is as small as $50$). FEL thus seems easy to use in terms of algorithmic configuration; in particular, merely setting $R_2=50$ appears doing well. 

% Compared to the bootstrap, FEL also appears to require less simulation budget to stabilize, or in other words, to achieve the benchmark coverage. 

To further illustrate the robustness of the proposed approach in terms of algorithmic configurations, relative to the bootstrap, we show in Table \ref{mm1_stable} the coverages as we increase the simulation budget. The first row shows the coverage estimates of the bootstrap and FEL under allocations that satisfy the same overall simulation budget. Both appear to be close to their respective benchmark coverages shown in Table \ref{mm1_small}. However, the coverages of the bootstrap could be illusory in this case since, as the bootstrap size $B$ increases with $R_b$ fixed, the coverage rises from $91\%$ to $95\%$ as shown in the following rows. These deviate from the benchmark coverages, and indicate that neither $B$ nor $R_b$ is large enough for the bootstrap to work properly. In contrast, the coverage of FEL appears quite stable and remains close to the benchmark when $R_1$ or $R_2$ increases. 
% Given the similar benchmark coverages between the bootstrap and our methods, FEL seems to elicit in overall a more robust and lighter simulation effort than the bootstrap.

 \begin{table}[h]
 \caption{$M/M/1$ queue. $n_1=30,n_2=25$.}
 \begin{center}
 \begin{tabular}{|l|l|l|l|}
 \hline
 \multicolumn{2}{|c|}{standard BT}&\multicolumn{2}{c|}{FEL}\\\hline
 parameters&\tabincell{l}{coverage\\ estimate}&parameters&\tabincell{l}{coverage\\ estimate}\\\hline
 $B=40,R_b=15$&$90.9\%$&$R_1=500,R_2=50$&$90.3\%$\\\hline
 $B=100,R_b=15$&$92.4\%$&$R_1=2000,R_2=50$&$91.9\%$\\\hline
 $B=200,R_b=15$&$93.6\%$&$R_1=500,R_2=200$&$90.2\%$\\\hline
 $B=500,R_b=15$&$94.7\%$&$R_1=2000,R_2=200$&$90.8\%$\\\hline
 \end{tabular}
 \end{center}
 \label{mm1_stable}
 \end{table}

% \begin{table}[h]
% \caption{System $\sharp 2$. $n_1=30,n_2=25,n_3=30,n_4=25$.}
% \begin{center}
% \begin{tabular}{|l|l|l|l|}
% \hline
% \multicolumn{2}{|c|}{bootstrap}&\multicolumn{2}{c|}{FEL}\\\hline
% parameters&\tabincell{l}{coverage\\ estimate}&parameters&\tabincell{l}{coverage\\ estimate}\\\hline
% $B=37,R_b=30$&$93.8\%$&$R_1=1000,R_2=50$&$92.7\%$\\\hline
% $B=100,R_b=30$&$96.6\%$&$R_1=4000,R_2=50$&$93.1\%$\\\hline
% $B=200,R_b=30$&$96.7\%$&$R_1=1000,R_2=200$&$91.5\%$\\\hline
% $B=400,R_b=30$&$96.9\%$&$R_1=4000,R_2=200$&$92.1\%$\\\hline
% \end{tabular}
% \end{center}
% \label{two_smalldata_comp}
% \end{table}

Compared to the nonparametric delta method, our optimization-based CIs possess better coverages, especially in the situation of limited input data size. When the data size is less than $30$ for each input model (Table \ref{mm1_small}), the coverage probabilities of the delta-method CIs are around $85\%$, while our methods are around $90\%$ to $96\%$, depending on the particular variants. The unsatisfactory coverage of the delta-method CI could be attributed to the overshoot issue. Table \ref{mm1_small} shows that frequently the delta-method CI exceeds the natural bounds of the target performance measure, which renders its effective length shorter and hence an inferior coverage. The coverage gets much better for the delta-method CI when input data size rises above $100$ (Table \ref{mm1_large}), which gets close to, but still falls short of, our optimization-based counterparts especially FEL.

\subsection{Stochastic Activity Networks} \label{sec:stochastic}
\begin{figure}[h]
    \begin{subfigure}[b]{.5\linewidth}
    \centering
    \begin{tikzpicture}[scale=0.75,auto=left,every node/.style={circle,draw=black}]
  \node (n1) at (1,3)  {1};
  \node (n2) at (4,5)  {2};
  \node (n3) at (4,1)  {3};
  \node (n4) at (7,3)  {4};

%   \draw [->] (n1) -- node[sloped,font=\small,left] {x} (n2);
%   \draw [->] (n1) -- (n3);
%   \draw [->] (n2) -- (n3);
%   \draw [->] (n2) -- (n4);
%   \draw [->] (n3) -- (n4);
  
\path[->,every node/.style={}]
    (n1) edge node [above] {$X_1$} (n2)
    (n2) edge node [left] {$X_2$} (n3)
    (n1) edge node [below] {$X_3$} (n3)
    (n2) edge node [above] {$X_4$} (n4)
    (n3) edge node [below] {$X_5$} (n4);
\end{tikzpicture}
\caption{$4$ nodes and $m=5$ tasks.}\label{m=5}
\end{subfigure}%
\hfill
    \begin{subfigure}[b]{.5\linewidth}
    \centering
\begin{tikzpicture}[scale=0.55,auto=left,every node/.style={circle,draw=black}]
  \node (n1) at (1,4)  {1};
  \node (n2) at (3,7)  {2};
  \node (n3) at (3,5)  {3};
  \node (n4) at (3,2)  {4};
  \node (n5) at (6,7)  {5};
  \node (n6) at (6,5)  {6};
  \node (n7) at (6,3)  {7};
  \node (n8) at (6,1)  {8};
  \node (n9) at (8.5,6)  {9};
  \node (n10) at (10,4)  {10};

%   \draw [->] (n1) -- node[sloped,font=\small,left] {x} (n2);
%   \draw [->] (n1) -- (n3);
%   \draw [->] (n2) -- (n3);
%   \draw [->] (n2) -- (n4);
%   \draw [->] (n3) -- (n4);
  
\path[->,every node/.style={}]
    (n1) edge node [left] {$X_1$} (n2)
    (n1) edge node [below] {$X_2$} (n3)
    (n1) edge node [left] {$X_3$} (n4)
    (n2) edge node [above] {$X_4$} (n5)
    (n2) edge node [above left] {$X_5$} (n6)
    (n3) edge node [below left] {$X_6$} (n5)
    (n3) edge node [below] {$X_7$} (n6)
    (n4) edge node [above] {$X_8$} (n7)
    (n4) edge node [below] {$X_9$} (n8)
    (n5) edge node [above] {$X_{10}$} (n9)
    (n6) edge node [below] {$X_{11}$} (n9)
    (n7) edge node [above] {$X_{12}$} (n10)
    (n8) edge node [below] {$X_{13}$} (n10)
    (n9) edge node [right] {$X_{14}$} (n10);
\end{tikzpicture}
\caption{$10$ nodes and $m=14$ tasks.}\label{m=14}
\end{subfigure}
\caption{Stochastic activity networks.}\label{network}
\end{figure}
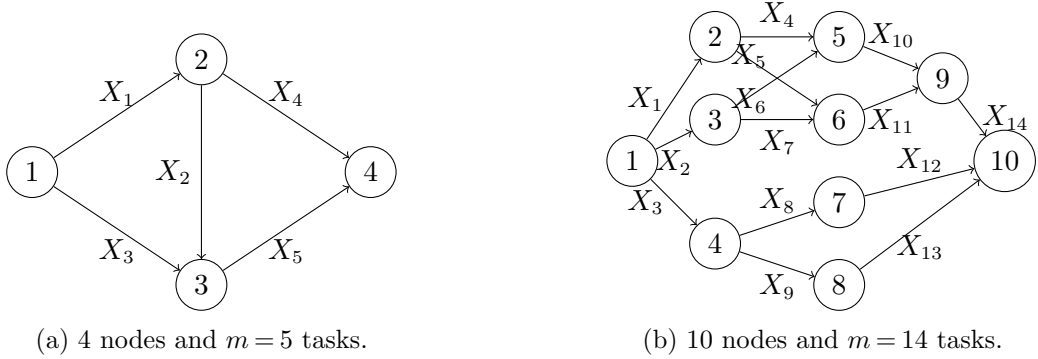
We consider a larger-scale problem and larger ranges of data sizes, in the setting of stochastic activity networks shown in Figure \ref{network}. The first network Figure \ref{m=5} is borrowed from \cite{yi2017efficient}. Each edge $i=1,\ldots,5$ of the network represents a task that can be completed in $X_i$ units of time. Assigning each $X_i$ to edge $i$ as its length, the total time to finish the project is the length of the longest path from node $1$ to node $4$, i.e. $h(X_1,\ldots,X_5)=\max\{X_1+X_2+X_5,X_1+X_4,X_3+X_5\}$. Assume that the unknown distributions of the $X_i$'s are exponential with rate $10,5,12,11,5$ for $i$ from $1$ to $5$, and we are interested in computing the expected time to finish the project $\mathbb E[h(X_1,\ldots,X_5)]$.

\begin{table}[h]
\caption{Stochastic activity network in Figure \ref{m=5}. $n_1=n_2=200,n_3=n_4=n_5=30$. Total simulation budget $8000$. Run times (second/CI): three EL methods $3.3\times 10^{-2}$, the bootstrap $1.7\times 10^{-2}$, delta method $3.2\times 10^{-2}$.}
\begin{center}
\begin{tabular}{|l|l|llll|}
\hline
\multicolumn{2}{|c|}{\tabincell{l}{methods \&\\parameters}}&\tabincell{l}{coverage\\ estimate}&\tabincell{l}{mean CI\\length}&\tabincell{l}{std. CI\\length}&\tabincell{l}{\% of\\overshoot}\\\hline
\multirow{4}{*}{BEL}&$R_1=4000,R_2=2000$&$92.7\%$&$0.17$&$0.03$&$0\%$\\\cline{2-2}
&$R_1=7000,R_2=500$&$91.9\%$&$0.17$&$0.04$&$0\%$\\\cline{2-2}
&$R_1=7800,R_2=100$&$84.9\%$&$0.18$&$0.06$&$0\%$\\\cline{2-2}
&$R_1=7900,R_2=50$&$81.7\%$&$0.18$&$0.07$&$0\%$\\\hline
\multirow{4}{*}{EEL}&$R_1=4000,R_2=2000$&$96.1\%$&$0.20$&$0.03$&$0\%$\\\cline{2-2}
&$R_1=7000,R_2=500$&$97.7\%$&$0.23$&$0.04$&$0\%$\\\cline{2-2}
&$R_1=7800,R_2=100$&$99.0\%$&$0.30$&$0.07$&$0\%$\\\cline{2-2}
&$R_1=7900,R_2=50$&$99.4\%$&$0.35$&$0.09$&$0\%$\\\hline
\multirow{4}{*}{FEL}&$R_1=4000,R_2=2000$&$92.2\%$&$0.17$&$0.03$&$0\%$\\\cline{2-2}
&$R_1=7000,R_2=500$&$93.2\%$&$0.18$&$0.04$&$0\%$\\\cline{2-2}
&$R_1=7800,R_2=100$&$94.6\%$&$0.22$&$0.06$&$0\%$\\\cline{2-2}
&$R_1=7900,R_2=50$&$94.5\%$&$0.25$&$0.08$&$0\%$\\\hline
\multirow{4}{*}{standard BT}&$B=50,R_b=160$&$94.0\%$&$0.21$&$0.04$&$0\%$\\\cline{2-2}
&$B=100,R_b=80$&$97.1\%$&$0.22$&$0.04$&$0\%$\\\cline{2-2}
&$B=400,R_b=20$&$99.7\%$&$0.33$&$0.04$&$0\%$\\\cline{2-2}
&$B=1000,R_b=8$&$100\%$&$0.47$&$0.05$&$0\%$\\\hline
\multirow{4}{*}{\tabincell{l}{adaptive BT\\(4 best combinations)}}&$B=300,n_0=15,r=1.2,M=1$&$94.9\%$&$0.22$&$0.04$&$0\%$\\\cline{2-2}
&$B=100,n_0=20,r=1.2,M=1$&$93.9\%$&$0.22$&$0.05$&$0\%$\\\cline{2-2}
&$B=400,n_0=10,r=1.2,M=3$&$95.6\%$&$0.24$&$0.04$&$0\%$\\\cline{2-2}
&$B=100,n_0=5,r=1.2,M=3$&$96.2\%$&$0.22$&$0.04$&$0\%$\\\hline
\tabincell{l}{nonparametric delta\\method}&$R_d=8000$&$94.9\%$&$0.18$&$0.03$&$0\%$\\\hline
\end{tabular}
\end{center}
\label{network_5_imbalance}
\end{table}

We test our method in cases where the data sizes for different input models vary significantly. Specifically we consider the case where $n_1=n_2=200$ and $n_3=n_4=n_5=30$, which produce a ratio of roughly $7$ between the maximal and minimal data sizes. Table \ref{network_5_imbalance} shows the results under a simulation budget of $8000$. All the methods seem to exhibit performances similar to the cases with more balanced observations in Tables \ref{mm1_small} and \ref{mm1_large}. For example, FEL and the adaptive bootstrap generate CIs with similar coverage probabilities (around the nominal level $95\%$), EEL and the standard bootstrap tend to over-cover, and BEL tends to under-cover especially for small values of $R_2$. In contrast to the last example, the nonparametric delta method in this case seems to have a good performance that is similar to our FEL. This could be because the performance function $h$ here is piecewise linear with only three pieces, hence can be well approximated by a single linear function and in turn leads to the better finite-sample performance of the delta method that relies crucially on linearization.

\begin{table}[h]
\caption{Stochastic activity network in Figure \ref{m=14}. $n_i=30$ for $1\leq i\leq 7$ and $25$ for $8\leq i\leq 14$. Total simulation budget $4000$. Run times (second/CI): three EL methods $2.7\times 10^{-2}$, the bootstrap $2.7\times 10^{-2}$, delta method $1.7\times 10^{-2}$.}
\begin{center}
\begin{tabular}{|l|l|llll|}
\hline
\multicolumn{2}{|c|}{\tabincell{l}{methods \&\\parameters}}&\tabincell{l}{coverage\\ estimate}&\tabincell{l}{mean CI\\length}&\tabincell{l}{std. CI\\length}&\tabincell{l}{\% of\\overshoot}\\\hline
\multirow{4}{*}{BEL}&$R_1=3000,R_2=500$&$91.6\%$&$0.24$&$0.04$&$0\%$\\\cline{2-2}
&$R_1=3500,R_2=250$&$90.4\%$&$0.24$&$0.05$&$0\%$\\\cline{2-2}
&$R_1=3800,R_2=100$&$89.1\%$&$0.24$&$0.06$&$0\%$\\\cline{2-2}
&$R_1=3900,R_2=50$&$85.0\%$&$0.24$&$0.09$&$0\%$\\\hline
\multirow{4}{*}{EEL}&$R_1=3000,R_2=500$&$97.3\%$&$0.31$&$0.05$&$0\%$\\\cline{2-2}
&$R_1=3500,R_2=250$&$96.9\%$&$0.33$&$0.06$&$0\%$\\\cline{2-2}
&$R_1=3800,R_2=100$&$98.3\%$&$0.39$&$0.08$&$0\%$\\\cline{2-2}
&$R_1=3900,R_2=50$&$98.9\%$&$0.45$&$0.11$&$0\%$\\\hline
\multirow{4}{*}{FEL}&$R_1=3000,R_2=500$&$93.3\%$&$0.25$&$0.04$&$0\%$\\\cline{2-2}
&$R_1=3500,R_2=250$&$93.2\%$&$0.26$&$0.05$&$0\%$\\\cline{2-2}
&$R_1=3800,R_2=100$&$93.3\%$&$0.29$&$0.07$&$0\%$\\\cline{2-2}
&$R_1=3900,R_2=50$&$94.9\%$&$0.32$&$0.09$&$0\%$\\\hline
\multirow{4}{*}{standard BT}&$B=50,R_b=80$&$94.9\%$&$0.31$&$0.06$&$0\%$\\\cline{2-2}
&$B=100,R_b=40$&$98.4\%$&$0.33$&$0.06$&$0\%$\\\cline{2-2}
&$B=400,R_b=10$&$99.9\%$&$0.50$&$0.08$&$0\%$\\\cline{2-2}
&$B=1000,R_b=4$&$100\%$&$0.73$&$0.10$&$0\%$\\\hline
\multirow{4}{*}{\tabincell{l}{adaptive BT\\(4 best combinations)}}&$B=100,n_0=15,r=1.8,M=1$&$95.0\%$&$0.30$&$0.06$&$0\%$\\\cline{2-2}
&$B=100,n_0=5,r=1.2,M=7$&$95.3\%$&$0.31$&$0.06$&$0\%$\\\cline{2-2}
&$B=100,n_0=10,r=1.8,M=1$&$94.1\%$&$0.31$&$0.06$&$0\%$\\\cline{2-2}
&$B=100,n_0=20,r=1.2,M=1$&$93.7\%$&$0.30$&$0.06$&$0\%$\\\hline
\tabincell{l}{nonparametric delta\\method}&$R_d=2000$&$93.8\%$&$0.26$&$0.04$&$0\%$\\\hline
\end{tabular}
\end{center}
\label{network_14_small}
\end{table}

Next we consider a bigger stochastic activity network, shown in Figure \ref{m=14}, that is borrowed from \cite{chu2014new} that consists of $14$ tasks. The time to completion $X_i$ of each task follows exponential distribution with rate $10,5,12,11,5,8,4,9,13,7,6,9,10,6$ for $i$ from $1$ to $14$. In addition to computing the expected time to complete the project (Table \ref{network_14_small}), which is represented by the length of the longest path from node $1$ to $10$, we also test our methods in estimating the tail probability that the time to finish the project exceeds $1.5$ units of time (Tables \ref{network_14_large_tail} and \ref{network_14_verylarge_tail}). The true value of the probability is $0.0747$  (estimated from abundunt simulation).

Table \ref{network_14_small} shows that our FEL and the adaptive bootstrap consistently exhibit satisfactory coverage levels when the number of input models is fairly big compared with the input data size (per input model). Here we use a simulation budget of $4000$, and a data size of $30$ for the first $7$ input models, and $25$ for the other $7$ inputs. The coverage probabilities and their trends in each method are similar to our observations before (e.g., in Tables \ref{mm1_large} and \ref{network_5_imbalance}). For example, the coverage of FEL stays around $94\%$, the standard bootstrap over-covers for small $R_b$, and BEL under-covers for small $R_2$.

% When the target quantity is a probability as small as $0.0747$, our FEL still seems quite competitive, compared to other methods. In  which we attribute to the target probability being too small than before which could also be due to the small magnitude of the target probability we do point out that to a lower coverage than the simulation output is the hence in Algorithm \ref{algo3} a relatively large number of simulation replications is needed to  probability is small

Table \ref{network_14_large_tail} shows the tail probability estimation results, with a data size around $100$ per input model. Table \ref{network_14_verylarge_tail} considers a bigger data size of $400$-$500$. The simulation budgets are $16000$ and $60000$ respectively. FEL and the delta method seem to have accurate coverage probabilities ($93\%$ in Table \ref{network_14_large_tail} and $94\%$ in Table \ref{network_14_verylarge_tail}). EEL continues to over-cover. Notably, BEL suffers from severe under-coverage issues, while the standard bootstrap suffers from severe over-coverage issues. Though FEL gives accurate CIs in most cases, the simple budget allocation strategy of setting $R_2=50$ and investing the remainder to $R_1$ appears to perform less well than using a larger $R_2$ such as $100,250$. This could be because of the highly skewed performance function, which requires more $R_2$ to invoke the central limit behavior needed in the CI construction. Our suggestion is to use $R_2$ in the range of hundreds in FEL for tail estimation problems.

\begin{table}[h]
\caption{Tail probability of stochastic activity network in Figure \ref{m=14}. $n_i=120$ for $1\leq i\leq 7$ and $100$ for $8\leq i\leq 14$. Total simulation budget $16000$. Run times (second/CI): three EL methods $0.11$, the bootstrap $0.03$, delta method $0.10$.}
\begin{center}
\begin{tabular}{|l|l|llll|}
\hline
\multicolumn{2}{|c|}{\tabincell{l}{methods \&\\parameters}}&\tabincell{l}{coverage\\ estimate}&\tabincell{l}{mean CI\\length}&\tabincell{l}{std. CI\\length}&\tabincell{l}{\% of\\overshoot}\\\hline
\multirow{4}{*}{BEL}&$R_1=15000,R_2=500$&$86.0\%$&$0.064$&$0.020$&$0\%$\\\cline{2-2}
&$R_1=15500,R_2=250$&$80.0\%$&$0.064$&$0.026$&$0\%$\\\cline{2-2}
&$R_1=15800,R_2=100$&$70.3\%$&$0.064$&$0.040$&$0\%$\\\cline{2-2}
&$R_1=15900,R_2=50$&$57.8\%$&$0.062$&$0.055$&$0\%$\\\hline
\multirow{4}{*}{EEL}&$R_1=15000,R_2=500$&$98.5\%$&$0.110$&$0.023$&$0\%$\\\cline{2-2}
&$R_1=15500,R_2=250$&$98.8\%$&$0.130$&$0.031$&$1.2\%$\\\cline{2-2}
&$R_1=15800,R_2=100$&$98.7\%$&$0.166$&$0.046$&$30\%$\\\cline{2-2}
&$R_1=15900,R_2=50$&$97.5\%$&$0.205$&$0.067$&$65\%$\\\hline
\multirow{4}{*}{FEL}&$R_1=15000,R_2=500$&$93.2\%$&$0.079$&$0.020$&$0\%$\\\cline{2-2}
&$R_1=15500,R_2=250$&$93.0\%$&$0.090$&$0.027$&$0\%$\\\cline{2-2}
&$R_1=15800,R_2=100$&$93.2\%$&$0.120$&$0.044$&$0\%$\\\cline{2-2}
&$R_1=15900,R_2=50$&$91.4\%$&$0.155$&$0.062$&$3.8\%$\\\hline
\multirow{4}{*}{standard BT}&$B=50,R_b=320$&$97.1\%$&$0.090$&$0.018$&$0\%$\\\cline{2-2}
&$B=100,R_b=160$&$99.2\%$&$0.104$&$0.017$&$0\%$\\\cline{2-2}
&$B=400,R_b=40$&$100\%$&$0.170$&$0.026$&$0\%$\\\cline{2-2}
&$B=1000,R_b=16$&$100\%$&$0.230$&$0.038$&$0\%$\\\hline
% \multirow{4}{*}{\tabincell{l}{adaptive BT\\(4 best combinations)}}&$B=400,n_0=5,r=1.2,M=9$&$96.3\%$&$0.21$&$0.04$&$0\%$\\\cline{2-2}
% &$B=500,n_0=10,r=1.4,M=1$&$96.7\%$&$0.24$&$0.03$&$0\%$\\\cline{2-2}
% &$B=300,n_0=10,r=1.6,M=3$&$96.3\%$&$0.19$&$0.04$&$0\%$\\\cline{2-2}
% &$B=100,n_0=15,r=1.4,M=5$&$93.9\%$&$0.19$&$0.04$&$0\%$\\\hline
\multirow{4}{*}{\tabincell{l}{adaptive BT\\(3 best combinations)}}
&$B=100,n_0=80,r=1.1,M=5$&$89.0\%$&$0.093$&$0.026$&$0\%$\\\cline{2-2}
&$B=100,n_0=100,r=1.1,M=4$&$92.3\%$&$0.089$&$0.023$&$0\%$\\\cline{2-2}
&$B=100,n_0=100,r=1.2,M=2$&$91.4\%$&$0.091$&$0.024$&$0\%$\\\hline
\tabincell{l}{nonparametric delta\\method}&$R_d=16000$&$93.2\%$&$0.070$&$0.011$&$0\%$\\\hline
\end{tabular}
\end{center}
\label{network_14_large_tail}
\end{table}

\begin{table}[h]
\caption{Tail probability of stochastic activity network in Figure \ref{m=14}. $n_i=480$ for $1\leq i\leq 7$ and $400$ for $8\leq i\leq 14$. Total simulation budget $60000$. Run times (second/CI): three EL methods $1.4$, the bootstrap $0.08$, delta method $1.3$.}
\begin{center}
\begin{tabular}{|l|l|llll|}
\hline
\multicolumn{2}{|c|}{\tabincell{l}{methods \&\\parameters}}&\tabincell{l}{coverage\\ estimate}&\tabincell{l}{mean CI\\length}&\tabincell{l}{std. CI\\length}&\tabincell{l}{\% of\\overshoot}\\\hline
\multirow{4}{*}{BEL}&$R_1=59000,R_2=500$&$73.3\%$&$0.032$&$0.017$&$0\%$\\\cline{2-2}
&$R_1=59500,R_2=250$&$63.1\%$&$0.033$&$0.024$&$0\%$\\\cline{2-2}
&$R_1=59800,R_2=100$&$50.6\%$&$0.032$&$0.038$&$0\%$\\\cline{2-2}
&$R_1=59900,R_2=50$&$43.0\%$&$0.032$&$0.054$&$0\%$\\\hline
\multirow{4}{*}{EEL}&$R_1=59000,R_2=500$&$99.1\%$&$0.078$&$0.018$&$0\%$\\\cline{2-2}
&$R_1=59500,R_2=250$&$98.6\%$&$0.097$&$0.025$&$0\%$\\\cline{2-2}
&$R_1=59800,R_2=100$&$97.9\%$&$0.132$&$0.040$&$15\%$\\\cline{2-2}
&$R_1=59900,R_2=50$&$94.9\%$&$0.172$&$0.061$&$58\%$\\\hline
\multirow{4}{*}{FEL}&$R_1=59000,R_2=500$&$93.4\%$&$0.055$&$0.017$&$0\%$\\\cline{2-2}
&$R_1=59500,R_2=250$&$94.1\%$&$0.071$&$0.025$&$0\%$\\\cline{2-2}
&$R_1=59800,R_2=100$&$94.0\%$&$0.104$&$0.041$&$0\%$\\\cline{2-2}
&$R_1=59900,R_2=50$&$93.2\%$&$0.141$&$0.061$&$28\%$\\\hline
\multirow{4}{*}{standard BT}&$B=50,R_b=1200$&$97.6\%$&$0.047$&$0.007$&$0\%$\\\cline{2-2}
&$B=100,R_b=600$&$99.3\%$&$0.054$&$0.006$&$0\%$\\\cline{2-2}
&$B=400,R_b=150$&$100\%$&$0.090$&$0.007$&$0\%$\\\cline{2-2}
&$B=1000,R_b=60$&$100\%$&$0.134$&$0.012$&$0\%$\\\hline
\multirow{4}{*}{\tabincell{l}{adaptive BT\\(4 best combinations)}}&$B=400,n_0=5,r=1.2,M=9$&$96.3\%$&$0.21$&$0.04$&$0\%$\\\cline{2-2}
&$B=500,n_0=10,r=1.4,M=1$&$96.7\%$&$0.24$&$0.03$&$0\%$\\\cline{2-2}
&$B=300,n_0=10,r=1.6,M=3$&$96.3\%$&$0.19$&$0.04$&$0\%$\\\cline{2-2}
&$B=100,n_0=15,r=1.4,M=5$&$93.9\%$&$0.19$&$0.04$&$0\%$\\\hline
\tabincell{l}{nonparametric delta\\method}&$R_d=60000$&$94.3\%$&$0.035$&$0.003$&$0\%$\\\hline
\end{tabular}
\end{center}
\label{network_14_verylarge_tail}
\end{table}

\subsection{Summary and Comparisons with the Bootstrap}\label{sec:bootstrap}
Based on the findings in Sections \ref{sec:queue} and \ref{sec:stochastic}, we provide some general comparisons between our optimization-based approach and the standard bootstrap in terms of the required simulation burden, the ease of implementation and the computation cost.
% , and explain how the comparison is supported by our numerical experiments.

Because of the nested simulation, the total simulation load of the standard bootstrap is $BR_b$. To ensure the stochastic noise is negligible relative to input uncertainty, one would need $R_b\gg n$ (where ``$\gg$" means ``of larger order than"). On the other hand, Theorems \ref{erfreeCI2} and \ref{erfreeCI3} suggest that, in the optimization-based approach, one can choose $R_1\gg n,R_2\gg 1$. Thus, the bootstrap requires $BR_b\gg Bn$ total simulation load, whereas ours requires $R_1+2R_2\gg n$ simulation load. Since $B$ is typically a big number (in the experiments we use $B$ between $50$ and $1000$), our method seems to be more efficient in terms of simulation cost. In Tables \ref{mm1_small} and \ref{mm1_large}, we have observed that under the same total simulation budget FEL consistently possess coverage probabilities close to the benchmark coverage while the bootstrap very often significantly exceeds the benchmark level.
% , which validates the lighter simulation requirement of our approach in achieving stable coverage.

% Closely related to the lighter simulation requirement is the benefit in 
We also notice that our optimization-based approach is more robust with respect to the algorithmic parameter configuration. Given a fixed total simulation budget, it could be challenging to figure out a good choice of $B$ and $R_b$ for the bootstrap, as it can highly depend on the input data sizes and the magnitude of the simulation error. Indeed, our experiments indicate that the coverage of the bootstrap CIs is quite sensitive to the allocations of $B$ and $R_b$. When $B$ and $R_b$ are not appropriately chosen, the bootstrap CI tends to over-cover the truth. On the other hand, in the optimization-based method, particularly FEL, setting $R_2$ to be a fixed moderately large number (say $50$) and investing the remaining budget to $R_1$ seems to be quite stable regardless of the data size. Nonetheless, we have seen that if the performance measure is a small probability, choosing a larger $R_2$ would improve the coverages.

% and Although our optimization-based approach enjoys the above two advantages, 
Despite the simulation savings and stability, the optimization-based approach calls for a heavier computation overhead than the bootstrap beyond the simulation effort. In the bootstrap, the extra numerical computation other than simulation runs is negligible. In our approach, we need to estimate  gradient information (the influence function) in \eqref{gradientest} in Step 1, and solve the optimization  pair in Step 2. Computation of the score function $S_{i,j}(\mathbf X_i^r)$ for all $i,j$ and $r=1,\ldots,R_1$ requires $O((\sum_{i=1}^mT_i)R_1)$ time, by counting the occurrence of each $X_{i,j}$ in the generated input variates. The sample covariance between the output $h$ and the score function is computed in $O((\sum_{i=1}^mn_i)R_1)$ time. Thus the total computation in Step 1 has a complexity $O((\sum_{i=1}^mn_i+\sum_{i=1}^mT_i)R_1)$. Using the approach suggested by Proposition \ref{optroutine}, the optimization pair \eqref{ciopt4min} can be solved in $O(c^{bi}(\sum_{i=1}^mn_ic_i^{nt}))$ time, where $c^{bi}$ is the number of bisection iterations on $\beta$ and $c_i^{nt}$ is the number of Newton iterations to obtain each $\lambda_i(\beta)$. The global linear convergence of bisection and Newton's method in our setting suggest that, to achieve a given tolerance level, typically $c^{bi}$ and each $c_i^{nt}$ only need to be logarithmically large. Ignoring logarithmic factors, we see that the computation cost of Step 2 is roughly $O(\sum_{i=1}^mn_i)$. Thus the cost of Step 1 dominates Step 2, leading to a total overhead cost $O((\sum_{i=1}^mn_i+\sum_{i=1}^mT_i)R_1)$. In the case of large data size, these overhead costs of our method can be substantial, which is reflected by the significantly longer run times of the EL methods compared to the bootstrap in Tables \ref{network_14_large_tail} and \ref{network_14_verylarge_tail}.

\section{Conclusion}
We have proposed an optimization-based approach to construct CIs for simulation output performance measures that account for the input uncertainty from finite data. This approach relies on solving a pair of optimization programs posited over distributions supported on the data, with a constraint expressed in terms of the weighted average of empirically defined Burg-entropy divergences. It then uses the solutions to define probability weights that subsequently drive simulation runs. We present several related procedures under this approach and analyze their statistical performances using a generalization of the EL method. Compared to the bootstrap, our approach requires less simulation budget to achieve stable coverage and is less sensitive to the allocation choices, as explained both theoretically and shown by our numerical experiments. The numerical results also reveal that our approach tends to curb the under-coverage issues encountered in the delta method. The last of our procedures, FEL, seems particularly attractive compared to both the bootstrap and the delta method in terms of finite-data finite-simulation performance.

% Future work consists of several lines of work. We plan to extend our approach to quantify input uncertainty for other common types of performance measures than the expectation type, such as quantile and conditional value-at-risk. We also plan to study methodologies that can, in certain sense, break down the overall uncertainty into stochastic and input uncertainties, in cases where simulation replications are limited. %Such methodologies provide measurements of the relative amount of input uncertainty, which constitutes another core problem in simulation uncertainty quantification.

% Acknowledgments here
\ACKNOWLEDGMENT{A preliminary conference version of this paper will appear in the Winter Simulation Conference. We gratefully acknowledge support from the National Science Foundation under grants CMMI-1400391/1542020 and CMMI-1436247/1523453.}

% References here (outcomment the appropriate case)

% CASE 1: BiBTeX used to constantly update the references
%   (while the paper is being written).
\bibliographystyle{informs2014} % outcomment this and next line in Case 1
\bibliography{reference} % if more than one, comma separated

\ECSwitch

%\ECDisclaimer
%%%%%%%%%%%%%%%%%%%%%%%%%%%%%%%%%%%%%%%%%%%%%%%%%%%%%%%%%%

%%% Main head for the e-companion
\ECHead{Proofs of Statements}

We introduce some notations. Given a positive semi-definite matrix $\Sigma\in\R^{d\times d}$, $\mathcal N(\mathbf{0},\Sigma)$ denotes the multivariate normal distribution on $\R^d$ with mean zero and covariance matrix $\Sigma$. In particular, $\mathcal N(0,1)$ denotes the univariate standard normal. $\Phi(\cdot)$ is the cumulative distribution function of $\mathcal N(0,1)$. $\Rightarrow$ denotes weak convergence of probability measures. Given the data $\{X_{i,j}\}_{i,j}$ and the optimal probability weights $(\mathbf w_1^{\min},\ldots,\mathbf w_m^{\min})$ and $(\mathbf w_1^{\max},\ldots,\mathbf w_m^{\max})$ in Step 2 of our algorithms, let $\sigma_{\min}^2=\mathrm{Var}_{\mathbf w_1^{\min},\ldots,\mathbf w_m^{\min}}(h(\mathbf X_1,\ldots,\mathbf X_m))$, $\sigma_{\max}^2=\mathrm{Var}_{\mathbf w_1^{\max},\ldots,\mathbf w_m^{\max}}(h(\mathbf X_1,\ldots,\mathbf X_m))$ be the variances of the simulation output driven by input models under the weighted empirical distributions. We denote $\mathbb E[\cdot]$/$P(\cdot)$ as the expectation/probability with respect to the randomness in the data, and also all the simulation runs when the quantity in consideration involves them. We use $\mathbb E_{\xi_2}[\cdot]:=\mathbb E[\cdot\vert \text{data and Step 1 of the algorithms}]$ to represent the expectation conditioned on the input data and the simulation in Step 1 (i.e., the expectation is only on the randomness of the simulation in Step 3), and $\mathbb E_{\xi_1}[\cdot]:=\mathbb E[\cdot\vert \text{data}]$ the expectation conditioned on the input data. $\mathbb E_{D}[\cdot]$ is the expectation with respect to the input data, and therefore $\mathbb E[\cdot]=\mathbb E_D[\mathbb E_{\xi_1}[\mathbb E_{\xi_2}[\cdot]]]$. When applicable, we denote $\mathbb E_{D,\xi_1}[\cdot]$ as the expectation with respect to both the randomness in the data and the simulation in Step 1. Probabilities $P_{\xi_2}(\cdot),P_{\xi_1}(\cdot),P_D(\cdot)$ and variances $\mathrm{Var}_{\xi_2}(\cdot),\mathrm{Var}_{\xi_1}(\cdot),\mathrm{Var}_D(\cdot)$ are defined accordingly.

%  if the quantity/event depends on Step 1 or Step 3 or both of the algorithms At times, we use $D$ to stand for the data, $\xi_1$ for the simulation runs in Step 1 of our procedure, and $\xi_2$ for the simulation runs in Step 3. 

We present our proofs as follows. We first prove all the results in Section \ref{sec:theory}, organized via the subsections. Given these developments, we then prove the main results in Section \ref{sec:EL} including Theorems \ref{erfreeCI1}, \ref{erfreeCI2}, \ref{erfreeCI3}, and also Proposition \ref{optroutine}.

\section{Proofs of Results in Section \ref{sec:linearization}}\label{proof:section 4.2}
\proof{Proof of Proposition \ref{derivative}.}Let $\mathbf x_i=(x_{i,1},\ldots,x_{i,T_i})$. First we rewrite the performance measure as an integral
\begin{align}
&Z((1-\epsilon)Q_1^1+\epsilon Q_1^2,\ldots,(1-\epsilon)Q_m^1+\epsilon Q_m^2)\notag\\
=&\int h(\mathbf x_1,\ldots,\mathbf x_m) \prod_{i=1}^m\prod_{t=1}^{T_i}d(Q_i^1+\epsilon (Q_i^2-Q_i^1))(x_{i,t})\notag\\
=&Z(Q_1^1,\ldots,Q_m^1)+\sum_{i=1}^m\sum_{t=1}^{T_i}\epsilon\int h(\mathbf x_1,\ldots,\mathbf x_m)\prod_{r\neq i\;\text{or}\;s\neq t}dQ_r^1(x_{r,s})\cdot d(Q_i^2-Q_i^1)(x_{i,t})+\mathcal R\label{interim1 revised}
\end{align}
by expanding out all the $Q_i^1$ and $\epsilon (Q_i^2-Q_i^1)$ in the product measure,  and the remainder $\mathcal R$ includes all the terms that have an $\epsilon^k$ with $k\geq 2$. The integrability condition guarantees that all the integral terms above, including those in $\mathcal R$, are finite. Note that
% As $\epsilon\to 0+$, the limit of the finite difference can be identified as the first order remainder, i.e.
\begin{align*}
&\sum_{i=1}^m\sum_{t=1}^{T_i}\int h(\mathbf x_1,\ldots,\mathbf x_m)\prod_{r\neq i\;\text{or}\;s\neq t}dQ_r^1(x_{r,s})\cdot d(Q_i^2-Q_i^1)(x_{i,t})\\
=&\sum_{i=1}^m\sum_{t=1}^{T_i}\prth{\int h(\mathbf x_1,\ldots,\mathbf x_m)\prod_{r\neq i\;\text{or}\;s\neq t}dQ_r^1(x_{r,s})\cdot dQ_i^2(x_{i,t})-Z(Q_1^1,\ldots,Q_m^1)}\\
=&\sum_{i=1}^m\sum_{t=1}^{T_i}\int\prth{\int h(\mathbf x_1,\ldots,\mathbf x_m)\prod_{r\neq i\;\text{or}\;s\neq t}dQ_r^1(x_{r,s})-Z(Q_1^1,\ldots,Q_m^1)} dQ_i^2(x_{i,t})\\
=&\sum_{i=1}^m\sum_{t=1}^{T_i}\int\prth{\int h(\mathbf x_1,\ldots,\mathbf x_i^{(t)},\ldots,\mathbf x_m)\prod_{r\neq i\;\text{or}\;s\neq t}dQ_r^1(x_{r,s})-Z(Q_1^1,\ldots,Q_m^1)} dQ_i^2(x_i)\\
=&\sum_{i=1}^m\int\sum_{t=1}^{T_i}\prth{\int h(\mathbf x_1,\ldots,\mathbf x_i^{(t)},\ldots,\mathbf x_m)\prod_{r\neq i\;\text{or}\;s\neq t}dQ_r^1(x_{r,s})-Z(Q_1^1,\ldots,Q_m^1)} dQ_i^2(x_i)\\
=&\sum_{i=1}^m\int G_i^{Q_1^1,\ldots,Q_m^1}(x_i)dQ_i^2(x_i)=\sum_{i=1}^m\mathbb E_{Q_i^2} [G_i^{Q_1^1,\ldots,Q_m^1}(X_i)],
\end{align*}
where the second equality holds because $dQ_i^2$ is a probability measure, and the third equality is a notational replacement of $x_{i,t}$ by $x_i$, with $\mathbf x_i^{(t)}$ defined as $\mathbf x_i$ but with $x_{i,t}$ replaced by $x_i$. This and \eqref{interim1 revised} together show the derivative expression \eqref{derivative new}. The mean zero property of $G_i^{Q_1^1,\ldots,Q_m^1}$ follows from the tower property
\begin{equation*}
\mathbb E_{Q_i^1} \brac{\mathbb E_{Q_1^1,\ldots,Q_m^1}[h(\mathbf X_1,\ldots,\mathbf X_m)\vert X_{i}(t)]}=Z(Q_1^1,\ldots,Q_m^1)
\end{equation*}
for all $t=1,\ldots,T_i$.\hfill\Halmos
\endproof

\proof{Proof of Proposition \ref{linearization error}.}We first provide two lemmas.
\begin{lemma}\label{lowerbound}
Every feasible solution $\prth{\mathbf w_1,\ldots,\mathbf w_m}\in\mathcal U_{\alpha}$ satisfies
\begin{equation*}
\frac{l(\alpha)}{n_i}\leq w_{i,j}\leq \frac{u(\alpha)}{n_i},\forall\, i=1,\ldots,m,j=1,\ldots,n_i
\end{equation*}
where $0<l(\alpha)<1<u(\alpha)<+\infty$ are the two solutions of the equation $xe^{1+\frac{\mathcal X_{1,1-\alpha}^2}{2}-x}=1$.
\end{lemma}
\proof{Proof of Lemma \ref{lowerbound}.}Consider $\prth{\mathbf w_1,\ldots,\mathbf w_m}\in\mathcal U_{\alpha}$. By Jensen's inequality, for each $i$ we have
\begin{equation*}
-\sum_{j=1}^{n_i}\log(n_iw_{i,j})\geq -n_i\log\sum_{j=1}^{n_i}w_{i,j}=0,
\end{equation*}
and thus
\begin{equation*}
-2\sum_{j=1}^{n_i}\log(n_iw_{i,j})\leq -2\sum_{i=1}^{m}\sum_{j=1}^{n_i}\log(n_iw_{i,j})\leq \mathcal X_{1,1-\alpha}^2.
\end{equation*}
This implies for each $i=1,\ldots,m$
\begin{equation}\label{ineq1}
\prod_{j=1}^{n_i} n_iw_{i,j}\geq e^{-\frac{\mathcal X_{1,1-\alpha}^2}{2}}.
\end{equation}
For any $s=1,\ldots,n_i$, we shall show that $l(\alpha)\leq n_iw_{i,s}\leq u(\alpha)$. Taking $n_iw_{i,s}$ out of the product in \eqref{ineq1} and noticing the inequality $\prod_{j\neq s}n_iw_{i,j}\leq \big(\frac{n_i}{n_i-1}\sum_{j\neq s}w_{i,j}\big)^{n_i-1}=\big(\frac{n_i(1-w_{i,s})}{n_i-1}\big)^{n_i-1}$ gives
\begin{equation*}
n_iw_{i,s}\prth{1+\frac{1-n_iw_{i,s}}{n_i-1}}^{n_i-1}\geq n_iw_{i,s}\prod_{j\neq s}n_iw_{i,j}\geq e^{-\frac{\mathcal X_{1,1-\alpha}^2}{2}}.
\end{equation*}
Applying $e^x\geq 1+x$ to $1+\frac{1-n_iw_{i,s}}{n_i-1}$ gives
\begin{equation}
n_iw_{i,s}e^{1-n_iw_{i,s}}\geq e^{-\frac{\mathcal X_{1,1-\alpha}^2}{2}}.\label{bdofw}
\end{equation}
Simple calculations show that the function $xe^{1-x}$ strictly increases from $0$ to $1$ for $x\in(0,1)$ and decreases from $1$ to $0$ for $x\in(1,+\infty)$. So it follows from \eqref{bdofw} that $n_iw_{i,s}$ must fall between the two solutions of $xe^{1-x}=e^{-\frac{\mathcal X_{1,1-\alpha}^2}{2}}$.\hfill\Halmos
\endproof

\begin{lemma}\label{l2upbound}
Let $u(\alpha)$ be the constant from Lemma \ref{lowerbound}. Every feasible solution $(\mathbf w_1,\ldots,\mathbf w_m)\in \mathcal U_{\alpha}$ satisfies
\begin{equation*}
\sum_{i=1}^mn_i^2\sum_{j=1}^{n_i}(w_{i,j}-\frac{1}{n_i})^2\leq u(\alpha)^2\mathcal X_{1,1-\alpha}^2.
\end{equation*}
% Moreover, for those such that $-2\sum_{i=1}^m\sum_{j=1}^{n_i}\log(n_iw_{i,j})=\mathcal X_{1,1-\alpha}^2$, the following lower bound holds
% \begin{equation*}
% \sum_{i=1}^mn_i^2\sum_{j=1}^{n_i}(w_{i,j}-\frac{1}{n_i})^2\geq l(\alpha)^2\mathcal X_{1,1-\alpha}^2.
% \end{equation*} log-likelihood ratio constraint at the uniform weights,
\end{lemma}
\proof{Proof of Lemma \ref{l2upbound}.}Taylor expand each summand in the left hand side of the first constraint in $\mathcal U_\alpha$, around the uniform weights, and use the mean value theorem to get
\begin{eqnarray*}
-2\sum_{i=1}^{m}\sum_{j=1}^{n_i}\log(n_iw_{i,j})&=&\sum_{i=1}^{m}\sum_{j=1}^{n_i}\prth{0-2n_i(w_{i,j}-\frac{1}{n_i})+(\theta_{i,j}w_{i,j}+(1-\theta_{i,j})\frac{1}{n_i})^{-2}(w_{i,j}-\frac{1}{n_i})^2}\\
&=&\sum_{i=1}^{m}\sum_{j=1}^{n_i}(\theta_{i,j}w_{i,j}+(1-\theta_{i,j})\frac{1}{n_i})^{-2}(w_{i,j}-\frac{1}{n_i})^2
\end{eqnarray*}
where $\theta_{i,j}$ is some constant such that  $0\leq \theta_{i,j}\leq 1$, for each $i,j$. Lemma \ref{lowerbound} implies $\theta_{i,j}w_{i,j}+(1-\theta_{i,j})\frac{1}{n_i}\leq \frac{u(\alpha)}{n_i}$. Hence
\begin{equation*}
\sum_{i=1}^m\sum_{j=1}^{n_i}\frac{n_i^2}{u(\alpha)^2}(w_{i,j}-\frac{1}{n_i})^2\leq -2\sum_{i=1}^{m}\sum_{j=1}^{n_i}\log(n_iw_{i,j})\leq \mathcal X^2_{1,1-\alpha}.
\end{equation*}
Multiplying $u(\alpha)^2$ on both sides completes the proof.\hfill\Halmos
\endproof

Now we are ready to prove Proposition \ref{linearization error}.
Let $\mathbf x_i=(x_{i,1},\ldots,x_{i,T_i})$. We will first show the uniform error bound of the linear approximation $Z_L$, and then $\widehat{Z_L}$. We start the analysis by expressing $Z(\mathbf w_{1},\ldots,\mathbf w_{m})$ as
\begin{equation}\label{zw}
Z(\mathbf w_{1},\ldots,\mathbf w_{m})=\int h(\mathbf x_{1},\ldots,\mathbf x_m)\prod_{i=1}^m\prod_{t=1}^{T_i}d\mathbf w_i(x_{i,t})
\end{equation}
where we abuse notation to write $\mathbf w_i$ as a probability measure over the observations $\{X_{i,j}\}_{j=1,\ldots,n_i}$. Rewrite $d\mathbf w_i$ as $d(\mathbf w_i-\hat P_i+\hat P_i-P_i+P_i)$, where $\hat P_i$ is the empirical distribution of the $i$-th sample, and expand out $\mathbf w_i-\hat P_i$, $\hat P_i-P_i$ and $P_i$ in the product measure in \eqref{zw} to get
\begin{align}
\nonumber&Z(\mathbf w_{1},\ldots,\mathbf w_{m})\\
\nonumber=&\sum_{\mathcal T_i^1,\mathcal T_i^2}\int h(\mathbf X_{1},\ldots,\mathbf X_m)\prod_{i=1}^m\prod_{t\notin \mathcal T_i^1\cup \mathcal T_i^2}dP_i(x_{i,t})\prod_{i=1}^m\prod_{t\in \mathcal T_i^1}d(\hat P_i-P_i)(x_{i,t})\prod_{i=1}^m\prod_{t\in \mathcal T_i^2}d(\mathbf w_i-\hat P_i)(x_{i,t})\\
=&\sum_{d=0}^T\sum_{\sum_{i}(\vert \mathcal T_i^1\vert+\vert \mathcal T_i^2\vert)=d}\int h\prod_{i=1}^m\prod_{t\notin \mathcal T_i^1\cup \mathcal T_i^2}dP_i(x_{i,t})\prod_{i=1}^m\prod_{t\in \mathcal T_i^1}d(\hat P_i-P_i)(x_{i,t})\prod_{i=1}^m\prod_{t\in \mathcal T_i^2}d(\mathbf w_i-\hat P_i)(x_{i,t})\label{expansion}
\end{align}
where for each $i$, $\mathcal T_i^1,\mathcal T_i^2$ are two disjoint and ordered (possibly empty) subsets of $\set{1,2,\ldots,T_i}$ that specifies the second subscript $t$ of the argument $x_{i,t}$, $\abs{\cdot}$ denotes the cardinality of a set, and $T=\sum_{i=1}^mT_i$.

The desired conclusion can be achieved upon completing the following two tasks: (1) show that the terms with $d=0,1$ above give the linear approximation; (2) each term with $d\geq 2$ is of order $O(1/n^d)$ in terms of its mean square.

% induces a squared error of order
% (2) terms with $d=2$ constitute the leading error term that is of the desired order of magnitude,

\noindent\textbf{Task one: $d=0,1$}

The only summand with $d=0$ is
\begin{equation*}
\int h(\mathbf x_{1},\ldots,\mathbf x_m)\prod_{i=1}^m\prod_{t=1}^{T_i}dP_i(x_{i,t})=Z(P_1,\ldots,P_m)=Z^*,
\end{equation*}
and each summand with $d=1$ is one of the following two types
\begin{equation*}
\int h(\mathbf x_{1},\ldots,\mathbf x_m)\prod_{i\neq r\;\text{or}\;t\neq s}dP_i(x_{i,t})d(\hat P_r-P_r)(x_{r,s}),\ \text{for }r=1,\ldots,m,s=1,\ldots,T_i
\end{equation*}
or
\begin{equation*}
\int h(\mathbf x_{1},\ldots,\mathbf x_m)\prod_{i\neq r\;\text{or}\;t\neq s}dP_i(x_{i,t})d(\mathbf w_r-\hat P_r)(x_{r,s}),\ \text{for }r=1,\ldots,m,s=1,\ldots,T_i.
\end{equation*}
For each $r$ and $s$ the two types sum up to
\begin{equation*}
\int h(\mathbf x_{1},\ldots,\mathbf x_m)\prod_{i\neq r\;\text{or}\;t\neq s}dP_i(x_{i,t})d(\mathbf w_r-P_r)(x_{r,s}).
\end{equation*}
Summing over all $r,s$ gives
\begin{align*}
&\sum_{r=1}^m\sum_{s=1}^{T_r}\int h(\mathbf x_{1},\ldots,\mathbf x_m)\prod_{i\neq r\;\text{or}\;t\neq s}dP_i(x_{i,t})d(\mathbf w_r-P_r)(x_{r,s})\\
=&\sum_{r=1}^m\sum_{s=1}^{T_r}\int h(\mathbf x_{1},\ldots,\mathbf x_{r}^{(s)},\ldots\mathbf x_m)\prod_{i\neq r\;\text{or}\;t\neq s}dP_i(x_{i,t})d(\mathbf w_r-P_r)(x_{r}){}\\
&{}\text{\ \ by replacing $x_{r,s}$ with $x_r$, and denoting $\mathbf x_r^{(s)}$ as $\mathbf x_r$ but with $x_{r,s}$ replaced by $x_r$}\\
=&\sum_{r=1}^m\int\Big(\sum_{s=1}^{T_r}\int h(\mathbf x_{1},\ldots,\mathbf x_{r}^{(s)},\ldots,\mathbf x_m)\prod_{i\neq r\;\text{or}\;t\neq s}dP_i(x_{i,t})\Big)d(\mathbf w_r-P_r)(x_{r})\\
=&\sum_{r=1}^m\int\sum_{s=1}^{T_r}\Big(\int h(\mathbf x_{1},\ldots,\mathbf x_{r}^{(s)},\ldots,\mathbf x_m)\prod_{i\neq r\;\text{or}\;t\neq s}dP_i(x_{i,t})-Z(P_1,\ldots,P_m)\Big)d\mathbf w_r(x_{r})\\
=&\sum_{r=1}^m\sum_{j=1}^{n_i}w_{r,j}G_r(X_{r,j}).
\end{align*}
This concludes that the summands with $d=0,1$ sum up to the linear approximation $Z_L=Z^*+\sum_{i=1}^m\sum_{j=1}^{n_i}w_{i,j}G_i(X_{i,j})$.

\noindent\textbf{Task two: $d\geq 2$}

Now we deal with the terms in \eqref{expansion} with $d\geq 2$. Define
\begin{equation}\label{2moment}
\mathcal M:=\max_{I_1,\ldots,I_m}\mathbb E_{P_1,\ldots,P_m}[\abs{h(\mathbf X_{1,I_1},\ldots,\mathbf X_{m,I_m})}^2],
\end{equation}
where each $I_i\in \{1,2,\ldots,T_i\}^{T_i}$. Note that $\mathcal M$ is finite under Assumption \ref{8th moment} due to Jensen's inequality. Consider a generic summand from \eqref{expansion}
\begin{equation*}%\label{dgt2}
R_d(\mathcal T^1,\mathcal T^2)=\int h(\mathbf X_{1},\ldots,\mathbf X_m)\prod_{i=1}^m\prod_{t\notin \mathcal T_i^1\cup \mathcal T_i^2}dP_i(x_{i,t})\prod_{i=1}^m\prod_{t\in \mathcal T_i^1}d(\hat P_i-P_i)(x_{i,t})\prod_{i=1}^m\prod_{t\in \mathcal T_i^2}d(\mathbf w_i-\hat P_i)(x_{i,t})
\end{equation*}
where we denote $\mathcal T^1=(\mathcal T^1_1,\ldots,\mathcal T^1_m)$, $\mathcal T^2=(\mathcal T^2_1,\ldots,\mathcal T^2_m)$. Note that $\sum_{i=1}^m(\vert \mathcal T_i^1\vert +\vert \mathcal T_i^2\vert )=d$, and the subscript $d$ in $R_d(\mathcal T^1,\mathcal T^2)$ is used to emphasize this dependence. Let $\mathcal T_i^1(t)$ (or $\mathcal T_i^2(t)$) be the $t$-th element of $\mathcal T_i^1$ (or $\mathcal T_i^2$). Our goal is to show that
\begin{equation}\label{high-order}
\mathbb E\big[\sup_{(\mathbf w_1,\ldots,\mathbf w_m)\in\mathcal U_{\alpha}}\lvert R_d(\mathcal T^1,\mathcal T^2)\rvert^2\big]=O\big(\prod_{i=1}^mn_i^{-(\vert \mathcal T_i^1\vert+\vert \mathcal T_i^2\vert)}\big)=O(n^{-d}).
\end{equation}

First, we rewrite $R_d(\mathcal T^1,\mathcal T^2)$ as a sum and from there derive an upper bound \eqref{keyineq:cs} of its supremum. Define the conditional expectation of $h$ for given subscripts $\mathcal T^1=(\mathcal T^1_1,\ldots,\mathcal T^1_m),\mathcal T^2=(\mathcal T^2_1,\ldots,\mathcal T^2_m)$
\begin{equation*}
h_{\mathcal T^1,\mathcal T^2}(\mathbf x_{1,\mathcal T_1^1},\mathbf x_{1, \mathcal T_1^2},\ldots,\mathbf x_{m,\mathcal T_m^1},\mathbf x_{m,\mathcal T_m^2})=\mathbb E_{P_1,\ldots,P_m}[h(\mathbf X_1,\ldots,\mathbf X_m)\vert X_{i,t}=x_{i,t},\forall i\text{ and }t\in \mathcal T_i^1\cup \mathcal T_i^2]
\end{equation*}
% \begin{equation*}
% h_{\mathcal T^1,\mathcal T^2}(\mathbf x_1,\ldots,\mathbf x_m)=\mathbb E_{P_1,\ldots,P_m}\Big[h(\mathbf X_1,\ldots,\mathbf X_m)\Big\vert \begin{array}{l}X_{i}(\mathcal T_i^1(t))=x_{i,t},\forall i,1\leq t\leq \vert \mathcal T_i^1\vert\\X_{i}(\mathcal T_i^2(t-\vert \mathcal T_i^1\vert))=x_{i,t},\forall i,\vert \mathcal T_i^1\vert+1\leq t\leq \vert \mathcal T_i^1\vert+\vert \mathcal T_i^2\vert\end{array}\Big]
% \end{equation*}
where each $\mathbf x_{i,\mathcal T_i^1}=(x_{i,t})_{t\in \mathcal T_i^1}$ and $\mathbf x_{i,\mathcal T_i^2}=(x_{i,t})_{t\in \mathcal T_i^2}$. Considering all possible subsets $\tilde{\mathcal T}_i^1$ of $\mathcal T_i^1$ for each $i$ and denoting $\tilde{\mathcal T}^1=(\tilde{\mathcal T}_1^1,\ldots,\tilde{\mathcal T}_m^1)$, we define a centered conditional expectation (its property will be discussed momentarily)
\begin{eqnarray}
\notag&&\tilde{h}_{\mathcal T^1, \mathcal T^2}(\mathbf x_{1,\mathcal T_1^1},\mathbf x_{1, \mathcal T_1^2},\ldots,\mathbf x_{m,\mathcal T_m^1},\mathbf x_{m,\mathcal T_m^2})\\
&=&\sum_{\tilde{\mathcal T}_i^1\subset \mathcal T_i^1,\forall i}(-1)^{\sum_i(\lvert\mathcal T_i^1\rvert-\lvert\tilde{\mathcal T}_i^1\rvert)}h_{\tilde{\mathcal T}^1,\mathcal T^2}(\mathbf x_{1,\tilde{\mathcal T}_1^1},\mathbf x_{1, \mathcal T_1^2},\ldots,\mathbf x_{m,\tilde{\mathcal T}_m^1},\mathbf x_{m,\mathcal T_m^2}).\label{square-term}
\end{eqnarray}
% \begin{align}
% \nonumber&\tilde{h}_{\mathcal T^1, \mathcal T^2}(\mathbf X_{1},\ldots,\mathbf X_{m})\\
% =&\sum_{\tilde{\mathcal T}_i^1\subset \mathcal T_i^1}(-1)^{\sum_i\lvert\tilde{\mathcal T}_i^1\rvert}\mathbb E_{P_1,\ldots,P_m}[h_{\mathcal T^1, \mathcal T^2}(\mathbf X_{1},\ldots,\mathbf X_{m})\vert X_i(t)=x_{i,t},\forall i,t\text{ s.t. }t> \lvert\mathcal T_i^1\rvert\text{ or }\mathcal T_i^1(t)\notin \tilde{\mathcal T}_i^1].\label{square-term}
% \end{align}
By expanding out the product measure $\prod_{i=1}^m\prod_{t\notin \mathcal T_i^1\cup \mathcal T_i^2}dP_i(x_{i,t})\prod_{i=1}^m\prod_{t\in \mathcal T_i^1}d(\hat P_i-P_i)(x_{i,t})$ and noticing that each $\hat P_i$ is a probability measure, $R_d(\mathcal T^1,\mathcal T^2)$ can be expressed as
\begin{equation}\label{rd3}
\int \tilde{h}_{\mathcal T^1, \mathcal T^2}(\mathbf x_{1,\mathcal T_1^1},\mathbf x_{1, \mathcal T_1^2},\ldots,\mathbf x_{m,\mathcal T_m^1},\mathbf x_{m,\mathcal T_m^2})\prod_{i=1}^m\prod_{t\in \mathcal T_i^1}d\hat P_i(x_{i,t})\prod_{i=1}^m\prod_{t\in\mathcal T_i^2}d(\mathbf w_i-\hat P_i)(x_{i,t}).
% &=&\int \tilde{h}_{\mathcal T^1, \mathcal T^2}(\mathbf x_{1,\mathcal T_1^1\cup \mathcal T_1^2},\ldots,\mathbf x_{m,\mathcal T_m^1\cup \mathcal T_m^2})\prod_{i=1}^m\prod_{t=1}^{\lvert\mathcal T^1_i\rvert}d\hat P_i(x_{i,t})\prod_{i=1}^m\prod_{t= \lvert\mathcal T^1_i\rvert+1}^{\lvert\mathcal T^1_i\rvert+\lvert\mathcal T^2_i\rvert}d(\mathbf w_i-\hat P_i)(x_{i,t})
\end{equation}
From now on, we denote $X_{i,j},i=1,\ldots,m,j=1,\ldots,n_i$as the observations, and for each $i$ let
\begin{align*}
J^1_i&=(J^1_i(1),\ldots,J^1_i(\vert \mathcal T_i^1\vert))\in \set{1,2,\ldots,n_i}^{\vert \mathcal T_i^1\vert}\\
J^2_i&=(J^2_i(1),\ldots,J^2_i(\vert \mathcal T_i^2\vert))\in \set{1,2,\ldots,n_i}^{\vert \mathcal T_i^2\vert}
\end{align*}
be two sequences of indices (if $\mathcal T_i^1$ or $\mathcal T_i^2$ is empty, then $J^1_i$ or $J^2_i$ is empty accordingly) that specify the second subscript of data $X_{i,j}$. Then \eqref{rd3} can be written more explicitly as
\begin{equation*}
R_d(\mathcal T^1,\mathcal T^2)=\sum_{J^2_1,\ldots,J^2_m}\Big[\prod_{i,t}\big(w_{i,J^2_i(t)}-\frac{1}{n_i})\Big]\sum_{J^1_1,\ldots,J^1_m}\frac{1}{\prod_{i}n_i^{\vert \mathcal T^1_i\vert}}\tilde{h}_{\mathcal T^1,\mathcal T^2}(\mathbf X_{1,J^1_1},\mathbf X_{1,J^2_1},\ldots,\mathbf X_{m,J^1_m},\mathbf X_{m,J^2_m})%,\label{intsum}
\end{equation*}
where each $\mathbf X_{i,J^1_i}=(X_{i,J_i^1(1)},\ldots,X_{i,J_i^1(\lvert\mathcal T_i^1 \rvert)})$ contains the input data specified by $J^1_i$, and similarly $\mathbf X_{i,J^2_i}=(X_{i,J_i^2(1)},\ldots,X_{i,J_i^2(\lvert\mathcal T_i^2 \rvert)})$. We bound the supremum as follows
\begin{eqnarray*}
\lvert R_d(\mathcal T^1,\mathcal T^2)\vert^2&\leq& \Big[\sum_{J^2_1,\ldots,J^2_m}\prod_{i,t}\big(w_{i,J^2_i(t)}-\frac{1}{n_i}\big)^2\Big]\Big[\sum_{J^2_1,\ldots,J^2_m}\Big(\frac{1}{\prod_{i}n_i^{\vert \mathcal T^1_i\vert}}\sum_{J^1_1,\ldots,J^1_m}\tilde{h}_{\mathcal T^1,\mathcal T^2}\Big)^2\Big]\\
% \nonumber&=\sup_{\mathcal U_{\alpha}}\Big[\sum_{J^2_1,\ldots,J^2_m}\prod_{i,t}\big(w_{i,J^2_i(t)}-\frac{1}{n_i}\big)^2\Big]\Big[\sum_{J^2_1,\ldots,J^2_m}\Big(\frac{1}{\prod_{i}n_i^{\vert \mathcal T^1_i\vert}}\sum_{J^1_1,\ldots,J^1_m}\tilde{h}_{\mathcal T^1,\mathcal T^2}\Big)^2\Big]\\
&=& \prod_{i=1}^m\prth{\sum_{j=1}^{n_i}\prth{w_{i,j}-\frac{1}{n_i}}^2}^{\vert \mathcal T_i^2\vert}\Big[\sum_{J^2_1,\ldots,J^2_m}\Big(\frac{1}{\prod_{i}n_i^{\vert \mathcal T^1_i\vert}}\sum_{J^1_1,\ldots,J^1_m}\tilde{h}_{\mathcal T^1,\mathcal T^2}\Big)^2\Big]
\end{eqnarray*}
where we suppress the arguments of $\tilde{h}_{\mathcal T^1,\mathcal T^2}$, and use the Cauchy-Schwartz inequality. The upper bound from Lemma \ref{l2upbound} then implies that $\sum_{j=1}^{n_i}\prth{w_{i,j}-\frac{1}{n_i}}^2\leq u(\alpha)^2\mathcal X_{1,1-\alpha}^2/n_i^2$, and hence for some constant $C_1$ depending on $\alpha$ and $d$
\begin{equation}\label{keyineq:cs}
    \sup_{(\mathbf w_1,\ldots,\mathbf w_m)\in \mathcal U_{\alpha}}\lvert R_d(\mathcal T^1,\mathcal T^2)\vert^2\leq C_1\prod_{i=1}^mn_i^{-2\vert \mathcal T_i^2\vert}\cdot\Big[\sum_{J^2_1,\ldots,J^2_m}\Big(\frac{1}{\prod_{i}n_i^{\vert \mathcal T^1_i\vert}}\sum_{J^1_1,\ldots,J^1_m}\tilde{h}_{\mathcal T^1,\mathcal T^2}\Big)^2\Big].
\end{equation}

%where $\mathbf X_{i,J_i^1\cup J_i^2}=(x_{i,j})_{j\in J_i^1\cup J_i^2}$, and $\tilde{h}_{J^1,J^2}$ is the sum of $2^{\vert J^1\vert}$ functions that take the form of $\pm h_{J},J^2_i\subset J_i\subset J^1_i\cup J^2_i$ for each $i$. Here $\vert J^1\vert=\sum_{i=1}^m\vert J^1_i\vert$. Since $x_{i,t}$'s are formal arguments, \eqref{rd3} can be recast in a neat form
%\begin{equation}
%R(J^1,J^2)=\int \tilde{h}_{J^1,J^2}(\mathbf X_{1},\ldots,\mathbf X_{m})\prod_{i=1}^m\prod_{t=1}^{\abs{J^1_i}}d\hat P_i(x_{i,t})\prod_{i=1}^m\prod_{t=\abs{J^1_i}+1}^{\abs{J_i^1}+\abs{J_i^2}}d(\mathbf w_i-\hat P_i)(x_{i,t}),\label{rd4}
%\end{equation}
%where $\mathbf X_{i}=(x_{i,1},\ldots,x_{i,\vert J^1_i\vert +\vert J^2_i\vert})$. 

From \eqref{keyineq:cs}, the proof now boils down to bounding the expectation of
\begin{equation*}
\Big(\frac{1}{\prod_{i}n_i^{\vert \mathcal T^1_i\vert}}\sum_{J^1_1,\ldots,J^1_m}\tilde{h}_{\mathcal T^1,\mathcal T^2}(\mathbf X_{1,J^1_1},\mathbf X_{1,J^2_1},\ldots,\mathbf X_{m,J^1_m},\mathbf X_{m,J^2_m})\Big)^2
\end{equation*}
for each fixed $J^2_1,\ldots,J^2_m$. We need a few properties of $\tilde{h}_{\mathcal T^1,\mathcal T^2}$. The first property, which follows from its definition, is that, for any $i$ and $t\in\mathcal T_i^1$, the marginal expectation under the true input distributions is zero, i.e.
\begin{equation}
\int \tilde{h}_{\mathcal T^1,\mathcal T^2}(\mathbf x_{1,\mathcal T_1^1},\mathbf x_{1, \mathcal T_1^2},\ldots,\mathbf x_{m,\mathcal T_m^1},\mathbf x_{m,\mathcal T_m^2})dP_i(x_{i,t})=0.\label{cond0}
\end{equation}
The second property is a bound of the second moment that is uniform in $\mathcal T^1,\mathcal T^2$. By Jensen's inequality, one can show that for any $m$ sequences of indices $I_i=(I_i(1),\ldots,I_i(\lvert\mathcal T_i^1\rvert+\lvert\mathcal T_i^2\rvert))\in \set{1,2,\ldots,\lvert\mathcal T_i^1\rvert+\lvert\mathcal T_i^2\rvert}^{\lvert\mathcal T_i^1\rvert+\lvert\mathcal T_i^2\rvert}$ the conditional expectation $h_{\mathcal T^1,\mathcal T^2}$ satisfies
\begin{equation*}
\mathbb E_{P_1,\ldots,P_m}[ h_{\mathcal T^1,\mathcal T^2}^{2}(\mathbf X_{1,I_1},\ldots,\mathbf X_{m,I_m})]\leq \mathcal M%,\label{condbd}
\end{equation*}
where $\mathbf X_{i,I_i}=(X_{i}(I_i(1)),\ldots,X_{i}(I_i(\lvert\mathcal T_i^1\rvert+\lvert\mathcal T_i^2\rvert)))$ and $\mathcal M$ is the second moment bound defined in \eqref{2moment}. \eqref{square-term} tells us that $\tilde{h}_{\mathcal T^1,\mathcal T^2}$ is the sum of $2^{\lvert\mathcal T^1\rvert}$ conditional expectations of such type. By the Minkowski inequality we have
\begin{equation}
\mathbb E_{P_1,\ldots,P_m} [\tilde{h}_{\mathcal T^1,\mathcal T^2}^{2}(\mathbf X_{1,I_1},\ldots,\mathbf X_{m,I_m})]\leq 4^{\vert \mathcal T^1\vert}\mathcal M.\label{bound2k}
\end{equation}
Now we are able to proceed with
\begin{align}
\nonumber&\mathbb E \Big(\frac{1}{\prod_{i}n_i^{\vert \mathcal T^1_i\vert}}\sum_{J^1_1,\ldots,J^1_m}\tilde{h}_{\mathcal T^1,\mathcal T^2}(\mathbf X_{1,J^1_1},\mathbf X_{1,J^2_1},\ldots,\mathbf X_{m,J^1_m},\mathbf X_{m,J^2_m})\Big)^2\\
\nonumber=&\frac{1}{\prod_{i}n_i^{2\vert \mathcal T^1_i\vert}}\sum_{J^1_1,\ldots,J^1_m}\sum_{\tilde{J}^1_1,\ldots,\tilde{J}^1_m}\mathbb E[\tilde{h}_{\mathcal T^1,\mathcal T^2}(\mathbf X_{1,J^1_1},\mathbf X_{1,J^2_1},\ldots,\mathbf X_{m,J^1_m},\mathbf X_{m,J^2_m})\cdot\\
&\hspace{34ex}\tilde{h}_{\mathcal T^1,\mathcal T^2}(\mathbf X_{1,\tilde{J}^1_1},\mathbf X_{1,J^2_1},\ldots,\mathbf X_{m,\tilde{J}^1_m},\mathbf X_{m,J^2_m})].\label{2ke}
\end{align}
Note that because of property \eqref{cond0}, the expectation in \eqref{2ke} is zero if there is some index $i^*\in\{1,\ldots,m\}$ and $j^*\in\{1,\ldots,n_i\}$ such that $X_{i^*,j^*}$ does not appear in $\mathbf X_{i^*,J^2_{i^*}}$ and shows up exactly once among $\mathbf X_{i^*,J^{1}_{i^*}},\mathbf X_{i^*,\tilde{J}^{1}_{i^*}}$. Note that, for each fixed $i=1,\ldots,m$, the number of choices of $J^{1}_i,\tilde{J}^{1}_i$ that avoid this occurrence is no more than $C_2n_i^{\vert \mathcal T^1_i\vert}$, where $C_2$ is some constant depending on $d$ only. So the total number of choices of $J^{1}_i,\tilde{J}^{1}_i,i=1,\ldots,m$ that can possibly produce a nonzero expectation in \eqref{2ke} is at most
\begin{equation}
C_2^m\Big(\prod_{i=1}^mn_i^{\vert \mathcal T^1_i\vert}\Big).\label{nonzero}
\end{equation}
% Some basic combinatorial analysis shows 
On the other hand, applying the Cauchy-Schwartz inequality and the upper bound \eqref{bound2k} to the expectation in \eqref{2ke} gives
\begin{equation*}
\abs{\mathbb E[\tilde{h}_{\mathcal T^1,\mathcal T^2}(\mathbf X_{1,J^1_1},\mathbf X_{1,J^2_1},\ldots,\mathbf X_{m,J^1_m},\mathbf X_{m,J^2_m})\tilde{h}_{\mathcal T^1,\mathcal T^2}(\mathbf X_{1,\tilde{J}^1_1},\mathbf X_{1,J^2_1},\ldots,\mathbf X_{m,\tilde{J}^1_m},\mathbf X_{m,J^2_m})]}\leq 4^{\vert \mathcal T^1\vert}\mathcal M
\end{equation*}
for any $J^{1}_i,\tilde{J}^{1}_i,J^2_i,i=1,\ldots,m$. We conclude from \eqref{2ke}, \eqref{nonzero} and the above bound that
\begin{equation}
\mathbb E \Big(\frac{1}{\prod_{i}n_i^{\vert \mathcal T^1_i\vert}}\sum_{J^1_1,\ldots,J^1_m}\tilde{h}_{\mathcal T^1,\mathcal T^2}(\mathbf X_{1,J^1_1},\mathbf X_{1,J^2_1},\ldots,\mathbf X_{m,J^1_m},\mathbf X_{m,J^2_m})\Big)^2\leq \frac{4^{\vert \mathcal T^1\vert}C_2^m\mathcal M}{\prod_{i}n_i^{\lvert\mathcal T_i^1\rvert}}\label{2kbound}
\end{equation}
uniformly for all choices of $J^2_i,i=1,\ldots,m$.

Finally, we go back to the inequality \eqref{keyineq:cs} to arrive at
\begin{eqnarray}
\nonumber\mathbb E\big[\sup_{(\mathbf w_1,\ldots,\mathbf w_m)\in\mathcal U_{\alpha}}\lvert R_d(\mathcal T^1,\mathcal T^2)\rvert^2\big]& \leq& C_1\prod_{i=1}^mn_i^{-2\vert \mathcal T_i^2\vert}\cdot\Big[\sum_{J^2_1,\ldots,J^2_m}\mathbb E\Big(\prod_{i}\frac{1}{n_i^{\vert \mathcal T^1_i\vert}}\sum_{J^1_1,\ldots,J^1_m}\tilde{h}_{\mathcal T^1,\mathcal T^2}\Big)^2\Big]\\
\nonumber & \leq& C_1\prod_{i=1}^mn_i^{-2\vert \mathcal T_i^2\vert}\cdot\Big[\sum_{J^2_1,\ldots,J^2_m}\frac{4^{\vert \mathcal T^1\vert}C_2^m\mathcal M}{\prod_{i}n_i^{\lvert\mathcal T_i^1\rvert}}\Big]\\
&\leq &4^{\vert \mathcal T^1\vert}C_1C_2^m\mathcal M \prod_{i=1}^mn_i^{-(\vert \mathcal T_i^1\vert+\vert \mathcal T_i^2\vert)}.\label{generalbd}
\end{eqnarray}
This proves \eqref{high-order}. Note that, since $T$ is fixed, from \eqref{expansion}, 
$$\sup_{(\mathbf w_1,\ldots,\mathbf w_m)\in\mathcal U_{\alpha}}\big\lvert\sum_{\mathcal T^1,\mathcal T^2,d\geq 2}R_d(\mathcal T^1,\mathcal T^2)\big\rvert\leq \sum_{\mathcal T^1,\mathcal T^2,d\geq 2}\sup_{(\mathbf w_1,\ldots,\mathbf w_m)\in\mathcal U_{\alpha}}\lvert R_d(\mathcal T^1,\mathcal T^2)\rvert$$ 
% there are only finite many of terms like $R_d(\mathcal T^1,\mathcal T^2)$ in
and the Minkowski inequality we conclude that $\mathbb E\big[\sup_{(\mathbf w_1,\ldots,\mathbf w_m)\in\mathcal U_{\alpha}}\lvert\sum_{\mathcal T^1,\mathcal T^2,d\geq 2}R_d(\mathcal T^1,\mathcal T^2)\rvert^2\big]=O(n^{-2})$. This therefore shows that $\mathbb E\big[\sup_{(\mathbf w_1,\ldots,\mathbf w_m)\in\mathcal U_{\alpha}}\lvert Z-Z_L\rvert^2\big]=O(n^{-2})$ as the data size $n\to\infty$.

Now we prove the uniform approximation error of $\widehat{Z_L}$. The approach is to expand the integral form of $Z(\mathbf w_1,\ldots,\mathbf w_m)$ in a similar way to \eqref{expansion}, but around $\hat P_i$'s instead of $P_i$'s
\begin{equation}\label{hat expansion}
Z(\mathbf w_1,\ldots,\mathbf w_m)=\sum_{d=0}^T\sum_{\sum_i\vert \mathcal T_i^2\vert=d}\int h(\mathbf x_1,\ldots,\mathbf x_m)\prod_{i=1}^m\prod_{t\notin \mathcal T_i^2}d\hat P_i(x_{i,t})\prod_{i=1}^m\prod_{t\in \mathcal T_i^2}d(\mathbf w_i-\hat P_i)(x_{i,t})
\end{equation}
where each $\mathcal T_i^2$ is again an ordered subset of $\set{1,2,\ldots,T_i}$ that contains the second subscript $t$ of the argument $x_{i,t}$. Similar to above, summands with $d=0,1$ gives the linear approximation at the empirical distributions, i.e.~$\widehat{Z_L}(\mathbf w_1,\ldots,\mathbf w_m)$, and all summands with $d\geq 2$ will be the associated approximation error. To bound each summand with $d\geq 2$, we rewrite $\hat P_i$ as $\hat P_i-P_i+P_i$, and suitably expand out the product measure $\prod_{i=1}^m\prod_{t\notin \mathcal T_i^2}d\hat P_i(x_{i,t})$ in \eqref{hat expansion} to get
\begin{eqnarray*}
&&\int h(\mathbf x_1,\ldots,\mathbf x_m)\prod_{i=1}^m\prod_{t\notin \mathcal T_i^2}d\hat P_i(x_{i,t})\prod_{i=1}^m\prod_{t\in \mathcal T_i^2}d(\mathbf w_i-\hat P_i)(x_{i,t})\\
&=&\sum_{\mathcal T_i^1,\text{ s.t.}\mathcal T_i^1\cap \mathcal T_i^2=\emptyset}\int h(\mathbf x_1,\ldots,\mathbf x_m)\prod_{i=1}^m\prod_{t\notin \mathcal T_i^1\cup \mathcal T_i^2}d P_i(x_{i,t})\prod_{i=1}^m\prod_{t\in \mathcal T_i^1}d(\hat P_i-P_i)(x_{i,t})\prod_{i=1}^m\prod_{t\in \mathcal T_i^2}d(\mathbf w_i-\hat P_i)(x_{i,t})\\
&=&\sum_{\mathcal T_i^1,\text{ s.t.}\mathcal T_i^1\cap \mathcal T_i^2=\emptyset}R_{\lvert \mathcal T^1 \rvert+d}(\mathcal T^1,\mathcal T^2)
\end{eqnarray*}
where each $\mathcal T_i^1$ is the ordered set consisting of the second subscripts $t$ of all $x_{i,t}$'s to which $\hat P_i-P_i$ is distributed, and $R_{\lvert \mathcal T^1 \rvert+d}(\mathcal T^1,\mathcal T^2)$ is the remainder term defined before.
% , where $\mathcal T^1=(\mathcal T_1^1,\ldots,\mathcal T_m^1)$ and $\mathcal T^2=(\mathcal T_1^2,\ldots,\mathcal T_m^2)$, gives the following representation for the sum of terms with $d\geq2$
% \begin{equation*}
% \sum_{\sum_i\vert \mathcal T_i^2\vert=d}\int h(\mathbf x_1,\ldots,\mathbf x_m)\prod_{i=1}^m\prod_{t\notin \mathcal T_i^2}d\hat P_i(x_{i,t})\prod_{i=1}^m\prod_{t\in \mathcal T_i^2}d(\mathbf w_i-\hat P_i)(x_{i,t})=\sum_{\mathcal T^1,\mathcal T^2,\vert \mathcal T^2\vert=d}R_{\lvert \mathcal T^1 \rvert+d}(\mathcal T^1,\mathcal T^2).
% \end{equation*}
The desired conclusion then follows from \eqref{high-order} and an argument analogous to the first part of the theorem.
% and that the approximation error $Z-\widehat{Z_L}$ consists of finitely many terms like $R_{\lvert \mathcal T^1 \rvert+d}(\mathcal T^1,\mathcal T^2)$.
\hfill\Halmos
\endproof

\section{Proof of Results in Section \ref{sec:EL theorem}}\label{sec:proof elt}
\proof{Proof of Theorem \ref{elt}.}To simplify the proof, we first argue that one can assume $\mathrm{Var}(Y_i)>0$ and $\mathbb EY_i=0$ for all $i=1,\ldots,m$ without loss of generality. Let $I=\{i:\mathrm{Var}(Y_i)>0, i=1,\ldots,m\}$ be the set of indices whose corresponding $Y_i$'s have non-zero variances. Then for $i\notin I$ each $Y_{i,j}=\mathbb EY_i$ almost surely, hence
\begin{eqnarray*}
R(\mu_0)&=&\max\set{\prod_{i=1}^m\prod_{j=1}^{n_i}n_iw_{i,j}\bigg\vert \sum_{i=1}^m\sum_{j=1}^{n_i}Y_{i,j}w_{i,j}=\mu_0,\ \sum_{j=1}^{n_i}w_{i,j}=1\text{\ for all\ }i,\ w_{i,j}\geq 0\text{ for all }i,j}\\
&=&\max\set{\prod_{i=1}^m\prod_{j=1}^{n_i}n_iw_{i,j}\bigg\vert \sum_{i\in I}\sum_{j=1}^{n_i}Y_{i,j}w_{i,j}=\mu_0-\sum_{i\notin I}\mathbb EY_i,\ \sum_{j=1}^{n_i}w_{i,j}=1\text{\ for all\ }i,\ w_{i,j}\geq 0\text{ for all }i,j}\\
&=&\max\set{\prod_{i\in I}\prod_{j=1}^{n_i}n_iw_{i,j}\bigg\vert \sum_{i\in I}\sum_{j=1}^{n_i}Y_{i,j}w_{i,j}=\sum_{i\in I}\mathbb EY_i,\ \sum_{j=1}^{n_i}w_{i,j}=1\text{\ for\ }i\in I,\ w_{i,j}\geq 0\text{ for all }i\in I,j}\\
&=&\max\set{\prod_{i\in I}\prod_{j=1}^{n_i}n_iw_{i,j}\bigg\vert \sum_{i\in I}\sum_{j=1}^{n_i}(Y_{i,j}-\mathbb EY_i)w_{i,j}=0,\ \sum_{j=1}^{n_i}w_{i,j}=1\text{\ for\ }i\in I,\ w_{i,j}\geq 0\text{ for all }i\in I,j}\\
&=&R_I(0)
\end{eqnarray*}
where $R_I(0)$ is the analog of $R(\mu_0)$ defined for the translated observations $\{Y_{i,1}-\mathbb EY_i,\ldots,Y_{i,n_i}-\mathbb EY_i\},i\in I$, and in the third equality we put $w_{i,j}=1/n_i$ for $i\notin I$ into the objective, which can be easily seen to be the maximizing weights for $i\notin I$. Therefore, to prove the theorem for $R(\mu_0)$, one can work with $R_I(0)$ instead, and note that the change of $m$, the number of independent distributions, does not affect the limit chi-square distribution. 
% By letting $w_{i,j}=\frac{1}{n_i}$ for $i\notin I$, it is clear that $R_I(\sum_{i\in I}\mathbb EX_i)$ is always equal to $R(\sum_{i=1}^m\mathbb EX_i)$. So we can always assume that $\text{Var}(X_i)>0$ for each $i$, i.e. $I=\{1,2,\ldots,m\}$. 
% Moreover, one can assume each $\mathbb  EY_i=0$, hence $\mu_0=0$, by (theoretically) translating the sample $\{Y_{i,1},\ldots,Y_{i,n_i}\}$ to $\{Y_{i,1}-\mathbb EY_i,\ldots,Y_{i,n_i}-\mathbb EY_i\}$ for each $i$ and noting that the profile likelihood ratio remains the same, i.e.
% \begin{equation*}
%     R(\mu_0)=\max\set{\prod_{i=1}^m\prod_{j=1}^{n_i}n_iw_{i,j}\bigg\vert \sum_{i=1}^m\sum_{j=1}^{n_i}(Y_{i,j}-\mathbb EY_i)w_{i,j}=0,\ \sum_{j=1}^{n_i}w_{i,j}=1\text{\ for all\ }i,\ w_{i,j}\geq 0\text{ for all }i,j}.
% \end{equation*}

In view of the above, we shall assume $\mathrm{Var}(Y_i)>0$ and $\mathbb EY_i=0$ for each $i$, hence $R(\mu_0)$ is just $R(0)$. Introducing a slack variable $\mu_i$ for each $\sum_{j=1}^{n_i}Y_{i,j}w_{i,j}$ and taking the negative logarithm of the objective convert the defining maximization of $R(0)$ to the following convex program
\begin{equation}
\begin{aligned}
& \underset{\mathbf w_1,\ldots,\mathbf w_m,\bm{\mu}}{\text{min}}& & -\sum_{i=1}^m\sum_{j=1}^{n_i}\log(n_iw_{i,j}) \\
& \text{subject to}
& & \sum_{j=1}^{n_i}Y_{i,j}w_{i,j}=\mu_i,\ i=1,\ldots,m\\
&&&\sum_{j=1}^{n_i}w_{i,j}=1,\ i=1,\ldots,m\\
&&&\sum_{i=1}^m\mu_i=0\\
\end{aligned}\label{keyopt}
\end{equation}
where $\bm{\mu}=(\mu_1,\ldots,\mu_m)$. The non-negativity constraints $w_{i,j}\geq 0$ are dropped since they are implicitly imposed in the objective function.

\textbf{Step one:} We prove that, with probability tending to one, Slater's condition holds for \eqref{keyopt}. In other words, consider the event
\begin{equation*}
    \mathcal S = \set{Y_{i,j},i=1,\ldots,m,j=1,\ldots,n_i\Big\vert\text{\begin{tabular}{l}
        \eqref{keyopt} has at least one feasible solution   \\
        $(\mathbf w_1,\ldots,\mathbf w_m,\bm{\mu})$ such that $w_{i,j}>0$ for all $i,j$
    \end{tabular}}}
\end{equation*}
and we prove $P(\mathcal S)\to 1$ as $n\to\infty$. To this end, consider the following events indexed by $i$
% \begin{equation*}
%     \tilde{\mathcal S} = \set{Y_{i,j},i=1,\ldots,m,j=1,\ldots,n_i:\text{\begin{tabular}{l}
%         for each $i$ there exists a $\mathbf w_i$ such that\\
%          $w_{i,j}>0$ for all $j$, $\sum_{j=1}^{n_i}w_{i,j}=1$, and $\sum_{j=1}^{n_i}Y_{i,j}w_{i,j}=0$
%     \end{tabular}}}.
% \end{equation*}
\begin{equation*}
    \tilde{\mathcal S}_i = \set{Y_{i,j},j=1,\ldots,n_i\Big\vert\;\min_{j=1,\ldots,n_i}Y_{i,j}<0<\max_{j=1,\ldots,n_i}Y_{i,j}}.
\end{equation*}
We shall prove that $P(\tilde{\mathcal S}_i)\to 1$ for all $i$ and that $\cap_{i=1}^m\tilde{\mathcal S}_i\subseteq \mathcal S$, which imply that $P(\mathcal S)\to 1$ because
$$P(\mathcal S^c)\leq P((\cap_{i=1}^m\tilde{\mathcal S}_i)^c)=P(\cup_{i=1}^m\tilde{\mathcal S}_i^c)\leq \sum_{i=1}^m P(\tilde{\mathcal S}_i^c)=\sum_{i=1}^m(1- P(\tilde{\mathcal S}_i))\to 0.$$
Note that $\mathrm{Var}(Y_i)>0$ and $\mathbb EY_i=0$ imply $P(Y_i\geq 0)<1,P(Y_i\leq 0)<1$. Hence as $n\to \infty$
\begin{equation*}
    P\big(\min_{j=1,\ldots,n_i}Y_{i,j}\geq 0\big)=\prod_{j=1}^{n_i}P(Y_{i,j}\geq 0)=(P(Y_i\geq 0))^{n_i}\to 0
\end{equation*}
which is equivalently $P\big(\min_jY_{i,j}< 0\big)\to 1$. Similarly, $P\big(\max_jY_{i,j}> 0\big)\to 1$ holds. Combining these two limits gives
% By the strong law of large numbers, for each $i$
% \begin{align}
% \frac{1}{n_i}\sum_{j=1}^{n_i}\mathbf{1}\{ Y_{i,j}>0 \} &\to P(Y_i>0)>0\label{>0}\\
% \frac{1}{n_i}\sum_{j=1}^{n_i}\mathbf{1}\{ Y_{i,j}<0 \} &\to P(Y_i<0)>0\label{<0}
% \end{align}
% almost surely, where $\mathbf{1}\{ \cdot\}$ is the indicator function. Since $\sum_{j=1}^{n_i}\mathbf{1}\{ Y_{i,j}>0 \}>0$ implies $\max_{j}Y_{i,j}>0$ and $\sum_{j=1}^{n_i}\mathbf{1}\{ Y_{i,j}<0 \}>0$ implies $\min_{j}Y_{i,j}<0$, \eqref{>0} and \eqref{<0} imply that, for each $i$, $\mathbf{1}\{\tilde{\mathcal S}_i\}\to 1$ almost surely 
$P(\tilde{\mathcal S}_i)\to 1$.
% Hence with probability tending to one
% \begin{equation*}
% \begin{aligned}
% \frac{1}{n_i}\sum_{j=1}^{n_i}\mathbf{1}\{ X_{i,j}>\mathbb EX_i\} >0,\\
% \frac{1}{n_i}\sum_{j=1}^{n_i}\mathbf{1}\{ X_{i,j}<\mathbb EX_i\} >0.\\
% \end{aligned}
% \end{equation*}
% This implies that $\mathbb EX_i$ lies in the open interval $(\min_{j}X_{i,j},\max_{j}X_{i,j})$, and 
To show $\cap_{i=1}^m\tilde{\mathcal S}_i\subseteq \mathcal S$, note that if $\tilde{\mathcal S}_i$ happens then there must exist convex-combination weights $w_{i,j}>0,\sum_{j=1}^{n_i}w_{i,j}=1$ such that $\sum_{j=1}^{n_i}Y_{i,j}w_{i,j}=0$. When all $\tilde{\mathcal S}_i$'s happen, one can take such weights and $\mu_i=0$ for each $i$ to see that $\mathcal S$ also happens.

\textbf{Step two:} We derive the KKT conditions for \eqref{keyopt}, conditioned on Slater's condition $\mathcal S$. Notice that each $-\log(n_iw_{i,j})$ is bounded below by $-\log n_i$, and when $w_{i,j}\to 0$ for some $i,j$ the corresponding $-\log(n_iw_{i,j})\to+\infty$, hence the objective $-\sum_{i,j}\log(n_iw_{i,j})\to+\infty$ as $\min_{i,j}w_{i,j}\to 0$. Therefore, the optimal solution, if it exists, must lie in the region where $\min_{i,j}w_{i,j}\geq\epsilon$ for some small $\epsilon>0$ that depends on $n_i$'s. Since the set $\{(\mathbf w_1,\ldots,\mathbf w_m):\sum_{i=1}^m\sum_{j=1}^{n_i}Y_{i,j}w_{i,j}=0,\sum_{j=1}^{n_i}w_{i,j}=1,w_{i,j}\geq \epsilon\text{ for all }i,j\}$ is compact, an optimal solution $(\mathbf w_1^*,\ldots,\mathbf w_m^*,\bm{\mu}^*)$ exists for \eqref{keyopt}. Moreover, strict convexity of the objective forces the optimal solution to be unique. By Corollary 28.3.1 of \cite{rockafellar2015convex}, there must exist Lagrange multipliers $(\bm{\lambda_1}^*,\bm{\lambda_2}^*,\lambda^*)$, where $\bm{\lambda_1}^*=(\lambda_{1,1},\ldots,\lambda_{1,m})$ is associated with the first $m$ constraints, $\bm{\lambda_2}^*=(\lambda_{2,1},\ldots,\lambda_{2,m})$ with the second $m$ constraints, and $\lambda^*$ with the last constraint in \eqref{keyopt}, such that together with the optimal solution $(\mathbf w_1^*,\ldots,\mathbf w_m^*,\bm{\mu}^*)$ satisfy the following KKT conditions
\begin{align}
\nonumber\sum_{j=1}^{n_i}Y_{i,j}w_{i,j}^*&=\mu_i^*,\text{ for }i=1,\ldots,m\\
\nonumber\sum_{j=1}^{n_i}w_{i,j}^*&=1,\text{ for }i=1,\ldots,m\\
\nonumber\sum_{i=1}^m\mu_i^*&=0\\
-\frac{1}{w_{i,j}^*}+Y_{i,j}\lambda_{1,i}^*+\lambda_{2,i}^*&=0,\text{ for all } i,j\label{wij}\\
\nonumber-\lambda_{1,i}^*+\lambda^*&=0,\text{ for }i=1,\ldots,m.
\end{align}
Some basic algebra shows $\lambda_{2,i}^*=n_i-\lambda_{1,i}^*\mu_i^*,\lambda^*=\lambda_{1,i}^*$ for all $i$, hence it follows from \eqref{wij} that
\begin{equation}\label{formulawij}
w_{i,j}^*=\frac{1}{n_i+\lambda^*(Y_{i,j}-\mu_i^*)}
\end{equation}
% where we substitute $\lambda_3^*$ with $\lambda^*$ for convenience, 
and $\lambda^*,\mu_i^*$ satisfy
\begin{align}
\sum_{j=1}^{n_i}\frac{Y_{i,j}-\mu_i^*}{n_i+\lambda^*(Y_{i,j}-\mu_i^*)}&=0,\text{ for }i=1,\ldots,m\label{meancon}\\
\sum_{i=1}^m\mu_i^*&=0.\label{zerocon}
\end{align}

\textbf{A note on Slater's condition:} 
% the rest of the proof relies on the existence of $\lambda^*,\mu_i^*,i=1,\ldots,m$ and the relations \eqref{formulawij}, \eqref{meancon} and \eqref{zerocon}. because the derivation builds upon \eqref{formulawij}, \eqref{meancon} and \eqref{zerocon}That being said,
Note that $\lambda^*,\mu_i^*,i=1,\ldots,m$ are guaranteed to exist and defined as above only when Slater's condition $\mathcal S$ holds. In the rest of the proof, we set $\lambda^*,\mu_i^*,i=1,\ldots,m$ as defined by \eqref{meancon} and \eqref{zerocon} when $\mathcal S$ happens, and arbitrarily defined otherwise (e.g., simply let them all be $0$). Every intermediate inequality/equality below related to $\lambda^*,\mu_i^*,i=1,\ldots,m$ is interpreted as restricted to the event of $\mathcal S$. For example, $a\leq b$ and $a=b$ should be interpreted as $a\cdot \mathbf{1}\{\mathcal S\}\leq b\cdot \mathbf{1}\{\mathcal S\}$ and $a\cdot \mathbf{1}\{\mathcal S\}= b\cdot \mathbf{1}\{\mathcal S\}$. All asymptotic statements or quantities that rely on stochastic orders $o_p,O_p$ and convergence in distribution, remain valid via a decomposition of the involved probability into $\mathcal S$ and $\mathcal S^c$ and using  $P(\mathcal S)\to 1$. To demonstrate this argument concretely, we will show as an example in \eqref{remove_slater} how it works. But to avoid adding overwhelming complexities to our proof, we will keep this aspect silent until then. 

% statements are free of the concern of conditioning on $\mathcal S$.
% For simplicity, the multiplication with $\mathbf{1}\{\mathcal S\}$ will be omitted until the last step. 

\textbf{Step three:} We show that the Lagrange multiplier $\lambda^*$ has a magnitude of $O_p(n^{1/2})$. Write \eqref{formulawij} as
\begin{equation}
\frac{1}{n_i+\lambda^*(Y_{i,j}-\mu_i^*)}=\frac{1}{n_i}\prth{1-\frac{\frac{\lambda^*}{n_i}(Y_{i,j}-\mu_i^*)}{1+\frac{\lambda^*}{n_i}(Y_{i,j}-\mu_i^*)}}\label{formula2wij}
\end{equation}
and substituting \eqref{formula2wij} into \eqref{meancon} gives
\begin{equation}
\bar{Y}_i-\mu_i^*=\frac{1}{n_i}\sum_{j=1}^{n_i}\frac{\frac{\lambda^*}{n_i}(Y_{i,j}-\mu_i^*)^2}{1+\frac{\lambda^*}{n_i}(Y_{i,j}-\mu_i^*)},\label{forlambda}
\end{equation}
where $\bar{Y}_i=\frac{1}{n_i}\sum_{j=1}^{n_i}Y_{i,j}$. Multiply both sides by $\text{sign}(\lambda^*)$ to make the right hand side positive
\begin{equation}
\text{sign}(\lambda^*)(\bar{Y}_i-\mu_i^*)=\frac{1}{n_i}\sum_{j=1}^{n_i}\frac{\frac{\lvert\lambda^*\rvert}{n_i}(Y_{i,j}-\mu_i^*)^2}{1+\frac{\lambda^*}{n_i}(Y_{i,j}-\mu_i^*)}.\label{lambda1}
\end{equation}
This is because, since each $w_{i,j}^*$ is strictly positive, from \eqref{formulawij} we must have $1+\frac{\lambda^*}{n_i}(Y_{i,j}-\mu_i^*)>0,\,\forall i,j$. Also note that $\lvert\mu_i^*\rvert=\abs{\sum_{j=1}^{n_i}Y_{i,j}w_{i,j}^*}\leq \sum_{j=1}^{n_i}w_{i,j}^*\abs{Y_{i,j}}\leq\max_{j=1,\ldots,n_i}\abs{Y_{i,j}}$. Let $Z_N=\max_{i=1,\ldots,m,j=1,\ldots,n_i}\abs{Y_{i,j}}$. A lower bound of the right hand side of \eqref{lambda1} can be derived as follows
\begin{eqnarray}
\nonumber\frac{1}{n_i}\sum_{j=1}^{n_i}\frac{\frac{\lvert\lambda^*\rvert}{n_i}(Y_{i,j}-\mu_i^*)^2}{1+\frac{\lambda^*}{n_i}(Y_{i,j}-\mu_i^*)}&\geq&\frac{1}{n_i}\sum_{j=1}^{n_i}\frac{\frac{\lvert\lambda^*\rvert}{n_i}(Y_{i,j}-\mu_i^*)^2}{1+\frac{\lvert\lambda^*\rvert}{n_i}\abs{Y_{i,j}-\mu_i^*}}\\
\nonumber&\geq&\frac{1}{n_i}\sum_{j=1}^{n_i}\frac{\frac{\lvert\lambda^*\rvert}{n_i}(Y_{i,j}-\mu_i^*)^2}{1+\frac{\lvert\lambda^*\rvert}{n_i}\cdot 2\max_{j=1,\ldots,n_i}\abs{Y_{i,j}}}\\
\nonumber&\geq&\frac{1}{n_i}\sum_{j=1}^{n_i}\frac{\frac{\lvert\lambda^*\rvert}{n\cdot\overline{c}/\underline{c}}(Y_{i,j}-\mu_i^*)^2}{1+\frac{\lvert\lambda^*\rvert}{n\cdot \underline{c}/\overline{c}}\cdot 2Z_N}\text{ where }\underline{c},\overline{c}\text{ are from Assumption \ref{balanced data}}\\
\nonumber&=&\frac{\frac{\lvert\lambda^*\rvert}{n\cdot \overline{c}/\underline{c}}}{1+\frac{\lvert\lambda^*\rvert}{n\cdot \underline{c}/\overline{c}}\cdot 2Z_N}\prth{\hat{\sigma}_i^2-2\bar{Y}_i\mu_i^*+{\mu_i^*}^2}\text{ where } \hat{\sigma}_i^2=\frac{1}{n_i}\sum_{j=1}^{n_i}Y_{i,j}^2\\
&\geq&\frac{\frac{\lvert\lambda^*\rvert}{n\cdot \overline{c}/\underline{c}}}{1+\frac{\lvert\lambda^*\rvert}{n\cdot \underline{c}/\overline{c}}\cdot 2Z_N}\prth{\hat{\sigma}_i^2-2\bar{Y}_i\mu_i^*}\label{lowbd1}
\end{eqnarray}
Applying Lemma 11.2 in \cite{owen2001empirical} to $\{Y_{i,1},\ldots,Y_{i,n_i}\}$ reveals that, almost surely, $\max_{j=1,\ldots,n_i}\abs{Y_{i,j}}=o(n_i^{\frac{1}{2}})$ as $n_i\to\infty$ for each $i$, hence $Z_N=o(n^{\frac{1}{2}})$ and $\mu_i^*=o(n^{\frac{1}{2}})$ almost surely. By the central limit theorem, each $\bar{Y}_i=O_p(n_i^{-\frac{1}{2}})=O_p(n^{-\frac{1}{2}})$. Substituting the lower bound \eqref{lowbd1} into \eqref{lambda1} and multiplying each side by $1+\frac{\lvert\lambda^*\rvert}{n\cdot \underline{c}/\overline{c}}\cdot 2Z_N$ give
\begin{eqnarray}
\nonumber\prth{1+\frac{\lvert\lambda^*\rvert}{n\cdot \underline{c}/\overline{c}}\cdot 2Z_N}\text{sign}(\lambda^*)\prth{\bar{Y}_i-\mu_i^*}&\geq& \frac{\lvert\lambda^*\rvert}{n\cdot \overline{c}/\underline{c}}\prth{\hat{\sigma}_i^2-2\bar{Y}_i\mu_i^*}\\
&\geq& \frac{\lvert\lambda^*\rvert}{n\cdot \overline{c}/\underline{c}}(\hat{\sigma}_i^2+O_p(n^{-\frac{1}{2}})o(n^{\frac{1}{2}}))\\
&=&\frac{\lvert\lambda^*\rvert}{n\cdot \overline{c}/\underline{c}}(\hat{\sigma}_i^2+o_p(1)).\label{indineq}
\end{eqnarray}
Summing up both sides of \eqref{indineq} over $i=1,\ldots,m$, and using \eqref{zerocon} and $Z_N=o(n^{\frac{1}{2}})$ we have
\begin{equation}\label{lambda:ineq}
\prth{1+\frac{\lvert\lambda^*\rvert}{n}o(n^{\frac{1}{2}})}\text{sign}(\lambda^*)\sum_{i=1}^m\bar{Y}_i\geq \frac{\lvert\lambda^*\rvert}{n\cdot \overline{c}/\underline{c}}\prth{\sum_{i=1}^m\hat{\sigma}_i^2+o_p(1)}.
\end{equation}
Rearranging the terms gives
\begin{equation}
\frac{\lvert\lambda^*\rvert}{n}\prth{\frac{\underline{c}}{\overline{c}}\sum_{i=1}^m\hat{\sigma}_i^2+o_p(1)+o(n^{\frac{1}{2}})\sum_{i=1}^m\bar{Y}_i}\leq \abs{\sum_{i=1}^m\bar{Y}_i}.\label{lambdaineq}
\end{equation}
Note that $\hat{\sigma}_i^2\to \sigma_i^2:=Var(Y_i)$ almost surely by the strong law of large numbers, and
$\sum_{i=1}^m\bar{Y}_i=\sum_{i=1}^mO_p(n_i^{-\frac{1}{2}})=O_p(n^{-\frac{1}{2}})$. By the assumption $\sum_{i=1}^m\sigma_i^2>0$, \eqref{lambdaineq} implies
\begin{equation*}
\frac{\lvert\lambda^*\rvert}{n}\leq \frac{O_p(n^{-\frac{1}{2}})}{\frac{\underline{c}}{\overline{c}}\sum_{i=1}^m\sigma_i^2+o_p(1)}.
\end{equation*}
That is, $\frac{\abs{\lambda^*}}{n}=O_p(n^{-\frac{1}{2}})$.

\textbf{Step four:} We show the convergence of $\mu_i^*$ to the true mean $0$, i.e., $\mu_i^*=o_p(1)$. From \eqref{formula2wij} it follows that
\begin{eqnarray}
\nonumber\bar{Y}_i-\mu_i^*&=&\sum_{j=1}^{n_i}\prth{\frac{1}{n_i}-w_{i,j}^*}Y_{i,j}\\
&=&\frac{1}{n_i}\sum_{j=1}^{n_i}\frac{\frac{\lambda^*}{n_i}(Y_{i,j}-\mu_i^*)}{1+\frac{\lambda^*}{n_i}(Y_{i,j}-\mu_i^*)}Y_{i,j}.\label{u-ex}
\end{eqnarray}
We have shown in the Step three that $Z_N=o(n^{\frac{1}{2}})$, $|\mu_i^*|\leq Z_N$ and $\frac{|\lambda^*|}{n}=O_p(n^{-\frac{1}{2}})$. Hence $\max_{j}\big\lvert\frac{\lambda^*}{n_i}(Y_{i,j}-\mu_i^*)\big\rvert=O\big(\frac{\lvert 2\lambda^*\rvert}{n}Z_N\big)=O_p(n^{-\frac{1}{2}})o(n^{\frac{1}{2}})=o_p(1)$. Therefore
\begin{eqnarray*}
\abs{\bar{Y}_i-\mu_i^*}&\leq &\frac{1}{n_i}\sum_{j=1}^{n_i}\abs{\frac{\frac{\lambda^*}{n_i}(Y_{i,j}-\mu_i^*)}{1+\frac{\lambda^*}{n_i}(Y_{i,j}-\mu_i^*)}}\abs{Y_{i,j}}\\
&\leq&\frac{1}{n_i}\sum_{j=1}^{n_i}\abs{\frac{\max_{j}\big\lvert\frac{\lambda^*}{n_i}(Y_{i,j}-\mu_i^*)\big\rvert}{1-\max_{j}\big\lvert\frac{\lambda^*}{n_i}(Y_{i,j}-\mu_i^*)\big\rvert}}\abs{Y_{i,j}}\mathbf{1}\set{\max_{j}\Big\lvert\frac{\lambda^*}{n_i}(Y_{i,j}-\mu_i^*)\Big\rvert<1}\\
&&\hspace{1ex}+\infty\cdot\mathbf{1}\set{\max_{j}\Big\lvert\frac{\lambda^*}{n_i}(Y_{i,j}-\mu_i^*)\Big\rvert\geq 1}\text{ where }\infty\cdot 0=0\\
&=&\abs{\frac{\max_{j}\big\lvert\frac{\lambda^*}{n_i}(Y_{i,j}-\mu_i^*)\big\rvert}{1-\max_{j}\big\lvert\frac{\lambda^*}{n_i}(Y_{i,j}-\mu_i^*)\big\rvert}}\mathbf{1}\set{\max_{j}\abs{\frac{\lambda^*}{n_i}(Y_{i,j}-\mu_i^*)}<1}\frac{1}{n_i}\sum_{j=1}^{n_i}\abs{Y_{i,j}}+o_p(1)\\
&\leq&\abs{\frac{o_p(1)}{1-o_p(1)}}\cdot\frac{1}{n_i}\sum_{j=1}^{n_i}\abs{Y_{i,j}}+o_p(1)\\
&=&o_p(1).
\end{eqnarray*}
The $o_p(1)$ in the first equality is valid because $\max_{j}\big\lvert\frac{\lambda^*}{n_i}(Y_{i,j}-\mu_i^*)\big\lvert=o_p(1)$, and so by definition $\infty\cdot \mathbf{1}\set{\max_{j}\big\lvert\frac{\lambda^*}{n_i}(Y_{i,j}-\mu_i^*)\big\rvert\geq 1}$ has an arbitrarily small stochastic order and hence is $o_p(1)$. 

On the other hand, $\bar{Y}_i=o_p(1)$ by the law of large numbers. Hence $\mu_i^*=o_p(1)$. 

% In fact we will show that $\mu_i^*-\mathbb EX_i=O_p(n^{-\frac{1}{2}})$, which is more than what the proof needs, but worth pointing out (IS THERE A REASON?).
%
%To further strengthen it to $\mu_i^*-\mathbb  EX_i=O_p(n^{-\frac{1}{2}})$, we exploit the result $\mu_i^*-\mathbb  EX_i=o_p(1)$ to give a tighter upper bound of \eqref{u-ex}
%\begin{eqnarray*}
%\abs{\bar{X_i}-\mu_i^*}&\leq&\frac{\abs{\lambda_i^*}/n_i}{1-\min(o_p(1),1)}\cdot\frac{1}{n_i}\sum_{j=1}^{n_i}\abs{(X_{i,j}-\mu_i^*)X_{i,j}}\\
%&\leq&\frac{\abs{\lambda_i^*}/n_i}{1-\min(o_p(1),1)}\cdot\brac{\frac{1}{n_i}\sum_{j=1}^{n_i}\abs{X_{i,j}-\mathbb EX_i}\abs{X_{i,j}}+\abs{\mathbb  EX_i-\mu_i^*}\frac{1}{n_i}\sum_{j=1}^{n_i}\abs{X_{i,j}}}\\
%&=&O_p(n^{-\frac{1}{2}})\brac{O_p(1)+o_p(1)O_p(1)}\\
%&=&O_p(n^{-\frac{1}{2}}).
%\end{eqnarray*}
%Since $\bar{X_i}-\mathbb EX_i=O_p(n^{-\frac{1}{2}})$, we arrive at $\mu_i^*-\mathbb EX_i=O_p(n^{-\frac{1}{2}})$.

\textbf{Step five:} We derive formula \eqref{formulalambda} for the Lagrange multiplier $\lambda^*$ in terms of the data. Rewrite \eqref{forlambda} as
\begin{eqnarray}
\nonumber\bar{Y}_i-\mu_i^*&=&\frac{1}{n_i}\sum_{j=1}^{n_i}\brac{\frac{\lambda^*}{n_i}(Y_{i,j}-\mu_i^*)^2-\frac{(\frac{\lambda^*}{n_i})^2(Y_{i,j}-\mu_i^*)^3}{1+\frac{\lambda^*}{n_i}(Y_{i,j}-\mu_i^*)}},\\
&=&\frac{\lambda^*}{n_i}\brac{\frac{1}{n_i}\sum_{j=1}^{n_i}(Y_{i,j}-\mu_i^*)^2}-\abs{\frac{\lambda^*}{n_i}}^2\frac{1}{n_i}\sum_{j=1}^{n_i}\frac{(Y_{i,j}-\mu_i^*)^3}{1+\frac{\lambda^*}{n_i}(Y_{i,j}-\mu_i^*)}.\label{solvelambda}
\end{eqnarray}
The second term in \eqref{solvelambda} can be bounded as
\begin{eqnarray}
\notag&\leq&\abs{\frac{\lambda^*}{n_i}}^2\frac{1}{n_i}\sum_{j=1}^{n_i}\frac{\abs{Y_{i,j}-\mu_i^*}^3}{\big\lvert1+\frac{\lambda^*}{n_i}(Y_{i,j}-\mu_i^*)\big\rvert}\\
\notag&\leq& \abs{\frac{\lambda^*}{n_i}}^2\cdot \frac{2Z_N}{1-\max_{j}\big\lvert\frac{\lambda^*}{n_i}(Y_{i,j}-\mu_i^*)\big\rvert}\cdot \frac{1}{n_i}\sum_{j=1}^{n_i}\abs{Y_{i,j}-\mu_i^*}^2\mathbf{1}\set{\max_{j}\Big\lvert\frac{\lambda^*}{n_i}(Y_{i,j}-\mu_i^*)\Big\rvert< 1}\\
\notag&&\hspace{1ex}+\infty\cdot \mathbf{1}\set{\max_{j}\Big\lvert\frac{\lambda^*}{n_i}(Y_{i,j}-\mu_i^*)\Big\rvert\geq  1}\text{ where }\infty\cdot 0=0\\
&=&O_p(n^{-1})\frac{o(n^{1/2})}{1-o_p(1)}O_p(1)+o_p(n^{-\frac{1}{2}})\label{arbitrary order}\\
\notag&=&o_p(n^{-\frac{1}{2}})
\end{eqnarray}
where in passing from $\frac{1}{n_i}\sum_{j=1}^{n_i}\abs{Y_{i,j}-\mu_i^*}^2$ to $O_p(1)$ we use
\begin{equation}\label{sigmahat}
    \frac{1}{n_i}\sum_{j=1}^{n_i}\abs{Y_{i,j}-\mu_i^*}^2=\frac{1}{n_i}\sum_{j=1}^{n_i}Y_{i,j}^2-2\bar{Y}_i\mu_i^*+{\mu_i^*}^2=\sigma_i^2+O_p\big(n^{-\frac{1}{2}}\big)o_p(1)+o_p(1)=\sigma_i^2+o_p(1)
\end{equation}
and the $o_p(n^{-1/2})$ term in \eqref{arbitrary order} is valid because $\max_{j}\big\lvert\frac{\lambda^*}{n_i}(Y_{i,j}-\mu_i^*)\big\lvert=o_p(1)$, and so $\infty\cdot \mathbf{1}\set{\max_{j}\big\lvert\frac{\lambda^*}{n_i}(Y_{i,j}-\mu_i^*)\big\rvert\geq 1}$ has an arbitrarily small stochastic order by definition and hence in particular is $o_p(n^{-1/2})$. \eqref{sigmahat} also implies that the first term in \eqref{solvelambda} is $\frac{\lambda^*}{n_i}(\sigma_i^2+o_p(1))$. 
% \begin{eqnarray}
% \nonumber\frac{1}{n_i}\sum_{j=1}^{n_i}(Y_{i,j}-\mu_i^*)^2&=&\frac{1}{n_i}\sum_{j=1}^{n_i}(X_{i,j}-\mathbb EX_i)^2+2(\bar{X_i}-\mathbb EX_i)(\mathbb EX_i-\mu_i^*)+(\mathbb EX_i-\mu_i^*)^2\\
% \nonumber&=&\hat{\sigma}_i^2+o_p(1)\\
% &=&\sigma_i^2+o_p(1)
% \end{eqnarray}
Hence \eqref{solvelambda} can be written as
\begin{equation}
\bar{Y}_i-\mu_i^*=\frac{\lambda^*}{n_i}\sigma_i^2+o_p(n^{-\frac{1}{2}}).\label{solvelambda2}
\end{equation}
Summing \eqref{solvelambda2} over $i=1,\ldots,m$ and using \eqref{zerocon} give
\begin{equation}\label{lambda:eq}
\sum_{i=1}^m\bar{Y}_i=\lambda^*\sum_{i=1}^m\frac{\sigma_i^2}{n_i}+o_p(n^{-\frac{1}{2}}).
\end{equation}
Therefore the expression for $\lambda^*$ is
\begin{equation}\label{formulalambda}
\lambda^*=\frac{\sum_{i=1}^m\bar{Y}_i+o_p(n^{-\frac{1}{2}})}{\sum_{i=1}^m\frac{\sigma_i^2}{n_i}}.
\end{equation}

\textbf{Step six:} We substitute $\mu_i^*=o_p(1)$ and \eqref{formulalambda} into \eqref{formulawij} to derive a formula for $w_{i,j}^*$, and from there we analyze the Taylor expansion of $-2\log R(0)$ to conclude the desired result. Each
\begin{equation*}
-\log (n_iw_{i,j}^*)=\log(1+\frac{\lambda^*}{n_i}(Y_{i,j}-\mu_i^*))=\frac{\lambda^*}{n_i}(Y_{i,j}-\mu_i^*)-\frac{{\lambda^*}^2}{2n_i^2}(Y_{i,j}-\mu_i^*)^2+\eta_{i,j},
\end{equation*}
where $\eta_{i,j}=\frac{1}{3(1+\theta_{i,j}\frac{\lambda^*}{n_i}(Y_{i,j}-\mu_i^*))^3}\prth{\frac{\lambda^*}{n_i}(Y_{i,j}-\mu_i^*)}^3$ for some $\theta_{i,j}\in (0,1)$, so the log profile likelihood ratio can be expressed as
\begin{eqnarray}
\nonumber-2\log R(0)&=&2\sum_{i=1}^m\sum_{j=1}^{n_i}\log(1+\frac{\lambda^*}{n_i}(Y_{i,j}-\mu_i^*))\\
\nonumber&=&2\sum_{i=1}^m\sum_{j=1}^{n_i}\prth{\frac{\lambda^*}{n_i}(Y_{i,j}-\mu_i^*)-\frac{{\lambda^*}^2}{2n_i^2}(Y_{i,j}-\mu_i^*)^2+\eta_{i,j}}\\
&=&2\sum_{i=1}^m\lambda^*(\bar{Y}_i-\mu_i^*)-\sum_{i=1}^m\frac{{\lambda^*}^2}{n_i}\cdot\frac{1}{n_i}\sum_{j=1}^{n_i}(Y_{i,j}-\mu_i^*)^2+\sum_{i=1}^m\sum_{j=1}^{n_i}2\eta_{i,j}\label{secondterm1}\\
&=&2\lambda^*\sum_{i=1}^m\bar{Y}_i-\sum_{i=1}^m\frac{{\lambda^*}^2}{n_i}(\sigma_i^2+o_p(1))+\sum_{i=1}^m\sum_{j=1}^{n_i}2\eta_{i,j}\label{secondterm2}
\end{eqnarray}
The equality between \eqref{secondterm1} and \eqref{secondterm2} follows from \eqref{zerocon} and \eqref{sigmahat}. To bound the last term in \eqref{secondterm2}
\begin{eqnarray*}
\abs{\sum_{i,j}2\eta_{ij}}&\leq&\frac{2}{3(1-\max_{i,j}\big\lvert\frac{\lambda^*}{n_i}(Y_{i,j}-\mu_i^*)\big\rvert)^3}\abs{\frac{\lambda^*}{\min_in_i}}^3\sum_{i,j}\abs{Y_{i,j}-\mu_i^*}^3\cdot \mathbf{1}\set{\max_{i,j}\big\lvert\frac{\lambda^*}{n_i}(Y_{i,j}-\mu_i^*)\big\rvert<1}\\
&&\hspace{1ex}+\infty\cdot \mathbf{1}\set{\max_{i,j}\big\lvert\frac{\lambda^*}{n_i}(Y_{i,j}-\mu_i^*)\big\rvert\geq 1}\\
&=&\frac{2}{3(1-o_p(1))^3}O_p\big(n^{-\frac{3}{2}}\big) \sum_{i=1}^m2n_iZ_N\sum_{j=1}^{n_i}\frac{1}{n_i}\abs{Y_{i,j}-\mu_i^*}^2+o_p(1)\\
&=&O_p\big(n^{-\frac{3}{2}}\big)\sum_{i=1}^m2n_iZ_NO_p(1)+o_p(1)\\
&=&O_p(n^{-\frac{3}{2}})no(n^{\frac{1}{2}})O_p(1)+o_p(1)\\
&=&o_p(1).
\end{eqnarray*}
Hence using the above bound and \eqref{formulalambda}, the log profile likelihood ratio \eqref{secondterm2} becomes
\begin{eqnarray}
\nonumber-2\log R(0)&=&2\lambda^*\sum_{i=1}^m\bar{Y}_i-{\lambda^*}^2\sum_{i=1}^m\frac{\sigma_i^2}{n_i}+o_p(1)\\
&=&\frac{\prth{\sum_{i=1}^m\bar{Y}_i}^2}{\sum_{i=1}^m\frac{\sigma_i^2}{n_i}}+o_p(1).\label{leading}
\end{eqnarray}
To resolve the issue caused by the possible absence of Slater's condition, note that the above result holds only in the event of $\mathcal S$, namely
\begin{equation*}
-2\log R(0)\cdot \mathbf{1}\{\mathcal S\}=\Big(\frac{\prth{\sum_{i=1}^m\bar{Y}_i}^2}{\sum_{i=1}^m\frac{\sigma_i^2}{n_i}}+o_p(1)\Big)\cdot \mathbf{1}\{\mathcal S\}
\end{equation*}
which implies
\begin{eqnarray}
\notag-2\log R(0)&=&\Big(\frac{\prth{\sum_{i=1}^m\bar{Y}_i}^2}{\sum_{i=1}^m\frac{\sigma_i^2}{n_i}}+o_p(1)\Big)\cdot \mathbf{1}\{\mathcal S\}-2\log R(0)\cdot \mathbf{1}\{\mathcal S^c\}\\
\notag&=&\frac{\prth{\sum_{i=1}^m\bar{Y}_i}^2}{\sum_{i=1}^m\frac{\sigma_i^2}{n_i}}+o_p(1)-\Big(\frac{\prth{\sum_{i=1}^m\bar{Y}_i}^2}{\sum_{i=1}^m\frac{\sigma_i^2}{n_i}}+o_p(1)\Big)\cdot \mathbf{1}\{\mathcal S^c\}-2\log R(0)\cdot \mathbf{1}\{\mathcal S^c\}\\
&=&\frac{\prth{\sum_{i=1}^m\bar{Y}_i}^2}{\sum_{i=1}^m\frac{\sigma_i^2}{n_i}}+o_p(1).\label{remove_slater}
\end{eqnarray}
\eqref{remove_slater} brings us back to \eqref{leading}. So by Slutsky's theorem it remains to show that the leading term in \eqref{leading} $\Rightarrow\mathcal X^2_1$. The leading term can be written as
\begin{equation}\label{leading2}
\brac{\sum_{i=1}^m\sum_{j=1}^{n_i}\frac{Y_{i,j}}{n_i\sqrt{\sum_{i=1}^m\frac{\sigma_i^2}{n_i}}}}^2.
\end{equation}
By the continuous mapping theorem it suffices to show that the sum in \eqref{leading2} $\Rightarrow\mathcal N(0,1)$. We check the Lindeberg condition for the triangular array
\begin{equation*}
(W_{N,1},\ldots,W_{N,N}):=\big(Y_{1,1},\ldots,Y_{1,n_1},\ldots,Y_{m,1},\ldots,Y_{m,n_m}\big)\Big/\Big(n_i\sqrt{\sum_{i=1}^m\frac{\sigma_i^2}{n_i}}\Big)
\end{equation*}
where $N=\sum_{i=1}^mn_i$. The independence and mean zero conditions are obviously met, and
\begin{equation*}
\sum_{k=1}^{N}\mathbb E W_{N,k}^2=\sum_{i=1}^{m}\sum_{j=1}^{n_i}\mathbb E\brac{\frac{Y_{i,j}^2}{n_i^2\sum_{i=1}^m\frac{\sigma_i^2}{n_i}}}=\sum_{i=1}^{m}\sum_{j=1}^{n_i}\frac{\sigma_i^2}{\sum_{i=1}^mn_i\sigma_i^2}=1.
\end{equation*}
For any $\epsilon>0$
\begin{eqnarray*}
&&\sum_{i=1}^m\sum_{j=1}^{n_i}\mathbb E\brac{\prth{\frac{Y_{i,j}}{{n_i\sqrt{\sum_{i=1}^m\frac{\sigma_i^2}{n_i}}}}}^2\cdot\mathbf{1}\set{\abs{\frac{Y_{i,j}}{{n_i\sqrt{\sum_{i=1}^m\frac{\sigma_i^2}{n_i}}}}}>\epsilon}}\\
&=&\sum_{i=1}^mn_i\mathbb E\brac{\prth{\frac{Y_{i,1}}{{n_i\sqrt{\sum_{i=1}^m\frac{\sigma_i^2}{n_i}}}}}^2\cdot\mathbf{1}\set{\abs{\frac{Y_{i,1}}{{n_i\sqrt{\sum_{i=1}^m\frac{\sigma_i^2}{n_i}}}}}>\epsilon}} \\
&\leq&\sum_{i=1}^mC_1\mathbb E\brac{Y_{i,1}^2\cdot\mathbf{1}\set{\abs{Y_{i,1}}>\epsilon C_2\sqrt{n}}}\text{ for some constants }C_1,C_2\\
&\to&\,0\text{\ \ by the dominated convergence theorem}.
\end{eqnarray*}
Therefore the Lindeberg condition holds for $W_{N,k}$. By the Lindeberg-Feller theorem (e.g., Theorem 3.4.5 in \citealt{durrett2010probability}), the sum in \eqref{leading2} $\Rightarrow\mathcal N(0,1)$ hence \eqref{leading2} itself $\Rightarrow\mathcal X_1^2$.\hfill\Halmos
\endproof

\section{Proofs of Results in Section \ref{sec:EL-based CI}}\label{proof:section 4.4}
\proof{Proof of Theorem \ref{opt2gua}.}From Theorem \ref{elt} we know $P(-2\log R(\mu_0)\leq \mathcal X^2_{1,1-\alpha})\to 1-\alpha$ as $n\to\infty$. That is, the set $\{\mu\in \R\vert -2\log R(\mu)\leq \mathcal X^2_{1,1-\alpha}\}$ contains the true value $\mu_0$ with probability $1-\alpha$ asymptotically. Note that this set can be identified as
\begin{equation*}
\mathcal V=\set{\sum_{i=1}^{m}\sum_{j=1}^{n_i}Y_{i,j}w_{i,j}\bigg\vert -2\sum_{i=1}^{m}\sum_{j=1}^{n_i}\log(n_iw_{i,j}) \leq \mathcal X_{1,1-\alpha}^2,\ \sum_{j=1}^{n_i}w_{i,j}=1\text{\ for all\ }i,w_{i,j}\geq 0\text{\ for all\ }i,j}.
\end{equation*}
It is obvious that $\underline{\mu}/\overline{\mu}=\min/\max\{\mu: \mu\in\mathcal V\}$, and they are attained because the feasible set $\mathcal U_{\alpha}$ is compact and the objective is linear hence continuous in $w_{i,j}$'s. So if the set $\mathcal V$ is convex, then $\mathcal V=[\underline{\mu},\overline{\mu}]$ which concludes the theorem. To show convexity, it is enough to notice that $\mathcal U_{\alpha}$ is convex, and the objective is linear in $w_{i,j}$. \hfill\Halmos
\endproof

\proof{Proof of Proposition \ref{connection to normal CI}.}We need the following corollary of Theorem \ref{elt}:
\begin{corollary}\label{deltabounds}
Let $\bar{Y}_i=\sum_{j=1}^{n_i}Y_{i,j}/n_i$ be the sample mean of the $i$-th sample, $\sigma_i^2=\mathrm{Var}(Y_i)$ be the true variance, and $z$ be a fixed constant. Under the same conditions of Theorem \ref{elt}, $-2\log R(\sum_{i=1}^m\bar{Y}_i+z\sqrt{\sum_{i=1}^m\sigma_i^2/n_i})\to z^2$ in probability as $n\to \infty$.
\end{corollary}
\proof{Proof of Corollary \ref{deltabounds}.}The proof closely follows that of Theorem \ref{elt}, and we only point out how each step should be modified in order to prove this corollary. Assuming $\mathrm{Var}(Y_i)>0$ and $\mathbb EY_i=0$ is still without loss of generality because, with $I=\{i:\mathrm{Var}(Y_i)>0\}$,
\begin{eqnarray*}
&&R\Big(\sum_{i=1}^m\bar{Y}_i+z\sqrt{\sum_{i=1}^m\frac{\sigma_i^2}{n_i}}\Big)\\
&=&\max\set{\prod_{i=1}^m\prod_{j=1}^{n_i}n_iw_{i,j}\bigg\vert \sum_{i=1}^m\sum_{j=1}^{n_i}Y_{i,j}w_{i,j}=\sum_{i=1}^m\bar{Y}_i+z\sqrt{\sum_{i=1}^m\frac{\sigma_i^2}{n_i}},\ \sum_{j=1}^{n_i}w_{i,j}=1\text{\ for all\ }i,\ w_{i,j}\geq 0\text{ for all }i,j}\\
&=&\max\set{\prod_{i=1}^m\prod_{j=1}^{n_i}n_iw_{i,j}\bigg\vert \sum_{i\in I}\sum_{j=1}^{n_i}Y_{i,j}w_{i,j}=\sum_{i\in I}\bar{Y}_i+z\sqrt{\sum_{i\in I}\frac{\sigma_i^2}{n_i}},\ \sum_{j=1}^{n_i}w_{i,j}=1\text{\ for all\ }i,\ w_{i,j}\geq 0\text{ for all }i,j}\\
&=&\max\set{\prod_{i\in I}\prod_{j=1}^{n_i}n_iw_{i,j}\bigg\vert \sum_{i\in I}\sum_{j=1}^{n_i}Y_{i,j}w_{i,j}=\sum_{i\in I}\bar{Y}_i+z\sqrt{\sum_{i\in I}\frac{\sigma_i^2}{n_i}},\ \sum_{j=1}^{n_i}w_{i,j}=1\text{\ for\ }i\in I,\ w_{i,j}\geq 0\text{ for all }i\in I,j}\\
&=&\max\Bigg\{\prod_{i\in I}\prod_{j=1}^{n_i}n_iw_{i,j}\bigg\vert \sum_{i\in I}\sum_{j=1}^{n_i}(Y_{i,j}-\mathbb EY_i)w_{i,j}=\sum_{i\in I}(\bar{Y}_i-\mathbb EY_i)+z\sqrt{\sum_{i\in I}\frac{\sigma_i^2}{n_i}},\\
&&\hspace{21ex}\sum_{j=1}^{n_i}w_{i,j}=1\text{\ for\ }i\in I,\ w_{i,j}\geq 0\text{ for all }i\in I,j\Bigg\}
\end{eqnarray*}
and the limit distribution, i.e., the point mass at $z^2$, does not depend on the number of distributions $m$. Next we consider the following counterpart of \eqref{keyopt}
\begin{equation}\label{keyopt_delta}
\begin{aligned}
& \underset{\mathbf w_1,\ldots,\mathbf w_m,\bm{\mu}}{\text{min}}& & -\sum_{i=1}^m\sum_{j=1}^{n_i}\log(n_iw_{i,j}) \\
& \text{subject to}
& & \sum_{j=1}^{n_i}Y_{i,j}w_{i,j}=\mu_i,\ i=1,\ldots,m\\
&&&\sum_{j=1}^{n_i}w_{i,j}=1,\ i=1,\ldots,m\\
&&&\sum_{i=1}^m\mu_i=\sum_{i=1}^m\bar{Y}_i+z\sqrt{\sum_{i=1}^m\frac{\sigma_i^2}{n_i}}.
\end{aligned}
\end{equation}

\textbf{Step one:} We show Slater's condition holds for \eqref{keyopt_delta} with a probability tending to one. Instead of $\tilde{\mathcal S}_i$, consider the event indexed by $i$
\begin{equation}\label{slater_delta}
    \min_{j=1,\ldots,n_i}Y_{i,j}<\bar{Y}_i+\frac{z}{m}\sqrt{\sum_{i=1}^m\frac{\sigma_i^2}{n_i}}<\max_{j=1,\ldots,n_i}Y_{i,j}.
\end{equation}
We need to show the probability that \eqref{slater_delta} happens goes to one. Note that $\bar{Y}_i+z/m\cdot\sqrt{\sum_{i=1}^m\sigma_i^2/n_i}=o_p(1)$, and for a small enough $\epsilon>0$ it holds $P(Y_i\geq -\epsilon)<1,P(Y_i\leq \epsilon)<1$. Hence
\begin{eqnarray*}
P\big(\min_{j=1,\ldots,n_i}Y_{i,j}\geq\bar{Y}_i+\frac{z}{m}\sqrt{\sum_{i=1}^m\frac{\sigma_i^2}{n_i}}\big)&\leq&P\big(\min_{j=1,\ldots,n_i}Y_{i,j}\geq -\epsilon\big)+P\big(\bar{Y}_i+\frac{z}{m}\sqrt{\sum_{i=1}^m\frac{\sigma_i^2}{n_i}}< -\epsilon\big)\\
&=&(P(Y_i\geq -\epsilon))^{n_i}+P(o_p(1)<-\epsilon)\to 0.
\end{eqnarray*}
This justifies the first inequality of \eqref{slater_delta}, and the second inequality can be treated in the same way. Applying the union bound shows that the probability of \eqref{slater_delta} approaches one. The rest of this step remains the same.

\textbf{Step two:} The only change is that one of the KKT conditions, \eqref{zerocon}, is replaced by
\begin{equation*}
    \sum_{i=1}^m\mu_i^*=\sum_{i=1}^m\bar{Y}_i+z\sqrt{\sum_{i=1}^m\frac{\sigma_i^2}{n_i}}.
\end{equation*}

\textbf{Step three:} \eqref{lambda:ineq} is replaced by
\begin{equation*}
-\prth{1+\frac{\lvert\lambda^*\rvert}{n}o(n^{\frac{1}{2}})}\text{sign}(\lambda^*)\cdot z\sqrt{\sum_{i=1}^m\frac{\sigma_i^2}{n_i}}\geq \frac{\lvert\lambda^*\rvert}{n\cdot \overline{c}/\underline{c}}\prth{\sum_{i=1}^m\hat{\sigma}_i^2+o_p(1)}
\end{equation*}
and \eqref{lambdaineq} becomes
\begin{equation*}
\frac{\lvert\lambda^*\rvert}{n}\prth{\frac{\underline{c}}{\overline{c}}\sum_{i=1}^m\hat{\sigma}_i^2+o_p(1)+o(n^{\frac{1}{2}})\cdot z\sqrt{\sum_{i=1}^m\frac{\sigma_i^2}{n_i}}}\leq \abs{z}\sqrt{\sum_{i=1}^m\frac{\sigma_i^2}{n_i}}.
\end{equation*}
The final bound $\lambda^*=O_p(n^{1/2})$ still holds by observing that $z\sqrt{\sum_{i=1}^m\sigma_i^2/n_i}=O(n^{-1/2})$ just like $\sum_{i=1}^m\bar{Y}_i$.

\textbf{Step four:} No changes needed.

\textbf{Step five:} \eqref{lambda:eq} needs to be replaced by
\begin{equation*}
-z\sqrt{\sum_{i=1}^m\frac{\sigma_i^2}{n_i}}=\lambda^*\sum_{i=1}^m\frac{\sigma_i^2}{n_i}+o_p(n^{-\frac{1}{2}}).
\end{equation*}
Hence \eqref{formulalambda} becomes
\begin{equation*}
\lambda^*=\frac{-z+o_p(1)}{\sqrt{\sum_{i=1}^m\frac{\sigma_i^2}{n_i}}}.
\end{equation*}

\textbf{Step six:} \eqref{secondterm2} and \eqref{leading} are replaced by
\begin{eqnarray*}
-2\log R\big(\sum_{i=1}^m\bar{Y}_i+z\sqrt{\sum_{i=1}^m\frac{\sigma_i^2}{n_i}}\big)
&=&-2\lambda^*z\sqrt{\sum_{i=1}^m\frac{\sigma_i^2}{n_i}}-\sum_{i=1}^m\frac{{\lambda^*}^2}{n_i}(\sigma_i^2+o_p(1))+\sum_{i=1}^m\sum_{j=1}^{n_i}2\eta_{i,j}\\
&=&z^2+o_p(1)
\end{eqnarray*}
and the desired conclusion follows.\hfill \Halmos
\endproof

% Define the profile likelihood ratio
% \begin{equation*}
% R_G(\mu)=\max\set{\prod_{i=1}^m\prod_{j=1}^{n_i}n_iw_{i,j}\bigg\vert \sum_{i=1}^m\sum_{j=1}^{n_i}w_{i,j}G_i(X_{i,j})=\mu,\ \sum_{j=1}^{n_i}w_{i,j}=1\text{\ for all\ }i,\ w_{i,j}\geq 0\text{ for all }i,j}.
% \end{equation*}
Now we are ready to prove Proposition \ref{connection to normal CI}. Recall the definition of profile likelihood ratio $R(\mu)$ in \eqref{profileratio}. Since $z_{1-\alpha/2}^2=\mathcal X_{1,1-\alpha}^2$, Corollary \ref{deltabounds} entails that for any fixed small $\epsilon>0$
\begin{align}
&P\Big(-2\log R\Big(\sum_{i=1}^m\bar{Y}_i-(z_{1-\alpha/2}-\epsilon)\sqrt{\sum_{i=1}^m\frac{\sigma_i^2}{n_i}}\Big)<\mathcal X_{1,1-\alpha}^2\Big)\to 1,\label{event1}\\
&P\Big(-2\log R\Big(\sum_{i=1}^m\bar{Y}_i-(z_{1-\alpha/2}+\epsilon)\sqrt{\sum_{i=1}^m\frac{\sigma_i^2}{n_i}}\Big)>\mathcal X_{1,1-\alpha}^2\Big)\to 1.\label{event2}
\end{align}
In the proof of Theorem \ref{opt2gua} it is shown that $\{\mu\in \R\vert -2\log R(\mu)\leq \mathcal X^2_{1,1-\alpha}\}=[\underline{\mu},\overline{\mu}]$. Therefore conditioned on the event in \eqref{event1} we must have $\sum_{i=1}^m\bar{Y}_i-(z_{1-\alpha/2}-\epsilon)\sqrt{\sum_{i=1}^m\sigma_i^2/n_i}\in [\underline{\mu},\overline{\mu}]$. Conditioned on the event in \eqref{event2} we have $\sum_{i=1}^m\bar{Y}_i-(z_{1-\alpha/2}+\epsilon)\sqrt{\sum_{i=1}^m\sigma_i^2/n_i}\notin [\underline{\mu},\overline{\mu}]$. Moreover, since the sum of sample means $\sum_{i=1}^m\bar{Y}_i\in[\underline{\mu},\overline{\mu}]$ almost surely and $\sum_{i=1}^m\bar{Y}_i-(z_{1-\alpha/2}+\epsilon)\sqrt{\sum_{i=1}^m\sigma_i^2/n_i}<\sum_{i=1}^m\bar{Y}_i$, it must be the case that $\sum_{i=1}^m\bar{Y}_i-(z_{1-\alpha/2}+\epsilon)\sqrt{\sum_{i=1}^m\sigma_i^2/n_i}<\underline{\mu}$. Applying the union bound we get
\begin{equation*}
P\Big(\Big\lvert\underline{\mu}-\Big(\sum_{i=1}^m\bar{Y}_i-z_{1-\alpha/2}\sqrt{\sum_{i=1}^m\frac{\sigma_i^2}{n_i}}\Big)\Big\rvert\leq \epsilon\sqrt{\sum_{i=1}^m\frac{\sigma_i^2}{n_i}}\Big)\to 1.
\end{equation*}
Sending $\epsilon$ to $0$ gives the desired conclusion for $\underline{\mu}$. The proof for $\overline{\mu}$ is similar.\hfill\Halmos
\endproof
\proof{Proof of Corollary \ref{CI: linear approximation truth}.}If we can show that $\mathrm{Var}(G_i(X_i))<\infty$ for all $i$, then this is a direct consequence of Theorem \ref{opt2gua} and Proposition \ref{connection to normal CI} with $Y_{i,j}=\frac{Z^*}{m}+G_i(X_{i,j})$ and the fact that $\mathbb E_{P_i}[G_i(X_{i,j})]=0$. Since Assumption \ref{8th moment} implies $\mathbb E[h^2(\mathbf X_1,\ldots,\mathbf X_m)]<\infty$, by Jensen's inequality any conditional expectation of $h(\mathbf X_1,\ldots,\mathbf X_m)$ also has a finite second moment. Note that $G_i(X_i)$ is the sum of $T_i$ conditional expectations of $h(\mathbf X_1,\ldots,\mathbf X_m)$. Therefore it has a finite second moment, hence a finite variance, by the Minkowski inequality.\hfill\Halmos
\endproof
\proof{Proof of Theorem \ref{ideal CI}.}We have
\begin{eqnarray*}
\mathscr L&=&\inf_{(\mathbf w_1,\ldots,\mathbf w_m)\in\mathcal U_{\alpha}}Z(\mathbf w_1,\ldots,\mathbf w_m)\\
&=&\inf_{(\mathbf w_1,\ldots,\mathbf w_m)\in\mathcal U_{\alpha}}\big[Z_L(\mathbf w_1,\ldots,\mathbf w_m)+\big(Z(\mathbf w_1,\ldots,\mathbf w_m)-Z_L(\mathbf w_1,\ldots,\mathbf w_m)\big)\big]\\
&\geq &\inf_{(\mathbf w_1,\ldots,\mathbf w_m)\in\mathcal U_{\alpha}}Z_L(\mathbf w_1,\ldots,\mathbf w_m)+\inf_{(\mathbf w_1,\ldots,\mathbf w_m)\in\mathcal U_{\alpha}}\big(Z(\mathbf w_1,\ldots,\mathbf w_m)-Z_L(\mathbf w_1,\ldots,\mathbf w_m)\big)\\
&\geq&\mathscr L_L-\sup_{(\mathbf w_1,\ldots,\mathbf w_m)\in\mathcal U_{\alpha}}\big\lvert Z(\mathbf w_1,\ldots,\mathbf w_m)-Z_L(\mathbf w_1,\ldots,\mathbf w_m)\big\rvert.
\end{eqnarray*}
Similarly it can be shown that $\mathscr L_L\geq \mathscr L-\sup_{(\mathbf w_1,\ldots,\mathbf w_m)\in\mathcal U_{\alpha}}\big\lvert Z(\mathbf w_1,\ldots,\mathbf w_m)-Z_L(\mathbf w_1,\ldots,\mathbf w_m)\big\rvert$. Therefore
\begin{equation}\label{error_lowerCB}
    \abs{\mathscr L-\mathscr L_L}\leq \sup_{(\mathbf w_1,\ldots,\mathbf w_m)\in\mathcal U_{\alpha}}\big\lvert Z(\mathbf w_1,\ldots,\mathbf w_m)-Z_L(\mathbf w_1,\ldots,\mathbf w_m)\big\rvert.
\end{equation}
By the error bound \eqref{error:truth approximation} in Proposition \ref{linearization error}, $\sup_{(\mathbf w_1,\ldots,\mathbf w_m)\in\mathcal U_{\alpha}}\big\lvert Z(\mathbf w_1,\ldots,\mathbf w_m)-Z_L(\mathbf w_1,\ldots,\mathbf w_m)\big\rvert=O_p(1/n)$ hence $\abs{\mathscr L-\mathscr L_L}=O_p(1/n)=o_p(1/\sqrt{n})$. Analogously $\abs{\mathscr U-\mathscr U_L}=o_p(1/\sqrt{n})$. In particular, the representation \eqref{equivalence to normal CI} holds for $\mathscr L,\mathscr U$ as well, i.e.
\begin{equation}\label{equivalence to normal CI:ideal CI}
\begin{aligned}
    \mathscr L&=Z^*+\sum_{i=1}^m\bar{G}_i- z_{1-\alpha/2}\sigma_I+o_p\big(\frac{1}{\sqrt{n}}\big)\\
    \mathscr U&=Z^*+\sum_{i=1}^m\bar{G}_i+ z_{1-\alpha/2}\sigma_I+o_p\big(\frac{1}{\sqrt{n}}\big).
\end{aligned}
\end{equation}
Now we show that \eqref{equivalence to normal CI:ideal CI} guarantees the asymptotic exactness of $[\mathscr L,\mathscr U]$ as a CI for $Z^*$. For convenience, assume $\mathrm{Var}(G_i(X_i))>0$ for all $i$ without loss of generality. The standard central limit theorem entails that $\frac{\bar{G}_i}{\sqrt{\mathrm{Var}(G_i(X_i))/n_i}}\Rightarrow \mathcal N(0,1)$. Since the data across different input models are independent, we have the joint convergence
\begin{equation*}
    \Big(\frac{\bar{G}_1}{\sqrt{\mathrm{Var}(G_1(X_1))/n_1}},\ldots,\frac{\bar{G}_m}{\sqrt{\mathrm{Var}(G_m(X_m))/n_m}}\Big)\Rightarrow \mathcal N(\mathbf 0,\mathbf I_m).
\end{equation*}
To proceed, we need the following result:
\begin{lemma}[Uniform convergence of measures, Theorem $4.2$ in \citealt{rao1962relations}]\label{unif_weak}
Let $\mu^*,\{\mu_n\}_{n=1}^{\infty}$ be probability measures on $\R^d$. If $\mu^*$ is absolutely continuous with respect to the Lebesgue measure on $\R^d$, then $\mu_n\Rightarrow\mu^*$ if and only if
\begin{equation*}
\lim_{n\to \infty}\sup_{C\in \mathscr C}\lvert \mu_n(C)-\mu^*(C)\rvert= 0,
\end{equation*}
where $\mathscr C$ denotes the set of all measurable convex sets.
\end{lemma}
Let $(W_1,\ldots,W_m)$ be an $m$ dimensional standard normal vector, then $\sum_{i=1}^m\frac{1}{\sigma_I}\sqrt{\mathrm{Var}(G_i(X_i))/n_i}W_i$ follows $\mathcal N(0,1)$. Hence
\begin{eqnarray}
    \notag&&\abs{P\Big(\frac{\sum_{i=1}^m\bar{G}_i}{\sigma_I}\leq z\Big)-\Phi(z)}\\
   \notag &= &\abs{P\Big(\sum_{i=1}^m\frac{\sqrt{\mathrm{Var}(G_i(X_i))/n_i}}{\sigma_I}\cdot \frac{\bar{G}_i}{\sqrt{\mathrm{Var}(G_i(X_i))/n_i}}\leq z\Big)-P\Big(\sum_{i=1}^m\frac{\sqrt{\mathrm{Var}(G_i(X_i))/n_i}}{\sigma_I}W_i\leq z\Big)}\\
   \notag &=&\abs{P\Big(\Big(\frac{\bar{G}_1}{\sqrt{\mathrm{Var}(G_1(X_1))/n_1}},\ldots,\frac{\bar{G}_m}{\sqrt{\mathrm{Var}(G_m(X_m))/n_m}}\Big)\in \tilde{C}\Big)-P\Big((W_1,\ldots,W_m)\in \tilde{C}\Big)}\\
    &&\text{\ \ where }\tilde{C}=\Big\{(x_1,\ldots,x_m)\in \R^m\Big\vert \sum_{i=1}^m\frac{\sqrt{\mathrm{Var}(G_i(X_i))/n_i}}{\sigma_I}x_i\leq z\Big\}.\label{ex:use uniform convergence}
\end{eqnarray}
Since the set $\tilde{C}$ is a half-space and in particular a convex set, Lemma \ref{unif_weak} implies
\begin{eqnarray*}
    &&\abs{P\Big(\frac{\sum_{i=1}^m\bar{G}_i}{\sigma_I}\leq z\Big)-\Phi(z)}\\
    &\leq&\sup_{C\in\mathscr C}\abs{P\Big(\Big(\frac{\bar{G}_1}{\sqrt{\mathrm{Var}(G_1(X_1))/n_1}},\ldots,\frac{\bar{G}_m}{\sqrt{\mathrm{Var}(G_m(X_m))/n_m}}\Big)\in C\Big)-P\Big((W_1,\ldots,W_m)\in C\Big)}\to 0.
\end{eqnarray*}
Therefore
\begin{equation}\label{influnece_normality}
    \frac{\sum_{i=1}^m\bar{G}_i}{\sigma_I}\Rightarrow \mathcal N(0,1).
\end{equation}
Now \eqref{equivalence to normal CI:ideal CI} forces
\begin{eqnarray*}
P(\mathscr L\leq Z^*)&=&P(\sum_{i=1}^m\bar{G}_i+o_p\big(\frac{1}{\sqrt{n}}\big)\leq z_{1-\alpha/2}\sigma_I)\\
&=&P\Big(\frac{\sum_{i=1}^m\bar{G}_i}{\sigma_I}+o_p(1)\leq z_{1-\alpha/2}\Big)\\
&\to& P(\mathcal N(0,1)\leq z_{1-\alpha/2})=1-\frac{\alpha}{2}\text{\ \ by Slutsky's theorem}.
\end{eqnarray*}
Similarly we have $P(\mathscr U\geq Z^*)\to 1-\alpha/2$. Moreover, $\mathscr U-\mathscr L=2z_{1-\alpha/2}\sigma_I+o_p(1/\sqrt{n})$ hence
$$P(\mathscr U<Z^*<\mathscr L)\leq P(\mathscr U<\mathscr L)=P(2z_{1-\alpha/2}\sigma_I+o_p(1/\sqrt{n})<0)=P(2z_{1-\alpha/2}<o_p(1))\to 0.$$
Combining the limit probabilities gives
\begin{eqnarray*}
P(\mathscr L\leq Z^*\leq \mathscr U)&=&P(\mathscr L\leq Z^*)+P(\mathscr U\geq Z^*)-P(\mathscr L\leq Z^* \text{ or }\mathscr U\geq Z^*)\\
&=&P(\mathscr L\leq Z^*)+P(\mathscr U\geq Z^*)-1+P(\mathscr U< Z^* <\mathscr L)\\
&\to&1-\frac{\alpha}{2}+1-\frac{\alpha}{2}-1+0=1-\alpha.
\end{eqnarray*}
This completes the proof.\hfill\Halmos
\endproof

\section{Proofs of Results in Section \ref{sec:nonconvexity}}\label{proof:section 4.5}
\proof{Proof of Proposition \ref{empirical_error}.}It suffices to show the first part $\mathbb E\big[\sup_{(\mathbf w_1,\ldots,\mathbf w_m)\in \mathcal U_{\alpha}}\big\lvert \widehat{Z_L}(\mathbf w_1,\ldots,\mathbf w_m)-\dbwidehat{Z_L}(\mathbf w_1,\ldots,\mathbf w_m)\big\rvert^2\big]=O\big( \frac{1}{R_1}\big)$ only, because the second part then follows from \eqref{error:empirical approximation} and the simple inequality $\sup\lvert Z-\dbwidehat{Z_L}\rvert\leq \sup\lvert Z-\widehat{Z_L}\rvert+\sup\lvert \widehat{Z_L}-\dbwidehat{Z_L}\rvert$. First we present two lemmas.
\begin{lemma}\label{var_error}
% Consider the maximal second moment $\mathcal M$ defined in \eqref{2moment}. If Assumption \ref{balanced data} holds and $\mathcal M<\infty$ (a weaker condition than Assumption \ref{8th moment}), then as $n\to\infty$ we have
% \begin{equation}
%     \mathbb E\big[\sup_{(\mathbf w_1,\ldots,\mathbf w_m)\in \mathcal U_{\alpha}}\abs{\mathbb E_{\mathbf w_1,\ldots,\mathbf w_m}[h(\mathbf X_1,\ldots,\mathbf X_m)]-\mathbb E_{P_1,\ldots,P_m}[h(\mathbf X_1,\ldots,\mathbf X_m)]}^2\big]= O\big(\frac{1}{n}\big).\label{1st_moment_error}
% \end{equation}
Under Assumptions \ref{balanced data} and \ref{8th moment}, as $n\to\infty$ for $k=1,2,3,4$ we have
\begin{equation}
    \mathbb E\big[\sup_{(\mathbf w_1,\ldots,\mathbf w_m)\in \mathcal U_{\alpha}}\abs{\mathbb E_{\mathbf w_1,\ldots,\mathbf w_m}[h^k(\mathbf X_1,\ldots,\mathbf X_m)]-\mathbb E_{P_1,\ldots,P_m}[h^k(\mathbf X_1,\ldots,\mathbf X_m)]}^2\big]= O\big(\frac{1}{ n}\big).\label{moments_error}
\end{equation}
In particular for $k=1,2,3,4$ it holds
\begin{equation}
    \mathbb E\big[\sup_{(\mathbf w_1,\ldots,\mathbf w_m)\in \mathcal U_{\alpha}}\abs{\mathbb E_{\mathbf w_1,\ldots,\mathbf w_m}[h^k(\mathbf X_1,\ldots,\mathbf X_m)]}^2\big]= O(1).\label{moments_bound}
\end{equation}
% If Assumption \ref{8th moment} is further assumed, we have
% \begin{equation}
% \mathbb E\big[\sup_{(\mathbf w_1,\ldots,\mathbf w_m)\in\mathcal U_{\alpha}}\lvert\sigma^2_{\mathbf w_1,\ldots,\mathbf w_m}-\sigma^2\rvert\big]=O\big(\frac{1}{\sqrt n}\big)\label{2nd_moment_error}
% \end{equation}
% where $\sigma^2=\mathrm{Var}_{P_1,\ldots,P_m}(h(\mathbf X_1,\ldots,\mathbf X_m))$ and $\sigma^2_{\mathbf w_1,\ldots,\mathbf w_m}=\mathrm{Var}_{\mathbf w_1,\ldots,\mathbf w_m}(h(\mathbf X_1,\ldots,\mathbf X_m))$.
\end{lemma}
\proof{Proof of Lemma \ref{var_error}.}\eqref{moments_error} is argued using the proof of Proposition \ref{linearization error}. Note that the proof for Proposition \ref{linearization error} goes through as long as the maximal second moment $\mathcal M$ defined in \eqref{2moment} is finite, a weaker condition than Assumption \ref{8th moment}. In particular, Assumption \ref{8th moment} remains valid if the target performance measure is changed to $\mathbb E_{P_1,\ldots,P_m}[h^k(\mathbf X_1,\ldots,\mathbf X_m)]$ for $k=2,3,4$, except that the maximal second moment $\mathcal M$ has to be replaced by the $4$-th, $6$-th and $8$-th moments respectively. Below we will argue for the case $k=1$ only, and the cases $k=2,3,4$ follow from the same reasoning. Recall the expansion \eqref{expansion}. The term with $d=0$ is simply $Z^*$. The argument leading to the bound \eqref{generalbd} works for all $d\geq 1$, and hence \eqref{generalbd} is valid for all $d\geq 1$. The leading remainders with $d=1$ then give rise to the order $O(1/n)$ in \eqref{moments_error}, as opposed to $d=2$ giving the order $O(1/n^2)$ in \eqref{error:truth approximation}.

To prove \eqref{moments_bound}, use the inequality
\begin{eqnarray*}
    &&\sup_{(\mathbf w_1,\ldots,\mathbf w_m)\in \mathcal U_{\alpha}}\abs{\mathbb E_{\mathbf w_1,\ldots,\mathbf w_m}[h^k(\mathbf X_1,\ldots,\mathbf X_m)]}\\
    &\leq& \sup_{(\mathbf w_1,\ldots,\mathbf w_m)\in \mathcal U_{\alpha}}\abs{\mathbb E_{\mathbf w_1,\ldots,\mathbf w_m}[h^k(\mathbf X_1,\ldots,\mathbf X_m)]-\mathbb E_{P_1,\ldots,P_m}[h^k(\mathbf X_1,\ldots,\mathbf X_m)]}+\mathbb E_{P_1,\ldots,P_m}[h^k(\mathbf X_1,\ldots,\mathbf X_m)]
\end{eqnarray*}
and the Minkowski inequality.\hfill\Halmos
\endproof

\begin{lemma}\label{graderror}
Under Assumptions \ref{balanced data} and \ref{8th moment}, as the input data size $n\to\infty$, the gradient estimator $\dbhat{G}_i(X_{i,j})$ in \eqref{gradientest} satisfies
\begin{equation*}
\mathbb E\big[ \sum_{i=1}^m\frac{1}{n_i^2}\sum_{j=1}^{n_i}(\dbhat{G}_i(X_{i,j})-\hat{G}_i(X_{i,j}))^2\big]=O\big(\frac{1}{R_1}\big)
\end{equation*}
% \begin{equation*}
% \mathbb E\big[ \sum_{i=1}^m\frac{1}{n_i^2}\sum_{j=1}^{n_i}(\dbhat{G}_i(X_{i,j})-\hat{G}_i(X_{i,j}))^2\big]\leq\frac{\sigma^2T}{R_1},
% \end{equation*}
where the expectation is taken with respect to the joint randomness from both input data and simulation.
\end{lemma}
\proof{Proof of Lemma \ref{graderror}.}We first note that due to the symmetry between the i.i.d.~data
\begin{equation*}
    \mathbb E\big[ \sum_{i=1}^m\frac{1}{n_i^2}\sum_{j=1}^{n_i}(\dbhat{G}_i(X_{i,j})-\hat{G}_i(X_{i,j}))^2\big]= \sum_{i=1}^m\frac{1}{n_i}\mathbb E\big[(\dbhat{G}_i(X_{i,1})-\hat{G}_i(X_{i,1}))^2\big],
\end{equation*}
and therefore it suffices to bound each $\mathbb E\big[(\dbhat{G}_i(X_{i,1})-\hat{G}_i(X_{i,1}))^2\big]$. Since $\dbhat{G}_i(X_{i,1})$ differs from the unbiased sample covariance by only a factor of $\frac{R_1-1}{R_1}$, its bias (conditioned on the input data) can be easily identified as $\hat{G}_i(X_{i,1})/R_1$. By the variance formula for the unbiased sample covariance, and suppressing the arguments in $h$ for notational simplicity, we have
\begin{eqnarray*}
&&\mathrm{Var}_{\xi_1}\big(\dbhat{G}_i(X_{i,1})\big)\\
&=&\frac{(R_1-1)^2}{R_1^3}\Big(\mathbb E_{\xi_1}\big[(h-\mathbb E_{\xi_1}[h])^2(S_{i,1}(\mathbf X_i))^2\big]+\frac{1}{R_1-1}\mathrm{Var}_{\xi_1}(h)\mathrm{Var}_{\xi_1}(S_{i,1}(\mathbf X_i))-\frac{R_1-2}{R_1-1}(\hat{G}_i(X_{i,1}))^2\Big).
\end{eqnarray*}
Hence the mean squared error
\begin{eqnarray}
\nonumber\mathbb E_{\xi_1}[(\dbhat{G}_i(X_{i,1})-\hat{G}_i(X_{i,1}))^2]&=&\mathrm{Var}_{\xi_1}\big(\dbhat{G}_i(X_{i,j})\big)+\big(\frac{\hat{G}_i(X_{i,1})}{R_1}\big)^2\\
\nonumber&\leq&\frac{1}{R_1}\mathbb E_{\xi_1}\big[(h-\mathbb E_{\xi_1}[h])^2(S_{i,1}(\mathbf X_i))^2\big]+\frac{1}{R_1^2}\mathrm{Var}_{\xi_1}(h)\mathrm{Var}_{\xi_1}(S_{i,1}(\mathbf X_i))\\
&\leq&\frac{1}{R_1}\mathbb E_{\xi_1}\big[(h-\mathbb E_{\xi_1}[h])^2(S_{i,1}(\mathbf X_i))^2\big]+\frac{n_iT_i}{R_1^2}\mathrm{Var}_{\xi_1}(h).\label{gradmse}
\end{eqnarray}
To tackle the first term in \eqref{gradmse}
\begin{eqnarray*}
&&\mathbb E_{\xi_1}\big[(h-\mathbb E_{\xi_1}[h])^2(S_{i,1}(\mathbf X_i))^2\big]\\
&=&\mathbb E_{\xi_1}\brac{(h-\mathbb E_{\xi_1}[h])^2\prth{T_i^2+n_i^2\prth{\sum_{t=1}^{T_i}\mathbf{1}\{X_i(t)=X_{i,1}\}}^2-2T_in_i\sum_{t=1}^{T_i}\mathbf{1}\{X_i(t)=X_{i,1}\}}}\\
&\leq&T_i^2\mathrm{Var}_{\xi_1}(h)+\mathbb E_{\xi_1} \brac{(h-\mathbb E_{\xi_1}[h])^2n_i^2\prth{\sum_{t=1}^{T_i}\mathbf{1}\{X_i(t)=X_{i,1}\}}^2}\\
&\leq&T_i^2\mathrm{Var}_{\xi_1}(h)+\mathbb E_{\xi_1} \brac{2h^2n_i^2\prth{\sum_{t=1}^{T_i}\mathbf{1}\{X_i(t)=X_{i,1}\}}^2}+\mathbb E_{\xi_1} \brac{2(\mathbb E_{\xi_1}[h])^2n_i^2\prth{\sum_{t=1}^{T_i}\mathbf{1}\{X_i(t)=X_{i,1}\}}^2}\\
&=&T_i^2\mathrm{Var}_{\xi_1}(h)+2n_i^2\mathbb E_{\xi_1} \brac{h^2\prth{\sum_{s, t=1}^{T_i}\mathbf{1}\{X_i(t)=X_i(s)=X_{i,1}\}}}+2(T_in_i+T_i(T_i-1))(\mathbb E_{\xi_1}[h])^2\\
&\leq&T_i^2\mathrm{Var}_{\xi_1}(h)+2n_i^2\sum_{s,t=1}^{T_i}\mathbb E_{\xi_1}[h^2\cdot\mathbf{1}\{X_i(t)=X_{i}(s)=X_{i,1}\}]+2(T_in_i+T_i^2)(\mathbb E_{\xi_1}[h])^2\\
&=&T_i^2\mathrm{Var}_{\xi_1}(h)+2(T_in_i+T_i^2)(\mathbb E_{\xi_1}[h])^2+2n_i\sum_{t=1}^{T_i}\mathbb E_{\xi_1}[h^2\vert X_i(t)=X_{i,1}]+\\
&&\hspace{5ex}2\sum_{s\neq t}\mathbb E_{\xi_1}[h^2\vert X_i(t)=X_i(s)=X_{i,1}].
% &=&n_i^2T_i\mathrm{Var}_{\xi}[h]+n_i\sum_{s\neq t}\mathbb E_{\xi_1}[(h-\mathbb E_{\xi_1}[h])^2\vert X_i(t)=X_{i}(s)]\\
% &\leq &n_i^2T_i\mathrm{Var}_{\xi}[h]+2n_i\sum_{s\neq t}\mathbb E_{\xi_1}[(h-Z^*)^2\vert X_i(t)=X_{i}(s)]+2n_iT_i(T_i-1)(\mathbb E_{\xi_1}[h]-Z^*)^2
\end{eqnarray*}
% where in the first inequality we leave out the negative term $-T_i^2n_i\mathrm{Var}_{\xi}[h]$ and in the second we use $(h-\mathbb E_{\xi_1}[h])^2\leq 2(h-Z^*)^2+2(\mathbb E_{\xi_1}[h]-Z^*)^2$ with $Z^*$ being the true performance measure $Z(P_1,\ldots,P_m)$. 
Using the notation in Assumption \ref{8th moment}, we can rewrite each conditional expectation $\mathbb E_{\xi_1}[h^2\vert X_i(t)=X_{i,1}]$ as
$$\mathbb E_{\xi_1}[h^2\vert X_i(t)=X_{i,1}]=\frac{1}{n_i^{T_i-1}\prod_{i'\neq i}n_{i'}^{T_{i'}}}\sum_{I_1,\ldots,I_m\text{ such that }I_i(t)=1}h^2(\mathbf X_{1,I_1},\ldots,\mathbf X_{m,I_m})$$
Therefore under Assumption \ref{8th moment} we have
\begin{equation*}
    \mathbb E_{D}\big[\mathbb E_{\xi_1}[h^2\vert X_i(t)=X_{i,1}]\big]\leq \mathcal M
\end{equation*}
where $\mathcal M$ is the maximal second moment defined in \eqref{2moment}. The same reasoning gives $\mathbb E_{D}\big[\mathbb E_{\xi_1}[h^2\vert X_i(t)=X_i(s)=X_{i,1}]\big]\leq \mathcal M$. Also note that $\mathrm{Var}_{\xi_1}(h)\leq \mathbb E_{\xi_1}[h^2]$ and $(\mathbb E_{\xi_1}[h])^2\leq \mathbb E_{\xi_1}[h^2]$ by Jensen's inequality.
Hence by \eqref{moments_bound} with $k=2$ from Lemma \eqref{var_error} it holds that $\mathbb E_{D}\big[\mathrm{Var}_{\xi_1}(h)\big]=O(1)$ and $\mathbb E_{D}\big[(\mathbb E_{\xi_1}[h])^2\big]=O(1)$.

% One can rewrite each conditional expectation as
% \begin{equation*}
% \mathbb E_{\xi_1}[(h-Z^*)^2\vert X_i(t)=X_{i}(s)]=\mathbb E_{\xi_1}[(h_{i}^{s,t}-Z^*)^2]
% \end{equation*}
% where $h_{i}^{s,t}=h(\mathbf X_{1,I_1},\ldots,\mathbf X_{m,I_m})$ with $I_r=(1,\ldots,T_r)$ for $r\neq i$ and $I_i=(1,\ldots,\max(s,t)-1,\min(s,t),\max(s,t),\ldots,T_i-1)$. 
% Apply the result \eqref{1st_moment_error} in Lemma \ref{var_error} to $(h_{i}^{s,t}-Z^*)^2$ and $h$ respectively to get
% \begin{eqnarray*}
% &&\big\lvert\mathbb E_{D}\big[\mathbb E_{\xi_1}[(h_{i}^{s,t}-Z^*)^2]\big]-\mathbb E_{P_1,\ldots,P_m}[(h_{i}^{s,t}-Z^*)^2]\big\rvert\\
% &\leq& \sqrt{\mathbb E_{D}\big[\lvert\mathbb E_{\xi_1}[(h_{i}^{s,t}-Z^*)^2]-\mathbb E_{P_1,\ldots,P_m}[(h_{i}^{s,t}-Z^*)^2]\rvert^2\big]}= O\big(\frac{1}{\sqrt n}\big),\\
% &&\mathbb E_{D}\big[(\mathbb E_{\xi_1}[h]-Z^*)^2\big]=O\big(\frac{1}{n}\big),
% \end{eqnarray*}
% where $\mathbb E_{D}$ denotes expectation taken with respect to the randomness in input data. 

% The variance convergence result \eqref{2nd_moment_error} in Lemma \ref{var_error} indicates that $\mathbb E_{D}[\mathrm{Var}_{\xi}[h]]=\sigma^2+O\big(\frac{1}{\sqrt n}\big)$.

Now we take expectation of \eqref{gradmse} with respect to the input data and use the upper bounds derived above to get
\begin{eqnarray*}
\mathbb E \big[(\dbhat{G}_i(X_{i,1})-\hat{G}_i(X_{i,1}))^2\big]&=&\mathbb E_{D}\bigg[\mathbb E_{\xi_1} \big[(\dbhat{G}_i(X_{i,1})-\hat{G}_i(X_{i,1}))^2\big]\bigg]\\
&=&\frac{1}{R_1}\mathbb E_{D}\big[\mathbb E_{\xi_1}\big[(h-\mathbb E_{\xi_1}[h])^2(S_{i,1}(\mathbf X_i))^2\big]\big]+\frac{n_iT_i}{R_1^2}O(1)\\
&=&\frac{1}{R_1}(T_i^2O(1)+(T_in_i+T_i^2)O(1)+O(n_iT_i)+O(T_i^2))+O\big(\frac{n_iT_i}{R_1^2}\big)\\
&=&O\big(\frac{T_i^2}{R_1}+\frac{n_iT_i}{R_1}+\frac{n_iT_i}{R_1^2}\big)\\
&=&O\big(\frac{n_i}{R_1}\big)\text{\ \ since each $T_i$ is treated as constant}.
\end{eqnarray*}
Dividing each side by $n_i$ and summing up over $i=1,\ldots,m$ gives the bound $O(1/R_1)$.\hfill\Halmos
\endproof

Now we can prove Proposition \ref{empirical_error}. We bound the maximal deviation as follows
\begin{eqnarray}
\notag&&\sup_{(\mathbf w_1,\ldots,\mathbf w_m)\in \mathcal U_{\alpha}}\abs{\widehat{Z_L}(\mathbf w_1,\ldots,\mathbf w_m)-\dbwidehat{Z_L}(\mathbf w_1,\ldots,\mathbf w_m)}\\
&\leq&\sup_{(\mathbf w_1,\ldots,\mathbf w_m)\in \mathcal U_{\alpha}}\abs{\sum_{i=1}^m\sum_{j=1}^{n_i}(\dbhat{G}_i(X_{i,j})-\hat{G}_i(X_{i,j}))w_{i,j}}+\abs{Z(\hat P_1,\ldots,\hat P_m)-\hat Z(\hat P_1,\ldots,\hat P_m)}.\label{sup_decompose}
\end{eqnarray}
On one hand, using conditioning and the moment bound \eqref{moments_bound} with $k=2$ from Lemma \ref{var_error}, we bound the second moment of the second term in \eqref{sup_decompose} as
\begin{eqnarray*}
    \mathbb E\big[\big\lvert Z(\hat P_1,\ldots,\hat P_m)-\hat Z(\hat P_1,\ldots,\hat P_m)\big\rvert^2\big]&=&\frac{1}{R_1}\mathbb E_{D}\big[\mathrm{Var}_{\xi_1}(h)\big]\\
    &\leq&\frac{1}{R_1}\mathbb E_{D}\big[\mathbb E_{\xi_1}[h^2]\big]\\
    &=&O\big(\frac{1}{R_1}\big).
\end{eqnarray*}
On the other hand, letting $Q_i^1=\hat P_i$ in Proposition \ref{derivative} reveals that $\sum_{j=1}^{n_i}\hat{G}_i(X_{i,j})=0$ for all $i$. Note that the estimator \eqref{gradientest} also has this property, i.e.~$\sum_{j=1}^{n_i}\dbhat{G}_i(X_{i,j})=0$ for all $i$. Hence the first term in \eqref{sup_decompose} can be bounded as
\begin{eqnarray*}
&&\sup_{(\mathbf w_1,\ldots,\mathbf w_m)\in \mathcal U_{\alpha}}\abs{\sum_{i=1}^m\sum_{j=1}^{n_i}(\dbhat{G}_i(X_{i,j})-\hat{G}_i(X_{i,j}))w_{i,j}}\\
&=&\sup_{(\mathbf w_1,\ldots,\mathbf w_m)\in \mathcal U_{\alpha}}\abs{\sum_{i=1}^m\sum_{j=1}^{n_i}(\dbhat{G}_i(X_{i,j})-\hat{G}_i(X_{i,j}))(w_{i,j}-\frac{1}{n_i})}\text{\ \ by }\sum_{j=1}^{n_i}\hat{G}_i(X_{i,j})=\sum_{j=1}^{n_i}\dbhat{G}_i(X_{i,j})=0\\
&=&\sup_{(\mathbf w_1,\ldots,\mathbf w_m)\in \mathcal U_{\alpha}}\abs{\sum_{i=1}^m\sum_{j=1}^{n_i}\frac{1}{n_i}(\dbhat{G}_i(X_{i,j})-\hat{G}_i(X_{i,j}))\cdot n_i(w_{i,j}-\frac{1}{n_i})}\\
&\leq&\sup_{(\mathbf w_1,\ldots,\mathbf w_m)\in \mathcal U_{\alpha}}\sqrt{\sum_{i=1}^m\sum_{j=1}^{n_i}\frac{1}{n_i^2}(\dbhat{G}_i(X_{i,j})-\hat{G}_i(X_{i,j}))^2\sum_{i=1}^m\sum_{j=1}^{n_i}n_i^2(w_{i,j}-\frac{1}{n_i})^2}\\
&\leq&\sqrt{u(\alpha)^2\mathcal X_{1,1-\alpha}^2\sum_{i=1}^m\sum_{j=1}^{n_i}\frac{1}{n_i^2}(\dbhat{G}_i(X_{i,j})-\hat{G}_i(X_{i,j}))^2}\text{\ \ by Lemma \ref{l2upbound}}.
\end{eqnarray*}
After combining the above bounds, the desired conclusion follows from an application of the Minkowski inequality to \eqref{sup_decompose} and using Lemma \ref{graderror}.\hfill\Halmos
\endproof

\proof{Proof of Theorem \ref{mainresult}.}In the proof of Theorem \ref{ideal CI}, if we replace the linear approximation $Z_L$ by $\dbwidehat{Z_L}$ then by exactly the same argument we have the following counterpart of \eqref{error_lowerCB} where on one hand
\begin{equation}\label{error_lowerCB_simulated}
    \big\lvert\mathscr L-\dbwidehat{Z_L}(\mathbf w_1^{\min},\ldots,\mathbf w_m^{\min})\big\rvert\leq \sup_{(\mathbf w_1,\ldots,\mathbf w_m)\in\mathcal U_{\alpha}}\big\lvert Z(\mathbf w_1,\ldots,\mathbf w_m)-\dbwidehat{Z_L}(\mathbf w_1,\ldots,\mathbf w_m)\big\rvert.
\end{equation}
On the other hand the following bound trivially holds
\begin{equation*}
\lvert Z^{\min}-\dbwidehat{Z_L}(\mathbf w_1^{\min},\ldots,\mathbf w_m^{\min})\rvert\leq \sup_{(\mathbf w_1,\ldots,\mathbf w_m)\in\mathcal U_{\alpha}}\big\lvert Z(\mathbf w_1,\ldots,\mathbf w_m)-\dbwidehat{Z_L}(\mathbf w_1,\ldots,\mathbf w_m)\big\rvert.
\end{equation*}
Therefore
\begin{equation*}
    \big\lvert\mathscr L-Z^{\min}\big\rvert\leq 2\sup_{(\mathbf w_1,\ldots,\mathbf w_m)\in\mathcal U_{\alpha}}\big\lvert Z(\mathbf w_1,\ldots,\mathbf w_m)-\dbwidehat{Z_L}(\mathbf w_1,\ldots,\mathbf w_m)\big\rvert.
\end{equation*}
The desired conclusion for $Z^{\min}$ then immediately follows from the maximal deviation result \eqref{error:linear approximation empirical simulated} in Proposition \ref{empirical_error}. The conclusion for $Z^{\max}$ can be established similarly.\hfill\Halmos
\endproof

% This is described in the following result (see Section \ref{proof:section 4.5} of the Appendix for the proof):
% are already able to establish a way of computing a valid CI. Instead of dealing with the theoretically perfect but practically inviable formulation \eqref{ciopt2}, one can  Note that the empirical output $Z(\hat{P}_1,\ldots,\hat{P}_m)$ in $\dbwidehat{Z_L}$ is also expensive to exactly evaluate, but can be well estimated by averaging the $R_1$ replications generated in the influence function estimator \eqref{gradientest}. Therefore there will be one more source of error in the CI bounds. 
The following result presents an alternate CI constructed directly from a linear approximation that is discussed at the end of Section \ref{sec:nonconvexity}.

\begin{theorem}\label{algoerror1}
Suppose Assumptions \ref{balanced data}, \ref{non-degeneracy} and \ref{8th moment} hold. Replace the outputs in Step 3 of Algorithm \ref{algo1} by
$$
L =\hat{Z}(\hat{P}_1,\ldots,\hat{P}_m)+\sum_{i=1}^{m}\sum_{j=1}^{n_i}\hat{\vphantom{\rule[4pt]{1pt}{5.5pt}}\smash{\hat{G}}}_{i}(X_{i,j})w_{i,j}^{\min},\ U =\hat{Z}(\hat{P}_1,\ldots,\hat{P}_m)+\sum_{i=1}^{m}\sum_{j=1}^{n_i}\hat{\vphantom{\rule[4pt]{1pt}{5.5pt}}\smash{\hat{G}}}_{i}(X_{i,j})w_{i,j}^{\max},$$
where $\hat{Z}(\hat{P}_1,\ldots,\hat{P}_m)$ is the same sample mean from Step 1. Then as $n\to\infty$ and $R_1\to\infty$
\begin{equation*}
\mathbb E[(L-\mathscr L)^2]=O\big(\frac{1}{n^2}+ \frac{1}{R_1}\big),\ \mathbb E[(U-\mathscr U)^2]=O\big(\frac{1}{n^2}+ \frac{1}{R_1}\big)
% \lim_{n_i\to\infty}P\prth{\mathscr L_{\alpha}+E_l\leq Z(P_1,\ldots,P_m)\leq \mathscr U_{\alpha}+E_u}= 1-\alpha,
\end{equation*}
where $\mathscr L,\mathscr U$ are the ideal confidence bounds defined in \eqref{original_pair} and the expectation is taken with respect to the joint randomness of the data and the simulation. Moreover, if $R_1$ satisfies $\frac{R_1}{n}\to\infty$ then
\begin{equation*}
    \lim_{n\to\infty,\frac{R_1}{n}\to\infty}P(L\leq Z^*\leq U)=1-\alpha.
\end{equation*}
\end{theorem}

\proof{Proof of Theorem \ref{algoerror1}.}The bound \eqref{error_lowerCB_simulated} derived in the proof of Theorem \ref{mainresult} is exactly $\lvert L-\mathscr L\rvert\leq \sup_{(\mathbf w_1,\ldots,\mathbf w_m)\in\mathcal U_{\alpha}}\big\lvert Z(\mathbf w_1,\ldots,\mathbf w_m)-\dbwidehat{Z_L}(\mathbf w_1,\ldots,\mathbf w_m)\big\rvert$. A direct application of result \eqref{error:linear approximation empirical simulated} from Proposition \ref{empirical_error} then gives $\mathbb E[(L-\mathscr L)^2]=O(1/n^2+1/R_1)$. The error bound of $U$ with respect to $\mathscr U$ can be obtained similarly. To establish the asymptotic exactness of $[L,U]$ when $R_1$ grows at a faster rate than $n$, note that when $R_1/n\to\infty$ we have $1/R_1=o(1/n)$ hence $L-\mathscr L=o_p(1/\sqrt n)$ and $U-\mathscr U=o_p(1/\sqrt n)$. In this case the representation \eqref{equivalence to normal CI:ideal CI} holds for $L,U$ as well. The rest of the proof is the same as that of Theorem \ref{ideal CI}.\hfill\Halmos
\endproof
% Proposition By using Proposition \ref{empirical_error} and \ref{linearize_error}, the lower bound error $E_l=Z(\mathbf w_1^{\min},\ldots,\mathbf w_m^{\min})-\mathcal L_{\alpha}$ now satisfies
% \begin{align*}
% \mathbb E[\lvert E_l\rvert]&\leq 2\mathbb E\big[\sup_{\mathcal U_{\alpha}}\lvert Z_L-\dbwidehat{Z_L}\rvert\big]+\mathbb E\big[\sup_{\mathcal U_{\alpha}}\lvert Z-Z_L\rvert\big]\\
% &\leq 2\mathbb E\big[\sup_{\mathcal U_{\alpha}}\lvert Z_L-\widehat{Z_L}\rvert+\sup_{\mathcal U_{\alpha}}\lvert \widehat{Z_L}-\dbwidehat{Z_L}\rvert\big]+\mathbb E\big[\sup_{\mathcal U_{\alpha}}\lvert Z-Z_L\rvert\big]\\
% &\leq 2\sqrt{\mathbb E\big[\sup_{\mathcal U_{\alpha}}\lvert Z_L-\widehat{Z_L}\rvert^2\big]}+2\sqrt{\mathbb E\big[\sup_{\mathcal U_{\alpha}}\lvert \widehat{Z_L}-\dbwidehat{Z_L}\rvert^2\big]}+\sqrt{\mathbb E\big[\sup_{\mathcal U_{\alpha}}\lvert Z-Z_L\rvert^2\big]}\\
% &\lessapprox C(\alpha)\Big(\sqrt{\sum_{i\neq i'}\frac{M_{i,i'}}{n_in_{i'}}+\sum_{i}\frac{M_i}{n_i^2}}+\sqrt{\frac{\sigma^2T}{R_1}}\Big).
% \end{align*}
% Again the same bound holds for $E_u=Z(\mathbf w_1^{\max},\ldots,\mathbf w_m^{\max})-\mathcal U_{\alpha}$.
\section{Proofs of Results in Section \ref{sec:evaluation}}\label{proof:section 4.6}
\proof{Proof of Proposition \ref{algoerror}.}
% We denote by $\mathbb E_{\xi_2}$ the expectation with respect to the $2R_2$ simulation runs in Step 3 conditioned on both the data and the $R_1$ simulation runs in Step 1. We also adopt the notations $\sigma_{\mathbf w_1,\ldots,\mathbf w_m}^2,\sigma^2$ from Lemma \ref{var_error} and $Z^{\min},Z^{\max}$ from Theorem \ref{mainresult}. Then 
We have
\begin{eqnarray*}
&&\mathbb E[(\mathscr L^{BEL}-\mathscr L)^2]\\
&=&\mathbb E[(\mathscr L^{BEL}-Z^{\min})^2]+2\mathbb E[(\mathscr L^{BEL}-Z^{\min})(Z^{\min}-\mathscr L)]+\mathbb E[(Z^{\min}-\mathscr L)^2]\\
&=&\mathbb E_{D,\xi_1}[\mathbb E_{\xi_2}[(\mathscr L^{BEL}-Z^{\min})^2]]+2\mathbb E_{D,\xi_1}\big[\mathbb E_{\xi_2}[(\mathscr L^{BEL}-Z^{\min})(Z^{\min}-\mathscr L)]\big]+O\big(\frac{1}{n^2}+\frac{1}{R_1}\big)\text{\ \ by Theorem \ref{mainresult}}\\
&=&\mathbb E_{D,\xi_1}\big[\frac{1}{R_2}\sigma_{\min}^2\big]+\mathbb E_{D,\xi_1}\big[(Z^{\min}-\mathscr L)\mathbb E_{\xi_2}[(\mathscr L^{BEL}-Z^{\min})]\big]+O\big(\frac{1}{n^2}+\frac{1}{R_1}\big)\\
&\leq&\frac{1}{R_2}\mathbb E_{D,\xi_1}\big[\mathbb E_{\mathbf w_1^{\min},\ldots,\mathbf w_m^{\min}}[h^2(\mathbf X_1,\ldots,\mathbf X_m)]\big]+0+O\big(\frac{1}{n^2}+\frac{1}{R_1}\big)\\
&=&O\big(\frac{1}{R_2}\big)+O\big(\frac{1}{n^2}+\frac{1}{R_1}\big)\text{\ \ by \eqref{moments_bound} with $k=2$ from Lemma \ref{var_error}}\\
&=&O\big(\frac{1}{n^2}+\frac{1}{R_1}+\frac{1}{R_2}\big).
\end{eqnarray*}
The bound for $\mathbb E[(\mathscr U^{BEL}-\mathscr U)^2]$ can be obtained by the same argument.\hfill\Halmos
\endproof

\proof{Proof of Proposition \ref{algo2/3 representation}.}We first establish the representations for $\mathscr L^{EEL},\mathscr U^{EEL}$. The uniform moment convergence result \eqref{moments_error} from Lemma \ref{var_error} implies that $\sigma_{\min}^2=\sigma^2+O_p(1/\sqrt n)$. By calculating the variance of sample variance, one can show that the $\hat \sigma_{\min}^2$ in Algorithm \ref{algo2} satisfies $\mathbb E_{\xi_2}[(\hat \sigma_{\min}^2-\sigma_{\min}^2)^2]\leq C\mathbb E_{\mathbf w_1^{\min},\ldots,\mathbf w_m^{\min}}[h^4(\mathbf X_1,\ldots,\mathbf X_m)]/R_2$ for some universal constant $C$. Using the result \eqref{moments_bound} with $k=4$ we have $\mathbb E\big[\mathbb E_{\mathbf w_1^{\min},\ldots,\mathbf w_m^{\min}}[h^4(\mathbf X_1,\ldots,\mathbf X_m)]\big]=O(1)$. Therefore we have $\mathbb E[(\hat \sigma_{\min}^2-\sigma_{\min}^2)^2]=\mathbb E\big[\mathbb E_{\xi_2}[(\hat \sigma_{\min}^2-\sigma_{\min}^2)^2]\big]=O(1/R_2)$, whereby
\begin{equation}\label{var_min_converge}
\hat \sigma_{\min}^2=\sigma_{\min}^2+O_p\big(\frac{1}{\sqrt{R_2}}\big)=\sigma^2+O_p\big(\frac{1}{\sqrt n}\big)+O_p\big(\frac{1}{\sqrt{R_2}}\big)=\sigma^2+o_p(1).
\end{equation}
Now the lower confidence bound $\mathscr L^{EEL}$ from Algorithm \ref{algo2} can be expressed as
\begin{eqnarray*}
\mathscr L^{EEL}&=&\hat Z^{\min}-z_{1-\alpha/2}\frac{\hat\sigma_{\min}}{\sqrt{R_2}}\\
&=&\mathscr L+(Z^{\min}-\mathscr L)+\hat Z^{\min}-Z^{\min}-z_{1-\alpha/2}\frac{\hat\sigma_{\min}}{\sqrt{R_2}}\\
&=&\mathscr L+O_p\big(\frac{1}{n}+\frac{1}{\sqrt{R_1}}\big)+\hat Z^{\min}-Z^{\min}-z_{1-\alpha/2}\frac{\sigma}{\sqrt{R_2}}+o_p\big(\frac{1}{\sqrt{R_2}}\big)\text{\ \ by \eqref{var_min_converge} and Theorem \ref{mainresult}}\\
&=&\mathscr L+\hat Z^{\min}-Z^{\min}-z_{1-\alpha/2}\frac{\sigma}{\sqrt{R_2}}+o_p\big(\frac{1}{\sqrt{n}}+\frac{1}{\sqrt{R_2}}\big)\text{\ \ because }\frac{R_1}{n}\to\infty\\
&=&Z^*+\sum_{i=1}^m\bar{G}_i- z_{1-\alpha/2}\sigma_I+\hat Z^{\min}-Z^{\min}-z_{1-\alpha/2}\frac{\sigma}{\sqrt{R_2}}+o_p\big(\frac{1}{\sqrt{n}}+\frac{1}{\sqrt{R_2}}\big)\text{\ \ because of \eqref{equivalence to normal CI:ideal CI}}.
\end{eqnarray*}
Rearranging the above gives the desired conclusion for $\mathscr L^{EEL}$. The representation for $\mathscr U^{EEL}$ can be obtained via a similar way.

To justify the representation for $\mathscr L^{FEL}$ and $\mathscr U^{FEL}$, we first need to establish the consistency of our input-induced variance estimate \eqref{input_var_est}. Specifically, we have:
\begin{lemma}\label{consis_varest}
Under Assumptions \ref{balanced data}, \ref{non-degeneracy} and \ref{8th moment}, as $n\to\infty$ and $R_1/n\to\infty$ the input-induced variance estimate \eqref{input_var_est} is relatively consistent, i.e., $\hat{\sigma}_I^2/\sigma_I^2\to 1$ in probability with respect to the joint randomness of both input data and simulation.
% \begin{equation*}
% \hat{\sigma}_I^2\Big/\sum_{i\in I}\frac{1}{n_i}\mathrm{Var}(G_i(X_i))\to 1\text{ in probability}.
% \end{equation*}
\end{lemma}
% \begin{lemma}\label{consis_vargrad}
% Under Assumption \ref{posvar} and Assumption \ref{4thmoment} with $k=2$, if $\min_{l\notin I}n_l=\omega(\sqrt{\min_{i\in I}n_i})$, where $I=\{i\vert \mathrm{Var}(G_i(X_i))>0\}$, then the sample influence function $\hat{G}_i(X_{i,j})$ yields a consistent variance estimate
% \begin{equation*}
% \sum_{i=1}^m\frac{1}{n_i^2}\sum_{j=1}^{n_i}\hat{G}_i(X_{i,j})^2\Big/\sum_{i\in I}\frac{1}{n_i}\mathrm{Var}(G_i(X_i))\to 1 \text{ in probability},\text{ as all }n_i\to \infty.
% \end{equation*}
% \end{lemma}
\proof{Proof of Lemma \ref{consis_varest}.}Since the input-induced variance $\sigma_I^2$ is of order $1/n$ and the strong law of large numbers ensures that $\big(\sum_{i=1}^m\sum_{j=1}^{n_i}\big(G_i(X_{i,j})\big)^2/n_i^2\big)/\sigma_I^2\to 1$ almost surely, it suffices to show
\begin{align}
    &\sum_{i=1}^m\frac{1}{n_i^2}\sum_{j=1}^{n_i}\big(\hat G_i(X_{i,j})\big)^2-\sum_{i=1}^m\frac{1}{n_i^2}\sum_{j=1}^{n_i} \big(G_i(X_{i,j})\big)^2=o_p\big(\frac{1}{n}\big),\label{input_variance_consistency1}\\
    &\hat\sigma_I^2-\sum_{i=1}^m\frac{1}{n_i^2}\sum_{j=1}^{n_i}\big(\hat G_i(X_{i,j})\big)^2=o_p\big(\frac{1}{n}\big).\label{input_variance_consistency2}
\end{align}

We bound the left hand side of \eqref{input_variance_consistency1} as
\begin{eqnarray*}
&&\big\lvert\text{left hand side of \eqref{input_variance_consistency1}}\big\rvert\\
&=&\Big\lvert\sum_{i=1}^m\frac{1}{n_i^2}\sum_{j=1}^{n_i}(2G_i(X_{i,j})(\hat G_i(X_{i,j})-G_i(X_{i,j}))+(\hat G_i(X_{i,j})-G_i(X_{i,j}))^2)\Big\rvert\\
&\leq &\sum_{i=1}^m\frac{1}{n_i^2}\sum_{j=1}^{n_i} (\hat G_i(X_{i,j})-G_i(X_{i,j}))^2+2\sqrt{\sum_{i=1}^m\frac{1}{n_i^2}\sum_{j=1}^{n_i} (G_i(X_{i,j}))^2\sum_{i=1}^m\frac{1}{n_i^2}\sum_{j=1}^{n_i}(\hat G_i(X_{i,j})-G_i(X_{i,j}))^2}.
\end{eqnarray*}
Hence it suffices to bound the error $(\hat G_i(X_{i,j})-G_i(X_{i,j}))^2$ for each $i,j$. Seeing that both $G_i$ and $\hat G_i$ take the form of a sum of conditional expectations, we can control this error via a similar analysis in proving Proposition \ref{linearization error}. In particular, for all $i,j$ we have $\mathbb E[(\hat G_i(X_{i,j})-G_i(X_{i,j}))^2]\leq C/n$ for some constant $C$ depending on $h$ (a similar observation has been proved in equation (EC.10) in Lemma EC.1 of \cite{lam2018subsampling}). Therefore $\big\lvert\text{left hand side of \eqref{input_variance_consistency1}}\big\rvert=O_p(1/n^2)+2\sqrt{O_p(1/n)O_p(1/n^2)}=O_p(1/n^{\frac{3}{2}})=o_p(1/n)$. Thus \eqref{input_variance_consistency1} follows.

\eqref{input_variance_consistency2} can be established in two steps. First we show that the bias correction term $\sum_{i=1}^m\frac{T_i\hat\sigma^2}{R_1}=o_p(1/n)$. Note that $\hat\sigma^2=\sigma^2+o_p(1)=O_p(1)$ can be proved via the same argument used to prove \eqref{var_min_converge} but with the minimal weights $\mathbf w_i^{\min},i=1,\ldots,m$ replaced by the uniform weights. When $R_1/n\to\infty$, we have each $\frac{T_i\hat\sigma^2}{R_1}=O_p(1/R_1)=o_p(1/n)$. Second, we examine the error 
% of the main term of the variance estimate
\begin{eqnarray*}
&&\Big\lvert\sum_{i=1}^m\frac{1}{n_i^2}\sum_{j=1}^{n_i}\big(\dbhat{G}_i(X_{i,j})\big)^2-\sum_{i=1}^m\frac{1}{n_i^2}\sum_{j=1}^{n_i} \big(\hat G_i(X_{i,j})\big)^2\Big\rvert\\
&\leq &\sum_{i=1}^m\frac{1}{n_i^2}\sum_{j=1}^{n_i} (\dbhat{G}_i(X_{i,j})-\hat G_i(X_{i,j}))^2+2\sqrt{\sum_{i=1}^m\frac{1}{n_i^2}\sum_{j=1}^{n_i} \big(\hat G_i(X_{i,j})\big)^2\sum_{i=1}^m\frac{1}{n_i^2}\sum_{j=1}^{n_i}(\dbhat{G}_i(X_{i,j})-\hat G_i(X_{i,j}))^2}\\
&=&O_p\big(\frac{1}{R_1}\big)+2\sqrt{O_p\big(\frac{1}{n}\big)O_p\big(\frac{1}{R_1}\big)}\text{\ \ by Lemma \ref{graderror}}\\
&=&o_p\big(\frac{1}{n}\big)+2\sqrt{O_p\big(\frac{1}{n}\big)o_p\big(\frac{1}{n}\big)}\\
&=&o_p\big(\frac{1}{n}\big).
\end{eqnarray*}
This concludes \eqref{input_variance_consistency2}.\hfill\Halmos
\endproof
Given the relative consistency of the input-induced variance estimate $\hat \sigma_I^2$ in estimating $\sigma_I^2$, if we couple the simulation runs of Algorithms \ref{algo2} and \ref{algo3}, then
\begin{eqnarray*}
\mathscr L^{FEL}&=&\mathscr L^{EEL}+z_{1-\alpha/2}\frac{\hat\sigma_{\min}}{\sqrt{R_2}}-z_{1-\alpha/2}\Big(\sqrt{\hat\sigma_I^2+\frac{\hat\sigma_{\min}^2}{R_2}}-\hat\sigma_I\Big)\\
&=&\mathscr L^{EEL}+z_{1-\alpha/2}\frac{\sigma}{\sqrt{R_2}}+o_p\big(\frac{1}{\sqrt{R_2}}\big)-z_{1-\alpha/2}\Big(\sqrt{\sigma_I^2+\frac{\sigma^2}{R_2}}-\sigma_I\Big)+o_p\big(\frac{1}{\sqrt n}+\frac{1}{\sqrt{R_2}}\big)\\
&=&\mathscr L^{EEL}-z_{1-\alpha/2}\Big(\sqrt{\sigma_I^2+\frac{\sigma^2}{R_2}}-\sigma_I-\frac{\sigma}{\sqrt{R_2}}\Big)+o_p\big(\frac{1}{\sqrt n}+\frac{1}{\sqrt{R_2}}\big)\\
&=&Z^*+\sum_{i=1}^m\bar{G}_i+\hat Z^{\min}-Z^{\min}-z_{1-\alpha/2}\sqrt{\sigma_I^2+\frac{\sigma^2}{R_2}}+o_p\big(\frac{1}{\sqrt{n}}+\frac{1}{\sqrt{R_2}}\big)
\end{eqnarray*}
where in the last equality we use the representation for $\mathscr L^{EEL}$. The representation for the upper bound $\mathscr U^{FEL}$ can be similarly obtained.\hfill\Halmos
\endproof
\section{Proofs of Proposition \ref{optroutine} and Theorems \ref{erfreeCI1}, \ref{erfreeCI2}, \ref{erfreeCI3}}\label{proof:section 3}
\proof{Proof of Proposition \ref{optroutine}.}It suffices to prove the theorem for the minimization problem. Since $w_{i,j}=\frac{1}{n_i}$ for each $i,j$ is a solution in the (relative) interior of the feasible set, Slater's conditions holds for \eqref{ciopt4min}. It is also clear, by a compactness argument, that the optimal value of the program is finite and attainable. By Corollary 28.3.1 of \cite{rockafellar2015convex}, $(\mathbf w_1^{\min},\ldots,\mathbf w_m^{\min})$ is a minimizer if and only if there exist Lagrange multipliers $\beta^*,\lambda_i^*\in \R,i=1,\ldots,m$ such that the following KKT conditions are satisfied
\begin{align*}
&2\sum_{i=1}^{m}\sum_{j=1}^{n_i}\log(n_iw_{i,j}^{\min})+\mathcal X_{1,1-\alpha}^2\geq 0,\;\beta^*\geq 0\\
&\beta^*\Big(2\sum_{i=1}^{m}\sum_{j=1}^{n_i}\log(n_iw_{i,j}^{\min})+\mathcal X_{1,1-\alpha}^2\Big)=0\\
&\sum_{j=1}^{n_i}w_{i,j}^{\min}=1\text{ for all }i=1,\ldots,m\\
&\dbhat{G}_i(X_{i,j})+\lambda_{i}^*-\frac{2\beta^*}{w_{i,j}^{\min}}=0\text{ for all }i,j.
\end{align*}
When $\dbhat{G}_{i_0}(X_{i_0,j_1})\neq \dbhat{G}_{i_0}(X_{i_0,j_2})$ for some $1\leq i_0\leq m$ and $1\leq j_1<j_2\leq n_{i_0}$, the objective is a non-constant linear function and thus any minimizer must lie on the (relative) boundary of the feasible set, i.e.~$2\sum_{i=1}^{m}\sum_{j=1}^{n_i}\log(n_iw_{i,j}^{\min})+\mathcal X_{1,1-\alpha}^2= 0$. Since the constraint $-2\sum_{i=1}^{m}\sum_{j=1}^{n_i}\log(n_iw_{i,j})\leq\mathcal X_{1,1-\alpha}^2$ is strictly convex, the minimizer must be unique. Moreover, we show that $\beta^*$ must be strictly positive in this case. Suppose $\beta^*=0$ then the last equation of KKT conditions requires $\dbhat{G}_i(X_{i,j})=-\lambda_{i}^*$ for all $i,j$, which is a contradiction. Note that the minimizer must have positive components $w_{i,j}^{\min}>0$ due to the logarithm in the constraint, hence
\begin{align}
&w_{i,j}^{\min}=\frac{2\beta^*}{\dbhat{G}_i(X_{i,j})+\lambda_i^*},\;\beta^*>0,\;\dbhat{G}_i(X_{i,j})+\lambda_i^*>0\text{ for all }i,j,\label{dual2primal}\\
&2\sum_{i=1}^{m}\sum_{j=1}^{n_i}\log \frac{2n_i\beta^*}{\dbhat{G}_i(X_{i,j})+\lambda_i^*}+\mathcal X_{1,1-\alpha}^2=0,\;\sum_{j=1}^{n_i}\frac{2\beta^*}{\dbhat{G}_i(X_{i,j})+\lambda_i^*}=1\text{ for all }i.\label{kkt}
\end{align}
To show that such $(\beta^*,\lambda_1^*,\ldots,\lambda_m^*)$ is also unique, let $i_0,j_1,j_2$ be the indices mentioned in the theorem. Then \eqref{dual2primal} stipulates $w_{i_0,j_1}^{\min}/w_{i_0,j_2}^{\min}=(\hat{\vphantom{\rule[4pt]{1pt}{5.5pt}}\smash{\hat{G}}}_{i_0,j_2}+\lambda_{i_0}^*)/(\hat{\vphantom{\rule[4pt]{1pt}{5.5pt}}\smash{\hat{G}}}_{i_0,j_1}+\lambda_{i_0}^*)$. Since the right hand side is strictly monotone in $\lambda_{i_0}^*$, the uniqueness of $w_{i,j}^{\min}$ implies the uniqueness of $\lambda_{i_0}^*$, which in turn implies the uniqueness of $\beta^*$ and other $\lambda_i^*$'s due to the second equation of line \eqref{kkt}.

We further show that $\beta^*$ must lie in the interval given in the proposition. We first argue that there is at least one $i\in\{1,\ldots,m\}$ such that
\begin{equation}\label{ineq:lambda*}
    \frac{\min_j\dbhat{G}_{i}(X_{i,j})+\lambda_{i}^*}{\max_j\dbhat{G}_{i}(X_{i,j})+\lambda_{i}^*}<e^{-\frac{\mathcal X^2_{1,1-\alpha}}{2N}}.
\end{equation}
Suppose $(\min_j\dbhat{G}_{i}(X_{i,j})+\lambda_{i}^*)/(\max_j\dbhat{G}_{i}(X_{i,j})+\lambda_{i}^*)\geq e^{-\frac{\mathcal X^2_{1,1-\alpha}}{2N}}$ for all $i$, then the equation $\sum_{j=1}^{n_i}2\beta^*/(\dbhat{G}_{i}(X_{i,j})+\lambda_{i}^*)=1$ implies that $2\beta^*/(\dbhat{G}_{i}(X_{i,j})+\lambda_{i}^*)\geq \frac{1}{n_i}e^{-\frac{\mathcal X^2_{1,1-\alpha}}{2N}}$ for all $i,j$ and the inequality must be strict for some $i,j$ because $e^{-\frac{\mathcal X^2_{1,1-\alpha}}{2N}}<1$. Therefore
\begin{equation*}
    2\sum_{i=1}^{m}\sum_{j=1}^{n_i}\log \frac{2n_i\beta^*}{\dbhat{G}_i(X_{i,j})+\lambda_i^*}+\mathcal X^2_{1,1-\alpha}>-2\sum_{i=1}^{m}\sum_{j=1}^{n_i}\frac{\mathcal X^2_{1,1-\alpha}}{2N}+\mathcal X^2_{1,1-\alpha}=0
\end{equation*}
which contradicts \eqref{kkt}. Now let $\lambda_{i'}^*$ be a multiplier that satisfies \eqref{ineq:lambda*}. Rearranging \eqref{ineq:lambda*} gives
\begin{equation}\label{upper:lambda*}
    \lambda_{i'}^*<\frac{e^{-\frac{\mathcal X^2_{1,1-\alpha}}{2N}}\max_j\dbhat{G}_{i'}(X_{i',j})-\min_j\dbhat{G}_{i'}(X_{i',j})}{1-e^{-\frac{\mathcal X^2_{1,1-\alpha}}{2N}}}.
\end{equation}
Hence
\begin{eqnarray*}
1=\sum_{j=1}^{n_{i'}}\frac{2\beta^*}{\dbhat{G}_{i'}(X_{i',j})+\lambda_{i'}^*}&\geq&\frac{2n_{i'}\beta^*}{\max_j\dbhat{G}_{i'}(X_{i',j})+\lambda_{i'}^*}\\
&>&\frac{2n_{i'}\beta^*(1-e^{-\frac{\mathcal X^2_{1,1-\alpha}}{2N}})}{\max_j\dbhat{G}_{i'}(X_{i',j})-\min_j\dbhat{G}_{i'}(X_{i',j})}\text{\ \ by using the upper bound \eqref{upper:lambda*}}\\
&\geq&\frac{2\min_in_i\beta^*(1-e^{-\frac{\mathcal X^2_{1,1-\alpha}}{2N}})}{\max\{\max_j\dbhat{G}_{i}(X_{i,j})-\min_j\dbhat{G}_{i}(X_{i,j})\vert i=1,\ldots,m\}}.
\end{eqnarray*}
Rearranging the above inequality gives the desired upper bound for $\beta^*$.

If $\dbhat{G}_i(X_{i,j})=c_i$ for some constant $c_i$, then the objective is the constant function $\sum_{i=1}^mc_i$, and any feasible solution is optimal.\hfill\Halmos
\endproof
% Up to now we have seen the asymptotic equivalence of confidence bounds based on EL and the delta method, and shown that consistent estimate of the input-induced variance is available. To finally justify the adjustment we need the following two standard results on weak convergence.
\proof{Proof of Theorem \ref{erfreeCI1}.}When $R_1/n\to\infty$ and $R_2/n\to\infty$, Proposition \ref{algoerror} stipulates that $\mathscr L^{BEL}=\mathscr L+o_p(1/\sqrt{n})$ and $\mathscr U^{BEL}=\mathscr U+o_p(1/\sqrt{n})$. Theorem \ref{ideal CI} then implies that the asymptotic representation \eqref{equivalence to normal CI:ideal CI} holds for $\mathscr L^{BEL}$ and $\mathscr U^{BEL}$. The rest of the proof is the same as that of Theorem \ref{ideal CI} from \eqref{equivalence to normal CI:ideal CI} onwards.\hfill\Halmos
\endproof

\proof{Proof of Theorems \ref{erfreeCI2} and \ref{erfreeCI3}.}For convenience, all limit statements are understood to be for $n,R_1,R_2\to\infty$ such that $\frac{R_1}{n}\to\infty,\frac{R_2}{n}\leq M$ (e.g., \eqref{joint_clt2} and \eqref{joint_clt}), unless stated otherwise. We need the Berry-Esseen Theorem stated as:
\begin{lemma}[Theorem 3.4.9 in \citealt{durrett2010probability}]\label{error_normal}
Let $\{\eta_i\}_{i=1}^{\infty}$ be a sequence of i.i.d.~random variables such that $\mathbb E[\eta_1]=0,\mathbb E[\eta_1^2]=\sigma_{\eta}^2,\mathbb E[\lvert \eta_1\rvert^3]=\rho_{\eta}<\infty$, and $S_n=\sum_{i=1}^n\eta_i/(\sigma_{\eta}\sqrt{n})$. Let $F_n(\cdot)$ be the cumulative distribution function of $S_n$. Then
\begin{equation*}
\sup_{x\in\R}\lvert F_n(x)-\Phi(x)\rvert\leq \frac{3\rho_{\eta}}{\sigma^3_{\eta}\sqrt{n}}.
\end{equation*}
\end{lemma}
% \begin{lemma}\label{error_normal}
% (Berry-Esseen bound, adapted from \cite{durrett2010probability}) Let $\{Y_i\}_{i=1}^{\infty}$ be a sequence of i.i.d.~variables such that $\mathbb E[Y_1]=0,\mathbb E[Y_1^2]=\sigma_Y^2,\mathbb E[\lvert Y_1\rvert^3]=\rho_Y<\infty$, and $S_N=\sum_{i=1}^NY_i$. Then for any $t\in \R$ and $N\geq 9\rho_Y^2t^2/(16\sigma_Y^6)$ it holds
% \begin{equation*}
% \lvert \mathbb E[e^{itS_n/(\sigma_Y\sqrt{N})}]-e^{-t^2/2}\rvert\leq \big(\frac{2\lvert t\rvert^3}{9}+\frac{t^4}{18}\big)e^{-t^2/4}\frac{3\rho_Y}{4\sigma_Y^3\sqrt{N}}.
% \end{equation*}
% \end{lemma}

We first show the following weak convergence to the joint standard normal
\begin{equation}\label{joint_clt2}
    \Big(\frac{\sum_{i=1}^m\bar{G}_i}{\sigma_I},\frac{\sqrt{R_2}(\hat{Z}^{\min}-Z^{\min})}{\sigma},\frac{\sqrt{R_2}(\hat{Z}^{\max}-Z^{\max})}{\sigma}\Big)\Rightarrow \mathcal N(\mathbf{0},\mathbf I_3).
\end{equation}
Since $\sigma^2_{\min}=\sigma^2+o_p(1)$ and $\sigma^2_{\max}=\sigma^2+o_p(1)$ as argued in \eqref{var_min_converge}, to show \eqref{joint_clt2} it suffices to show
\begin{equation}\label{joint_clt}
    \Big(\frac{\sum_{i=1}^m\bar{G}_i}{\sigma_I},\frac{\sqrt{R_2}(\hat{Z}^{\min}-Z^{\min})}{\sigma_{\min}},\frac{\sqrt{R_2}(\hat{Z}^{\max}-Z^{\max})}{\sigma_{\max}}\Big)\Rightarrow \mathcal N(\mathbf{0},\mathbf I_3)
\end{equation}
and then apply Slutsky's theorem. For any $(x,y,z)\in\R^3$, we compute the joint probability
\begin{eqnarray}
\notag&&P\Big(\frac{\sum_{i=1}^m\bar{G}_i}{\sigma_I}\leq x,\frac{\sqrt{R_2}(\hat{Z}^{\min}-Z^{\min})}{\sigma_{\min}}\leq y,\frac{\sqrt{R_2}(\hat{Z}^{\max}-Z^{\max})}{\sigma_{\max}}\leq z\Big)\\
\notag&=&\mathbb E\brac{\mathbf{1}\set{\frac{\sum_{i=1}^m\bar{G}_i}{\sigma_I}\leq x}\cdot\mathbf{1}\set{\frac{\sqrt{R_2}(\hat{Z}^{\min}-Z^{\min})}{\sigma_{\min}}\leq y}\cdot\mathbf{1}\set{\frac{\sqrt{R_2}(\hat{Z}^{\max}-Z^{\max})}{\sigma_{\max}}\leq z}}\\
\notag&=&\mathbb E_{D,\xi_1}\brac{\mathbf{1}\set{\frac{\sum_{i=1}^m\bar{G}_i}{\sigma_I}\leq x}\mathbb E_{\xi_2}\brac{\mathbf{1}\set{\frac{\sqrt{R_2}(\hat{Z}^{\min}-Z^{\min})}{\sigma_{\min}}\leq y}}\mathbb E_{\xi_2}\brac{\mathbf{1}\set{\frac{\sqrt{R_2}(\hat{Z}^{\max}-Z^{\max})}{\sigma_{\max}}\leq z}}}\\
\notag&&\text{\ \ by conditional independence of $\hat Z^{\min}$ and $\hat Z^{\max}$ given input data and Step 1}\\
\notag&=&\mathbb E_{D,\xi_1}\brac{\mathbf{1}\set{\frac{\sum_{i=1}^m\bar{G}_i}{\sigma_I}\leq x}(\Phi(y)+\epsilon^{\min})(\Phi(z)+\epsilon^{\max})}\text{\ \ for some error terms $\epsilon^{\min}$ and $\epsilon^{\max}$}\\
&=&P\Big(\frac{\sum_{i=1}^m\bar{G}_i}{\sigma_I}\leq x\Big)\Phi(y)\Phi(z)+\mathbb E_{D,\xi_1}\brac{\mathbf{1}\set{\frac{\sum_{i=1}^m\bar{G}_i}{\sigma_I}\leq x}(\Phi(y)\epsilon^{\max}+\Phi(z)\epsilon^{\min}+\epsilon^{\min}\epsilon^{\max})}.\label{joint_probability}
\end{eqnarray}
Denoting
\begin{align*}
    \rho_{\min}&=\mathbb E_{\mathbf w_1^{\min},\ldots,\mathbf w_m^{\min}}[\lvert h(\mathbf X_1,\ldots,\mathbf X_m)-Z^{\min}\rvert^3]\\
    \rho_{\max}&=\mathbb E_{\mathbf w_1^{\max},\ldots,\mathbf w_m^{\max}}[\lvert h(\mathbf X_1,\ldots,\mathbf X_m)-Z^{\max}\rvert^3]
\end{align*}
the errors $\epsilon^{\min},\epsilon^{\max}$ then satisfy $\lvert\epsilon^{\min}\rvert\leq \min\big\{1,\frac{3\rho_{\min}}{\sigma^3_{\min}\sqrt{R_2}}\big\},\lvert\epsilon^{\max}\rvert\leq \min\big\{1,\frac{3\rho_{\max}}{\sigma^3_{\max}\sqrt{R_2}}\big\}$. On one hand \eqref{moments_bound} entails that $\rho_{\min}=O_p(1)$ and $\rho_{\max}=O_p(1)$. On the other hand, $\sigma^2_{\min}=\sigma^2+o_p(1)$ and $\sigma^2_{\max}=\sigma^2+o_p(1)$ as mentioned before. These two facts together lead to $\epsilon^{\min}=O_p(1/\sqrt{R_2})$ and $\epsilon^{\max}=O_p(1/\sqrt{R_2})$. Since both errors do not exceed $1$, by the dominated convergence theorem, the second term in \eqref{joint_probability} converges to zero asymptotically. Moreover, the probability $P\big(\sum_{i=1}^m\bar{G}_i\leq x\sigma_I\big)\to\Phi(x)$ which has been shown in \eqref{influnece_normality}. Therefore the joint probability converges to $\Phi(x)\Phi(y)\Phi(z)$, hence weak convergence \eqref{joint_clt} holds by definition.

Secondly, we prove that $[\mathscr L^{FEL},\mathscr U^{FEL}]$ is asymptotically valid, i.e., the $\liminf$ part in Theorem \ref{erfreeCI3}. The $\liminf$ result for $[\mathscr L^{EEL},\mathscr U^{EEL}]$ is then a direct consequence of $[\mathscr L^{FEL},\mathscr U^{FEL}]$ by a coupling argument as follows. If Algorithms \ref{algo2} and \ref{algo3} use the same $R_1+2R_2$ simulation runs, then the two different adjustments in Step 3 satisfy $\frac{\hat\sigma_{\min}}{\sqrt{R_2}}\geq \sqrt{\hat\sigma_I^2+\frac{\hat\sigma_{\min}^2}{R_2}}-\hat\sigma_I$ almost surely, therefore $\mathscr L^{EEL}\leq \mathscr L^{FEL}$ and $\mathscr U^{EEL}\geq \mathscr U^{FEL}$ almost surely. To proceed, we write
% Denote $Z^*=Z(P_1,\ldots,P_m)$, $Z^{\min}=Z(\mathbf w_{1}^{\min},\ldots,\mathbf w_{m}^{\min})$, and $Z^{\max}=Z(\mathbf w_{1}^{\max},\ldots,\mathbf w_{m}^{\max})$. The coverage guarantee is based on the following relation
\begin{eqnarray}
\nonumber &&P(\mathscr L^{FEL}\leq Z^*\leq \mathscr U^{FEL})\\
\nonumber &=&P(\mathscr L^{FEL}\leq Z^*)+P(Z^*\leq \mathscr U^{FEL})-P(\mathscr L^{FEL}\leq Z^*\text{ or }Z^*\leq \mathscr U^{FEL})\\
&=&P(\mathscr L^{FEL}\leq Z^*)+P(Z^*\leq \mathscr U^{FEL})-1+P(\mathscr U^{FEL}<Z^*<\mathscr L^{FEL}).\label{coverage_eqn}
\end{eqnarray}
To compute the probabilities in \eqref{coverage_eqn}, we use the representation from Proposition \ref{algo2/3 representation} to get
\begin{eqnarray*}
P(\mathscr L^{FEL}\leq Z^*)&=&P\Big(\sum_{i=1}^m\bar{G}_i+\hat Z^{\min}-Z^{\min}-z_{1-\alpha/2}\sqrt{\sigma_I^2+\frac{\sigma^2}{R_2}}+o_p\big(\frac{1}{\sqrt{n}}+\frac{1}{\sqrt{R_2}}\big)\leq 0\Big)\\
&=&P\Big(\frac{1}{\sqrt{\sigma_I^2+\sigma^2/R_2}}\big(\sum_{i=1}^m\bar{G}_i+\hat Z^{\min}-Z^{\min}\big)+o_p(1)\leq z_{1-\alpha/2}\Big)\\
&=&P\Big(\frac{\sigma_I}{\sqrt{\sigma_I^2+\sigma^2/R_2}}\frac{\sum_{i=1}^m\bar{G}_i}{\sigma_I}+\frac{\sigma/\sqrt{R_2}}{\sqrt{\sigma_I^2+\sigma^2/R_2}}\frac{\sqrt{R_2}(\hat Z^{\min}-Z^{\min})}{\sigma}+o_p(1)\leq z_{1-\alpha/2}\Big)\\
&\to& 1-\frac{\alpha}{2}.
\end{eqnarray*}
The limit here is valid because, by rewriting the last probability above as the probability of a half-space of $\R^3$ like in \eqref{ex:use uniform convergence}, we can conclude from \eqref{joint_clt2} and Lemma \ref{unif_weak} that $$\frac{\sigma_I}{\sqrt{\sigma_I^2+\sigma^2/R_2}}\frac{\sum_{i=1}^m\bar{G}_i}{\sigma_I}+\frac{\sigma/\sqrt{R_2}}{\sqrt{\sigma_I^2+\sigma^2/R_2}}\frac{\sqrt{R_2}(\hat Z^{\min}-Z^{\min})}{\sigma}\Rightarrow\mathcal N(0,1)$$
which also holds with an additional $o_p(1)$ term on the left hand side by Slutsky's Theorem. Similary, one can show that $P(\mathscr U^{FEL}\geq Z^*)\to 1-\alpha/2$. Neglecting the last probability in \eqref{coverage_eqn} gives
\begin{equation*}
    P(\mathscr L^{FEL}\leq Z^*\leq \mathscr U^{FEL})\geq P(\mathscr L^{FEL}\leq Z^*)+P(Z^*\leq \mathscr U^{FEL})-1\to 2\big(1-\frac{\alpha}{2}\big)-1=1-\alpha
\end{equation*}
from which the $\liminf$ result follows. 

Thirdly, we prove the $\limsup$ results by further analyzing the last probability in \eqref{coverage_eqn}.
% Note that the $o_p(1)$ term can be absorbed into $(X_{G},X_{\min},X_{\max})$ without affecting its convergence to standard normal $\mathcal N(\mathbf{0},\mathbf{I}_3)$, so one can neglect it when computing limits of probabilities. Although the coefficients of $X_{G},X_{\min}$ can vary as $n_i,R_2$ grows, Lemma \ref{unif_weak} saves the validity of the following limit
% \begin{align*}
% P(\mathscr L_{\alpha}^{FEL}\leq Z^*)\to P\Big(\frac{\sigma_I}{\sqrt{\sigma_I^2+\sigma^2/R_2}}Z_1+\frac{\sigma/\sqrt{R_2}}{\sqrt{\sigma_I^2+\sigma^2/R_2}}Z_2\leq z_{1-\alpha/2}\Big)=P\Big(\mathcal N(0,1)\leq z_{1-\alpha/2}\Big)=1-\frac{\alpha}{2}
% \end{align*}
% where $Z_1,Z_2$ are independent standard normal variables. Similarly $P(Z^*\leq \mathcal U_{\alpha}^{FEL})\to 1-\alpha/2$. To treat the last probability in \eqref{coverage_eqn}, let $Z_1,Z_2,Z_3$ be independent standard normal variables
Using the representation from Proposition \ref{algo2/3 representation} again we have
\begin{eqnarray*}
&&P(\mathscr U^{FEL}<Z^*<\mathscr L^{FEL})\\
&=&P\Big(\frac{\sigma_I}{\sqrt{\sigma_I^2+\sigma^2/R_2}}\frac{\sum_{i=1}^m\bar{G}_i}{\sigma_I}+\frac{\sigma/\sqrt{R_2}}{\sqrt{\sigma_I^2+\sigma^2/R_2}}\frac{\sqrt{R_2}(\hat Z^{\min}-Z^{\min})}{\sigma}+o_p(1)> z_{1-\alpha/2}\text{ and }\\
&&\hspace{5ex}-\frac{\sigma_I}{\sqrt{\sigma_I^2+\sigma^2/R_2}}\frac{\sum_{i=1}^m\bar{G}_i}{\sigma_I}-\frac{\sigma/\sqrt{R_2}}{\sqrt{\sigma_I^2+\sigma^2/R_2}}\frac{\sqrt{R_2}(\hat Z^{\max}-Z^{\max})}{\sigma}+o_p(1)> z_{1-\alpha/2}\Big)\\
&=&P\Big(\frac{\sigma_I}{\sqrt{\sigma_I^2+\sigma^2/R_2}}\frac{\sum_{i=1}^m\bar{G}_i}{\sigma_I}+\frac{\sigma/\sqrt{R_2}}{\sqrt{\sigma_I^2+\sigma^2/R_2}}\Big(\frac{\sqrt{R_2}(\hat Z^{\min}-Z^{\min})}{\sigma}+o_p(1)\Big)> z_{1-\alpha/2}\text{ and }\\
&&\hspace{5ex}-\frac{\sigma_I}{\sqrt{\sigma_I^2+\sigma^2/R_2}}\frac{\sum_{i=1}^m\bar{G}_i}{\sigma_I}-\frac{\sigma/\sqrt{R_2}}{\sqrt{\sigma_I^2+\sigma^2/R_2}}\Big(\frac{\sqrt{R_2}(\hat Z^{\max}-Z^{\max})}{\sigma}+o_p(1)\Big)> z_{1-\alpha/2}\Big)
% &=&P\Big(\tilde{Z}_1>z_{1-\alpha/2},\tilde{Z}_2>z_{1-\alpha/2}\Big)
\end{eqnarray*}
where the second equality is valid because $\frac{R_2}{n}\leq M<\infty$ implies $\frac{\sigma/\sqrt{R_2}}{\sqrt{\sigma_I^2+\sigma^2/R_2}}\geq \epsilon>0$ for some fixed constant $\epsilon$.
% , therefore on one hand
% \begin{align*}
%     \frac{\sigma/\sqrt{R_2}}{\sqrt{\sigma_I^2+\sigma^2/R_2}}\frac{\sqrt{R_2}(\hat Z^{\min}-Z^{\min})}{\sigma}+o_p(1)&=\frac{\sigma/\sqrt{R_2}}{\sqrt{\sigma_I^2+\sigma^2/R_2}}\Big(\frac{\sqrt{R_2}(\hat Z^{\min}-Z^{\min})}{\sigma}+o_p(1)\Big)\\
%     -\frac{\sigma/\sqrt{R_2}}{\sqrt{\sigma_I^2+\sigma^2/R_2}}\frac{\sqrt{R_2}(\hat Z^{\max}-Z^{\max})}{\sigma}+o_p(1)&=-\frac{\sigma/\sqrt{R_2}}{\sqrt{\sigma_I^2+\sigma^2/R_2}}\Big(\frac{\sqrt{R_2}(\hat Z^{\max}-Z^{\max})}{\sigma}+o_p(1)\Big).
% \end{align*}
By Slutsky's theorem, if the three-dimensional random vector in \eqref{joint_clt2} is contaminated by a negligible noise of size $o_p(1)$ in each component, it still converges weakly to the joint standard normal. This convergence, together with Lemma \ref{unif_weak}, leads to the following limit
\begin{equation*}
    P(\mathscr U^{FEL}<Z^*<\mathscr L^{FEL})\to P(\tilde W_1>z_{1-\alpha/2},\tilde W_2>z_{1-\alpha/2})
\end{equation*}
where $(\tilde{W}_1,\tilde{W}_2)$ is the joint normal $\mathcal N\Big(\mathbf{0},   \Big[\begin{matrix} % or pmatrix or bmatrix or Bmatrix or ...
      1 & -\rho \\
      -\rho & 1 \\
   \end{matrix}\Big]\Big)$ and $\rho=\sigma_I^2/(\sigma_I^2+\sigma^2/R_2)>0$. To compute the limit probability, note that the conditional distribution $\tilde{W}_2\vert \tilde{W}_1$ is $\mathcal N(-\rho \tilde{W}_1,1-\rho^2)$, therefore
\begin{align*}
P(\tilde{W}_1>z_{1-\alpha/2},\tilde{W}_2>z_{1-\alpha/2})=\int_{z_{1-\alpha/2}}^{\infty}\phi(x)P(\mathcal N(-\rho x,1-\rho^2)>z_{1-\alpha/2})dx\leq \frac{\alpha}{2}\int_{z_{1-\alpha/2}}^{\infty}\phi(x)dx=\frac{\alpha^2}{4}.
\end{align*}
Here $\phi$ denotes the density of the standard normal, and the inequality follows since $-\rho x<0$ and $1-\rho^2<1$ and hence $P(\mathcal N(-\rho x,1-\rho^2)>z_{1-\alpha/2})\leq P(\mathcal N(0,1)>z_{1-\alpha/2})=\alpha/2$. This establishes
\begin{equation*}
\limsup P(\mathscr U^{FEL}<Z^*<\mathscr L^{FEL})\leq \frac{\alpha^2}{4}.
\end{equation*}
Substituting it into \eqref{coverage_eqn} gives the $\limsup$ statement of Theorem \ref{erfreeCI3}.

Following the above line of analysis, the $\limsup$ statement of Theorem \ref{erfreeCI2} can be derived. We use the representation from Proposition \ref{algo2/3 representation}. Since $\sigma_I+\frac{\sigma}{\sqrt{R_2}}\leq \sqrt{2}\sqrt{\sigma_I^2+\frac{\sigma^2}{R_2}}$, we have
\begin{equation*}
    \begin{aligned}
    \mathscr L^{EEL}&\geq \tilde{\mathscr L}:=Z^*+\sum_{i=1}^m\bar{G}_i+(\hat{Z}^{\min}-Z^{\min})- \sqrt{2}z_{1-\alpha/2}\sqrt{\sigma_I^2+\frac{\sigma^2}{R_2}}+o_p\big(\frac{1}{\sqrt{n}}+\frac{1}{\sqrt{R_2}}\big)\\
    \mathscr U^{EEL}&\leq \tilde{\mathscr U}:= Z^*+\sum_{i=1}^m\bar{G}_i+(\hat{Z}^{\max}-Z^{\max}) +\sqrt{2}z_{1-\alpha/2}\sqrt{\sigma_I^2+\frac{\sigma^2}{R_2}}+o_p\big(\frac{1}{\sqrt{n}}+\frac{1}{\sqrt{R_2}}\big)
    \end{aligned}
\end{equation*}
almost surely, where the $o_p\big(\frac{1}{\sqrt{n}}+\frac{1}{\sqrt{R_2}}\big)$ terms are those from Proposition \ref{algo2/3 representation}. Repeating the above analysis for $\tilde{\mathscr L},\tilde{\mathscr U}$ reveals that
\begin{equation*}
    \limsup P\prth{\tilde{\mathscr L}\leq Z^*\leq \tilde{\mathscr U}}\leq 1-\tilde\alpha+\frac{\tilde\alpha^2}{4}.
\end{equation*}
% _{n,R_1,R_2\to \infty\,\text{and}\, \frac{R_1}{n}\to\infty, \frac{R_2}{n}\,\text{bounded}}
The same $\limsup$ bound then holds for $\mathscr L^{EEL},\mathscr U^{EEL}$ because $\mathscr L^{EEL}\geq\tilde{\mathscr L}$ and $\mathscr U^{EEL}\leq\tilde{\mathscr U}$.

Lastly, when $R_2$ also grows at a faster rate than $n$, the adjustments in Algorithms \ref{algo2} and \ref{algo3} relative to Algorithm \ref{algo1} are of order $o_p(1/\sqrt{n})$, i.e., $\hat\sigma_{\min}/\sqrt{R_2}=o_p(1/\sqrt{n}),\hat\sigma_{\max}/\sqrt{R_2}=o_p(1/\sqrt{n})$ and $\sqrt{\hat\sigma_I^2+\hat\sigma_{\min}^2/R_2}-\hat\sigma_I=o_p(1/\sqrt{n}),\sqrt{\hat\sigma_I^2+\hat\sigma_{\max}^2/R_2}-\hat\sigma_I=o_p(1/\sqrt{n})$. Therefore, by coupling the simulation runs in Step 3 with Algorithm \ref{algo1}, the confidence bounds from Algorithms \ref{algo2} and \ref{algo3} differ from those from Algorithm \ref{algo1} by $o_p(1/\sqrt{n})$. Using the proof for Theorem \ref{erfreeCI1} concludes asymptotic exactness.\hfill\Halmos
\endproof
\end{document}